\documentclass[11pt,a4paper]{article}

\pdfoutput=1

\usepackage{jheppub}
\usepackage[normalem]{ulem}
\usepackage{slashed}
\usepackage{subcaption}
\usepackage{orcidlink}

\usepackage{multirow}

\def\beq{\begin{equation}}
\def\eeq{\end{equation}}
\def\bea{\begin{eqnarray}}
\def\eea{\end{eqnarray}}
\def\nn{\nonumber}

\def\roughly#1{\mathrel{\raise.3ex\hbox
{$#1$\kern-.75em\lower1ex\hbox{$\sim$}}}}

\def\BDtaunu{\bar{B} \to D \tau^{-} {\bar\nu}_\tau}
\def\BDlnu{\bar{B} \to D \ell^{-} {\bar\nu}_\ell}
\def\BDstartaunu{{\bar B}^0 \to D^{*+} \tau^{-} {\bar\nu}_\tau}
\def\BDstartauN{{\bar B}^0 \to D^{*+} \tau^{-} \bar{N}}

\def\BDstarlnu{{\bar B}^0 \to D^{*+} \ell^{-} {\bar\nu}_\ell}
\def \cB{{\cal B}}

\def \Ga{\Gamma}

\def \({\left(}
\def \){\right)}
\def \[{\left[}
\def \]{\right]}

\def \nuta{\nu_\tau}

\def \vp{{\vec p}}

\def \mA0T{\mathcal{A}^{D^*}_{0, T}}

\newcommand{\Amp}{\mathcal{M}}

\newcommand{\BDstaufull}{ \bar{B} \to D^* (\to D \pi) \tau (\to \pi \nu_\tau ) \bar X}
\newcommand{\tlnn}{\tau \to \pi \nu_\tau }

\newcommand{\eds}{\epsilon_{D^*}}
\newcommand{\lds}{{\lambda_{D^*}}}
\newcommand{\ew}{\epsilon_{W}}
\newcommand{\ra}{\rightarrow}

\allowdisplaybreaks

\title{\boldmath Massive right-handed neutrinos in $\bar{B} \to D^* \tau \bar X$ decay}

\author[a]{Nilakshi Das\,\orcidlink{0000-0002-5824-3094},}
\author[b]{Alakabha Datta\,\orcidlink{0000-0001-8713-2783},}
\author[c]{Tejhas Kapoor\,\orcidlink{0000-0001-5726-3037},}
\author[d]{Danny Marfatia,}
\author[e]{and Lopamudra Mukherjee\,\orcidlink{0000-0001-8765-7563}}

\affiliation[a]{Indian Institute of Technology Gandhinagar, Department of Physics, \\ Gujarat 382355, India}
\affiliation[b]{Department of Physics and Astronomy, 108 Lewis Hall, University of Mississippi, Oxford, MS 38677-1848, USA}
\affiliation[c]{LPC Caen, Normandie Univ, ENSICAEN, UNICAEN, CNRS/IN2P3, 6 boulevard Maréchal Juin, Caen, 14050 France}
\affiliation[d]{Department of Physics and Astronomy, University of Hawaii, Honolulu, HI 96822, USA}
\affiliation[e]{Department of Physics, University of Calcutta, 92 Acharya Prafulla Chandra Road, Kolkata 700009, India}

\emailAdd{nilakshi.das@iitgn.ac.in}
\emailAdd{datta@phy.olemiss.edu}
\emailAdd{kapoor@lpccaen.in2p3.fr}
\emailAdd{dmarf8@hawaii.edu}
\emailAdd{lopamudra.physics@gmail.com}

\abstract{
We explore signatures of a massive right-handed neutrino (RHN) in angular distributions of $\BDstaufull$ decays, where $X$ is an invisible state. We assume the new physics is described by the standard model effective field theory extended with an RHN in the MeV-GeV mass range. We calculate for the first time the full differential distributions in terms of the visible final states, including the decay of the $\tau$ lepton. We evaluate the sensitivity of various distributions to the new physics operators.

}

\begin{document}
\maketitle

\section{Introduction}
The question of whether there are additional quarks and leptons beyond those in the standard model (SM) is an interesting one. These additional states could be SM singlets, and a popular example of such states are right-handed neutrinos (RHNs)~\cite{Abazajian:2012ys}. Such sterile neutrinos could play a crucial role in understanding neutrino masses and mixing, and may have masses above or below the electroweak scale. In this work, we focus on MeV-GeV mass RHNs that may be accessible in $B$ decays.
RHNs may be produced in experiments if they interact with SM particles. The simplest mechanism to generate this interaction is through mixing with the active neutrinos of the SM. However, many extensions of the SM contain RHNs that have new interactions with the SM and other new particles. A general model-independent framework to study the interactions of RHNs is the SM effective field theory (SMEFT)~\cite{Buchmuller:1985jz, Grzadkowski:2010es} extended with RHNs, called the standard model neutrino effective field theory (SMNEFT)~\cite{delAguila:2008ir, Aparici:2009fh, Bhattacharya:2015vja, Liao:2016qyd, Bischer:2019ttk}. This effective theory systematically organizes the interactions of the RHNs with the SM in terms of higher-dimensional operators, starting with operators with mass-dimension six. We assume that the effective theory at the $m_b$ scale arises from SMNEFT. This low-energy EFT facilitates the calculation of the production and decay of RHNs involving SM particles.

We focus on a RHN produced in the charged current semileptonic $B$ decays with a $\tau$ lepton in the final state. 
These decays have been of recent interest in tests of lepton flavor universality via the ratios,
\begin{eqnarray*}
\label{babarnew}
R(D) &\equiv& \frac{{\cal B}(\BDtaunu)}
{{\cal B}(\BDlnu)}\,, \quad \quad
R(D^*) \equiv \frac{{\cal B}(\BDstartaunu)}
{{\cal B}(\BDstarlnu)}\,.
\label{RDexpt}
\end{eqnarray*}
Over the years, measurements of these ratios~\cite{babaranomaly1, babaranomaly2, belleanomaly1, belleanomaly2, belleanomaly3, belleanomaly4, belle2anomaly1, lhcbanomaly1,lhcbanomaly2} have shown enhancements relative to SM expectations~\cite{hflav}, with a combined significance of $3.8 \sigma$.
This may indicate new physics (NP) that is not lepton universal. If NP allows for decays to a light RHN, the decay rate for $\bar{B} \to D^* \tau \bar X$ (where $X$ is a left- or right-handed neutrino) is always enhanced because there is no interference with the SM amplitude in the limit of vanishing active neutrino mass. This feature naturally explains the enhancement in the $R(D^{(*)})$ measurements~\cite{He:2012zp, Cvetic:2017gkt, Asadi:2018wea, Greljo:2018ogz, Babu:2018vrl, Mandal:2020htr} assuming non-universal NP, typically assumed to only affect the $\tau$ lepton.
 
In this work we explore signatures of an RHN in the angular distributions of ${\bar B}^0 \to D^{*+} \tau^{-} \bar X$. It has been shown that observables in the angular distributions, such as polarization fractions or angular asymmetries, are useful tools for testing theoretical predictions of form factors \cite{Fedele:2023ewe} and distinguishing between the SM and various types of NP scenarios \cite{Blanke:2018yud,Blanke:2019qrx,Bhattacharya:2022bdk}. 
One of the issues with a $\tau$ lepton in the final state is that unlike the lighter charged leptons, the $\tau$ is reconstructed from its decay products. Since the decay of the $\tau$ involves a neutrino, there are at least two neutrinos in the final state so that the $\tau$ rest frame cannot be fully reconstructed. Hence, the angular distributions have to be expressed in terms of the visible daughter states from the $\tau$ decay. A step in this direction was taken in Refs.~\cite{Nierste:2008qe,Bhattacharya:2020lfm,Alonso:2016gym,Bhattacharya:2024zog} where $\tau$ decay was considered for $\BDstartaunu$ with general NP without an RHN. In our analysis, for the ${\bar B}^0 \to D^{*+} \tau^{-} \bar X$ angular distributions, we will consider $\tau$ to be reconstructed via the two-body decay  $\tau \to \pi \nu_\tau$. Our method can be easily extended to other $\tau$ decays like $ \tau \to \rho \nu_\tau \to 2 \pi \nu_\tau$ and  $ \tau \to a_1 \nu_\tau \to 3 \pi \nu_\tau$. 

We calculate the complete differential decay distribution in terms of the kinematic variables of the visible final states for vector, scalar and tensor operators. We assume that the $\tau$ decay is not affected by NP so that the $\tau$ cannot decay to $N$. We present univariate distributions for several values of the NP Wilson coefficients and heavy neutrino mass. 

\par
The paper is organized in the following manner. In Section~\ref{sec:effectivehamiltonian}, we present the effective Hamiltonian describing the new physics. In Section~\ref{sec:angdistfullbdtaunu}, we calculate the complete kinematic distribution for $\BDstaufull$. 
In Section~\ref{univ}, we present the angular and $q^2$ distributions for various NP scenarios. In Section~\ref{sec:statisticalanalysis} we perform a statistical analysis to assess the sensitivity of various distributions to the NP operators.
We conclude in Section~\ref{sec:conclusions}.

\section{Effective Hamiltonian}
\label{sec:effectivehamiltonian}
The EFT at the $m_b$ scale includes the dimension-six four-fermion operators that yield $b\to c \ell \bar X$: 
\begin{align}\label{eq:effective_hamiltonian}
\begin{aligned}
\mathcal{H}_{\text{eff}} 
&= \frac{4G_F}{\sqrt{2}} V_{cb} \Big[ 
 O_{V}^{LL} 
 + C_{V}^{RR} O_{V}^{RR}  
 + C_{S}^{LR} O_{S}^{LR} 
 + C_{S}^{RR} O_{S}^{RR} 
 + C_{T}^{RR} O_{T}^{RR} \Big]\,,
 \end{aligned}
\end{align}
where
\begin{align}
\begin{aligned}
O_{V}^{LL} &= \big( \bar{c} \gamma^{\mu} P_L b \big)\big( \bar{\ell} \gamma_{\mu} \nu_{\ell} \big) \,, \\
O_{V}^{RR} &= \big( \bar{c} \gamma^{\mu} P_R b \big)\big( \bar{\ell} \gamma_{\mu} P_R N\big) \,,\\
O_{S}^{LR} &= \big( \bar{c} P_Lb \big)\big( \bar{\ell}P_R N \big) \,,\\
O_{S}^{RR} &= \big( \bar{c}P_R b \big)\big( \bar{\ell} P_R N \big) \,,\\ 
O_{T}^{RR} &= (\bar{c} \sigma^{\mu\nu} P_R b)(\bar{{\ell}} \sigma_{\mu\nu} P_R N)  \,.\\
\end{aligned}
\end{align}
and the $C$'s are the corresponding Wilson coefficients, which we take to be real.
 Note that $O_{V}^{LL}$ corresponds to the SM contribution. The operator $O_V^{LR}$ does not have a gauge-invariant dimension-six four-fermion completion in SMNEFT because the $SU(2)_L$ quark current is neutral under hypercharge whereas the lepton current $\ell_R \gamma_\mu N_R$ is not. Hence, such a structure can arise only at dimension eight through two Higgs-field insertions and is  suppressed by a factor of $v^2/\Lambda^2$ relative to the dimension-six 
 operators~\cite{Robinson:2018gza}.

\section{Differential distribution for $\BDstaufull$ 
 }
 \label{sec:angdistfullbdtaunu}
We derive the differential distribution for $\BDstaufull$, following the method of Ref.~\cite{bdltaunudatta} for $\bar{B}\to D^*(\to D\pi)\tau(\to \pi\nu_\tau)\bar{\nu}_\tau$. Since the $\tau$ decay involves neutrinos, its rest frame and helicity angles cannot be reconstructed; we therefore work in the $W$ rest frame, where the pion angles are well defined. The differential decay rate is given by
\begin{align}\label{eq:decayrategeneral}
 d\Gamma = \frac{1}{2m_B}\int d\Pi_n\,|\Amp|^2\,,
\end{align}
with $\Amp$ the amplitude and $d\Pi_n$ the $n$-body phase space. Using these elements, we obtain the full measurable angular distribution of the $B$ decay after integrating over the non-observable degrees of freedom.

\subsection{Phase space}
\label{sec:phase-space}
\begin{figure}[t]
\setlength{\unitlength}{1mm}
  \centering
  \begin{picture}(140,60)
    \put(0,-1){
      \includegraphics*[width=130mm]{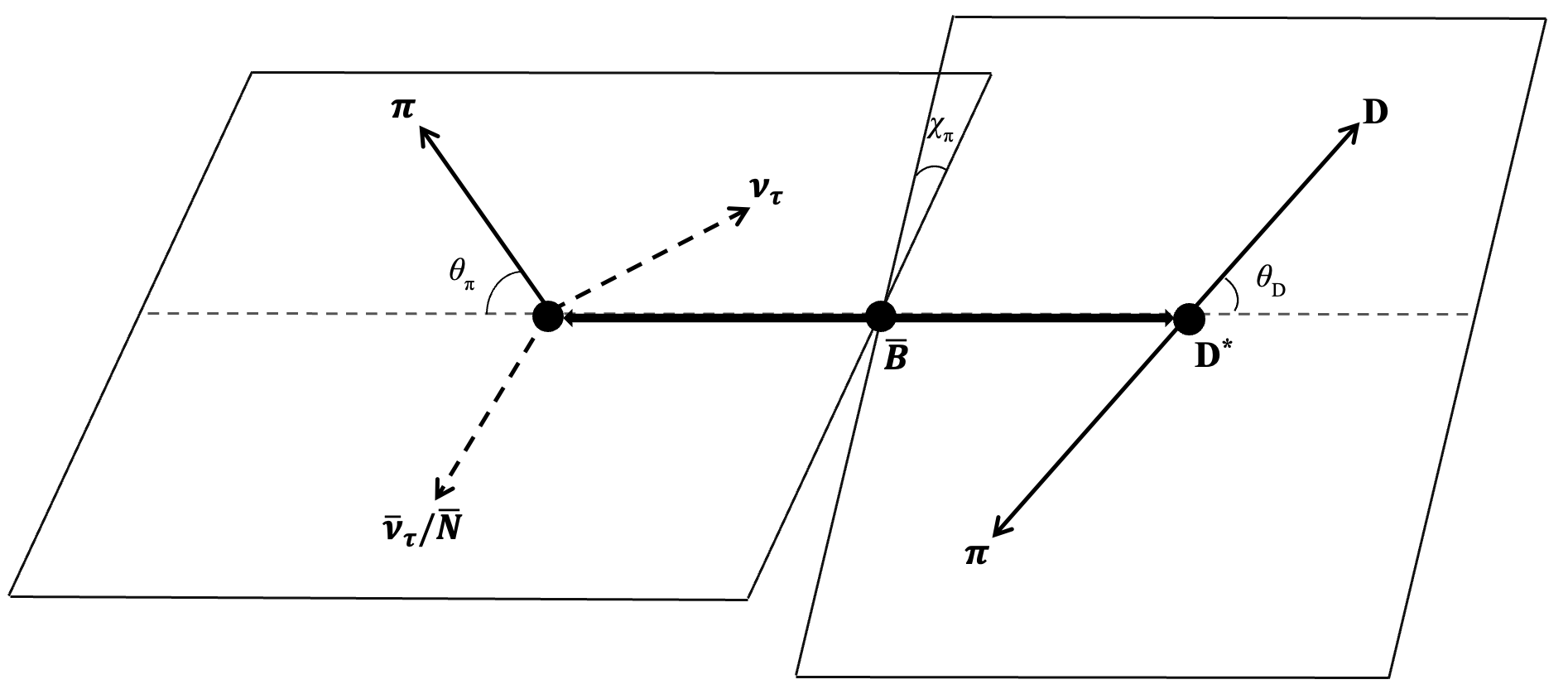}
    }
  \end{picture}
  \caption{Kinematic variables for $\BDstaufull$.  $\theta_D$ is the angle between the direction of the $D$ meson and the direction opposite to that of the $B$ meson in the $D^*$ rest frame. Similarly, $\theta_\pi$ is the angle between the direction of the $\pi$ meson and the direction opposite to that of the $B$ meson in the rest frame of the fictitious $W$. $\chi_{\pi}$ is the angle between the two decay planes. The fictitious $W$ decays to $\tau\bar{\nu}_\tau/\bar N$ and the $\tau$ decays to $\pi \nu_\tau$. 
  Note that the $\tau$ direction cannot be reconstructed. }
\label{fig:decayanglestau}
\end{figure}

The decay process $\BDstaufull$, which leads to a five-particle final state with two (unobservable) neutrinos, is illustrated in Fig.~\ref{fig:decayanglestau}. 

Conventionally, helicity angles are defined in the rest frame of the parent particle. For instance, the angle of the $D$ meson, $\theta_D$, is measured in the rest frame of its parent, the $D^*$. However, in the case of $\tlnn$, the $\tau$ direction cannot be reconstructed experimentally and the standard procedure cannot be applied. To overcome this limitation, we adopt the following strategy: the leptonic part of the decay is analyzed in the $q^2$ rest frame, which can be reconstructed from the hadronic side of the $B$ decay. The usual trick is to treat $q^2$ as the invariant mass of an off-shell particle  which helps to write the multi-body decay amplitude as products of two-body parts. In case of the SM, this fictitious particle is called a $W$ with the understanding that it is not the usual $W$ boson of the SM which has been integrated out to obtain the effective Hamiltonian. Another advantage of this formalism is that it can be easily generalized to NP operators, where the integrated mediators may be scalar, vector or tensor.
On the other hand, the hadronic part is studied in the $D^*$ rest frame. Thus, \emph{we work in the rest frame of the fictitious $W$ for the leptonic sector and in the $D^*$ rest frame for the hadronic sector}. 

 The 5-body phase space element in Eq.~\eqref{eq:decayrategeneral} is given by
\begin{align}\label{eq:phasespacerecursionrelation}
\begin{aligned}
d\Pi_5 &= \frac{dq^2}{2\pi}\frac{dp_{D^*}^2}{2\pi}\frac{dp_{\tau}^2}{2\pi} \\
&\times d\Pi_2^B (B \ra D^* W) d\Pi_2^{D^*}(D^* \ra D\pi) d\Pi_2^{W}(W \ra \tau \bar{N}) d\Pi_2^{\tau}(\tlnn)\,,
\end{aligned}
\end{align}
where $q^2 = p_W^2$, i.e., the square of the four-momentum of the fictitious $W$. Since each phase space element is a $1\to 2$ body process, consider the general decay $P_0 (p_0) \to P_1 (p_1) P_2 (p_2)$. In the $P_0$ center-of-mass frame, its phase space element can be written as~\cite{pdg24}
\begin{align}
\begin{aligned}
d\Pi_2^{P_0} &= \int \frac{d^3 p_1 }{(2\pi)^3 2 E_1 } \frac{d^3 p_2}{ (2\pi)^3 2 E_2} (2\pi)^4 \delta^4 (p_0 - p_1 - p_2) \\
&= \frac{1}{ 4 (2\pi)^2 } \int \frac{ |p_1| d |p_1| d\Omega_1}{ m_0 }  \delta\bigg( |\vec{p}_1| - \frac{\sqrt{\lambda(m_{0}^2, m_1^2, m_2^2)}}{2 m_0}\bigg)   \,,
\end{aligned}
\end{align}
where $\lambda(x,y,z) = x^2 + y^2 + z^2 - 2(xy + yz + zx)$ is the K\"all\'en function \cite{kallenfunction}.  
We use this to write the phase space elements for $B, D^*, W$ and $\tau$ decays.

The phase space of $B \to D^* W$  in the $B$ rest frame, where $q = (q_0, \vec{q})$ is the $W$ 4-momentum, is 
\begin{align}\label{eq:bphasespace}
\begin{aligned}
d\Pi_2^{B} &= \int \frac{d^3 p_{D^*}}{(2\pi)^3 2 E_{{D^*}}} \frac{d^3 q}{(2\pi)^3 2 q_0}  (2\pi)^4 \delta^4(p_B - p_{D^*} - q)\, \\
&=  \frac{1}{2(2\pi)}\int \frac{|\vec{p}_{D^*}|  d|\vec{p}_{D^*}| }{ m_B}  \delta\bigg(|\vec{p}_{D^*}| - \frac{\sqrt{\lambda(m_B^2, m_{D^*}^2,q^2)}}{2 m_B}\bigg)\,.
\end{aligned}
\end{align}

The phase space of $D^* \to D \pi $ is in the $D^*$ rest frame, 
\begin{align}\label{eq:dstarphasespace}
\begin{aligned}
d\Pi_2^{D^*} &= \int \frac{d^3 p_D }{(2\pi)^3 2 E_D } \frac{d^3 p_{\pi}}{ (2\pi)^3 2 E_{\pi}} (2\pi)^4 \delta^4 (p_{D^*} - p_{D} - p_{\pi}) \\
&= \frac{1}{ 4 (2\pi) } \int \frac{ |\vec{p}_D| d |\vec{p}_D| d\cos\theta_D}{  m_{D^*} }  \delta\bigg( |\vec{p}_D| - \frac{\sqrt{\lambda(m_{D^*}^2, m_D^2, m_{\pi}^2)}}{2 m_{D^*}}\bigg)\,,
\end{aligned}
\end{align}
where we have integrated the azimuthal angle $\chi_D$, and $\theta_D$ is shown in Fig.~\ref{fig:decayanglestau}.

The phase space of $W \to \tau \bar{N}$ is evaluated in the $W$ rest frame, where its four-momentum is $q = (\sqrt{q^2},0,0,0)$:
\begin{align}
\begin{aligned}
d\Pi_2^W &= \int \frac{d^3p_{\tau}}{(2\pi)^3 2 E_{\tau}} \frac{d^3p_{\bar{N}}}{(2\pi)^3 2 E_{\bar{N}}} (2\pi)^4 \delta^4 (q - p_{\tau} - p_{\bar{N}}) \\
&=\frac{1}{4 {(2\pi)}^2} \int \frac{|\vec{p}_{\tau}|  d|\vec{p}_{\tau}| d\Omega_{\tau}}{\sqrt{q^2}}
\delta\bigg( |\vec{p}_{\tau}| - \frac{ \sqrt{\lambda(q^2,m_{\tau}^2, m_N^2)}}{ 2\sqrt{q^2}}\bigg)\,. 
\end{aligned}  
\end{align}                       
Measuring angles from the $\pi$ direction, the differential angle element $d\Omega_\tau$ becomes $d\Omega_{\tau\pi}$, and we rewrite the phase space element as
\begin{align}\label{eq:wphasespace}
\begin{aligned}
d\Pi_2^W =\frac{1}{4 {(2\pi)}^2} \int \frac{|\vec{p}_{\tau}|  d|\vec{p}_{\tau}| d\cos\theta_{\tau\pi} d\chi_{\tau\pi}}{\sqrt{q^2}}
\delta\bigg( |\vec{p}_{\tau}| - \frac{ \sqrt{\lambda(q^2,m_{\tau}^2, m_N^2)}}{ 2\sqrt{q^2}}\bigg)\,. 
\end{aligned}
\end{align}
We will find this transformation to be useful in the subsequent steps. 

Since the $\tau$ rest frame is unknown, we calculate the phase space in the $W$ rest frame: 
\begin{align}
\begin{aligned}\label{eq:tauphasespace}
   d\Pi_2^\tau &= \int \frac{d^3p_{\pi}}{(2\pi)^3 2 E_{\pi}} \frac{d^3p_{\bar{\nu}_\tau}}{(2\pi)^3 2 E_{\bar{\nu}_\tau}} (2\pi)^4 \delta^4 (p_\tau - p_{\pi} - p_{\bar{\nu}_\tau}) \\
   &= \frac{1}{4 {(2\pi)}^2} \int \frac{d^3 p_{\pi}}{E_{\pi} |\vec{p}_\tau -p_{\pi}| } \delta(E_{\tau} - E_{\pi} - |\vec{p}_\tau -\vec{p}_{\pi}|)\\
   &= \frac{1}{4 {(2\pi)}^2}  \int   \frac{d{E_{\pi}} d\cos\theta_\pi d\chi_\pi }{|\vec{p}_{\tau}|} \delta\Big( \cos\theta_{\tau\pi} - \frac{2 E_{\tau} E_{\pi} -m_{\tau}^2 -m_{\pi}^2}{2 
|\vec{p}_{\tau}| |\vec{p}_{\pi}|}\Big) \,.
\end{aligned}
\end{align}
Notice that the delta function above helps integration over  $\cos\theta_{\tau\pi}$ in Eq.~\eqref{eq:wphasespace} (and $|\vec{p}_\tau|$ cancels too). Multiplying Eq.~\eqref{eq:wphasespace} with Eq.~\eqref{eq:tauphasespace} gives
\begin{align}\label{eq:wandtauphasespace}
d\Pi_2^W  d\Pi_2^\tau &=  \frac{1}{ {(4\pi)}^4} \int  \frac{d|\vec{p}_{\tau}|}{\sqrt{q^2}} d\cos{\theta}_{\tau\pi}
d\chi_{\tau\pi} dE_{\pi} d\cos\theta_{\pi}d\chi_{\pi}\,\\
&\delta\Bigg( |\vec{p}_{\tau}| - \frac{ \sqrt{\lambda(q^2,m_{\tau}^2, m_N^2)}}{ 2\sqrt{q^2}}\Bigg) 
\delta\Big( \cos\theta_{\tau\pi} - \frac{2 E_{\tau} E_{\pi} -m_{\tau}^2 -m_{\pi}^2}{2 
|\vec{p}_{\tau}| |\vec{p}_{\pi}|}\Big) 
\end{align}
Finally, by putting Eqs.~\eqref{eq:bphasespace},~\eqref{eq:dstarphasespace},~\eqref{eq:wandtauphasespace} in Eq.~\eqref{eq:phasespacerecursionrelation}, we get the total phase space element $d\Pi_5$, which when inserted in Eq.~\eqref{eq:decayrategeneral} yields the differential decay rate, 
\begin{align}\label{eq:decayratewithphasespace}
\begin{aligned}
\frac{d\Gamma}{dq^2 dE_\pi d\cos\theta_D d\cos\theta_\pi d\chi_\pi} = \int \frac{dp_{D^*}^2 dp_\tau^2 |\vec{p}_{D^*}||\vec{p}_D|d\chi_{\tau\pi}}{2^{17}\pi^9 m_B^2 m_{D^*}\sqrt{q^2}} |\Amp|^2 \,,
\end{aligned}
\end{align}
where
\beq
E_\tau =\left[\frac{\lambda(q^2,m_\tau^2,m_N^2)}{4q^2}
+m_\tau^2\right]^{1/2}\,,
\qquad
|\vec p_\tau| =
\frac{\sqrt{\lambda(q^2,m_\tau^2,m_N^2)}}{2\sqrt{q^2}}\,.
\eeq
Also,
\beq
\cos\theta_{\tau\pi} =
\frac{2E_\tau E_\pi-m_\tau^2-m_\pi^2}
{2|\vec p_\tau||\vec p_\pi|}\,,
\label{cth}
\eeq
and the kinematical range of $E_\pi$ corresponding to  $-1 \leq \cos\theta_{\tau\pi} \leq 1$ is
\beq
E_\pi^{\max,\min} =
\frac{
E_\tau(m_\pi^2+m_\tau^2)
\pm
|\vec p_\tau|(m_\tau^2-m_\pi^2)
}{2 m_\tau^2}\,.
\eeq

\subsection{Decay amplitude }
\label{sec:amplitudecalculation}
The total decay amplitude is given by the sum of the SM and NP contributions, $\Amp = \Amp^{\rm SM} + \Amp^{\rm NP}$, with no interference between them.  The SM calculation can be found in Ref.~\cite{Bhattacharya:2020lfm}. Here, we present the calculation for the NP amplitude. 

\begin{figure}[t]
\setlength{\unitlength}{1mm}
\centering
\includegraphics*[width=110mm]{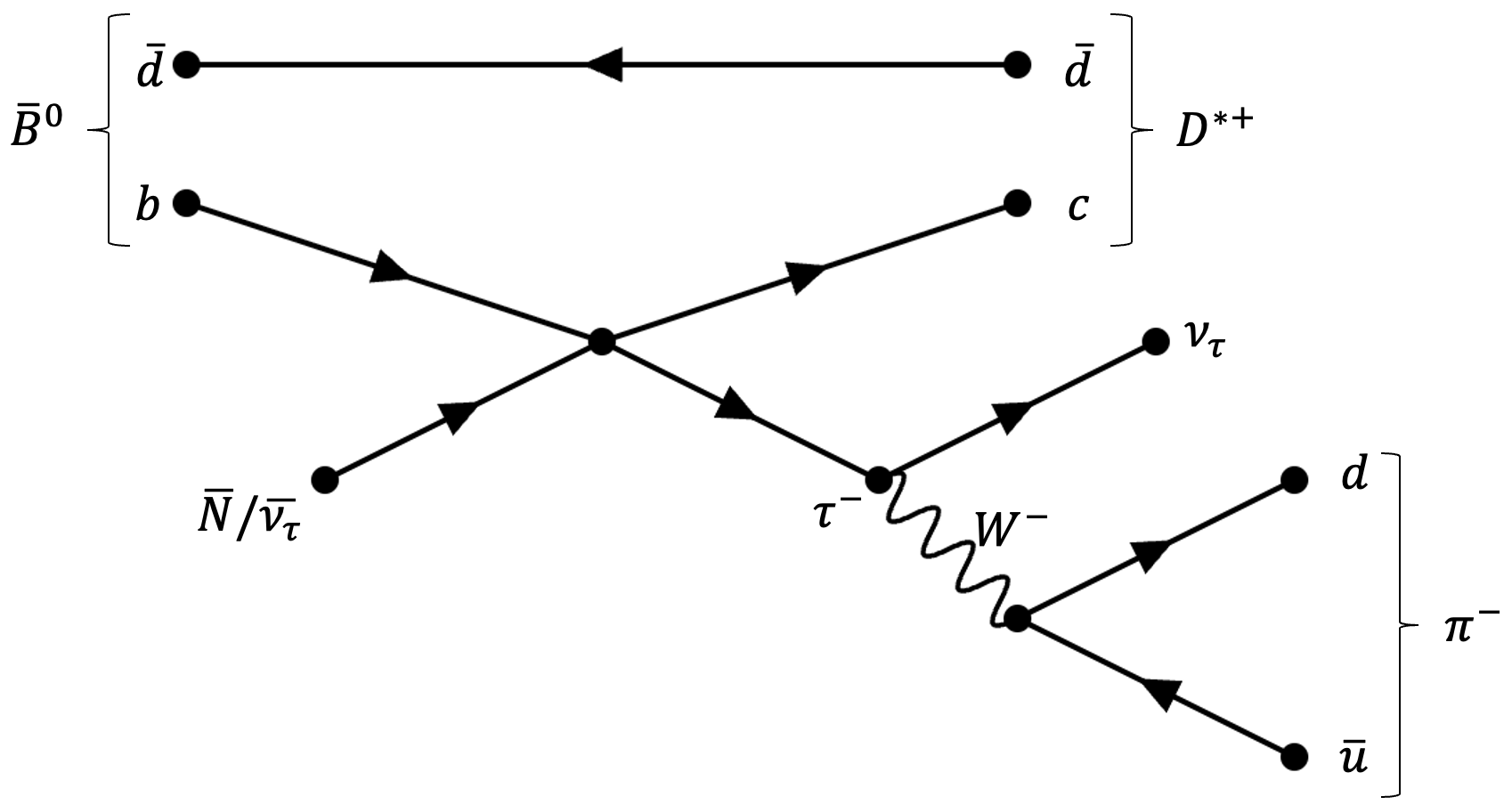}
\caption{Feynman diagram for $\bar{B}^0 \to D^{*+} \tau^-\bar{X}$, with $\tau^- \to \pi^- \nu_\tau$. }
\label{fig:tautomunuquark}
\end{figure}

The process, $\bar{B}^0 \to D^{*+} W^-(\to \tau^-\bar{N})$ with $\tau^- \to \pi^- \nu_\tau$, is shown in Fig.~\ref{fig:decayanglestau}, and the quark-level diagram with a $d$-quark spectator is shown in Fig.~\ref{fig:tautomunuquark}. The corresponding amplitude is given by

\begin{align}
\begin{aligned}\label{eq:amplitude}
   \mathcal{M}^{\rm NP} &= \sum\limits_{\lambda_{D^*}} \frac{G_F V_{cb}}{\sqrt{2}}
    \left\langle D\pi|D^*(\lambda_{D^*})
   \right\rangle \left\langle D^*(\lambda_{D^*})| \bar{c}\gamma^{\mu} 
   (1-\gamma^5)b|\bar{B} \right\rangle  (\bar{u}_{\tau}\gamma_{\mu} (1+\gamma^5) v_{{\bar{N}}} ) \\  
   & \times 
   \frac{G_F V_{ud}^*}{\sqrt{2}}  
   (\bar{u}_{\nu_\tau} \gamma_{\rho} (1-\gamma^5) u_{\tau})  \left\langle \pi|\bar{d}\gamma^\rho (1-
   \gamma^5) u|0
   \right\rangle\,.
    \\
\end{aligned}
\end{align}
The first and second lines are the amplitudes for $\BDstartauN$ and $\tau^- \to \pi^- \nu_\tau$, respectively, and the pion decay constant 
$f_\pi$ is given by
\begin{align}\label{eq:piondecayconst}
i f_\pi p_\pi^{\rho} = \left\langle \pi|\bar{d}\gamma^\rho (1-\gamma^5) u|0 \right\rangle\,.
\end{align}
Putting Eq.~\eqref{eq:piondecayconst} in Eq.~\eqref{eq:amplitude}, and rearranging to separate out the hadronic and leptonic parts of the amplitude, we get
\begin{align}
\begin{aligned}
   \mathcal{M^{\rm NP}} &= \sum\limits_{\lambda_{D^*}} \frac{G_F V_{cb}}{\sqrt{2}} 
    \Big\{ \left\langle D\pi|D^*(\lambda_{D^*})
   \right\rangle \left\langle D^*(\lambda_{D^*})| \bar{c}\gamma^{\mu} 
   (1-\gamma^5)b|\bar{B} \right\rangle  \Big\} \\  
   & \times 
   \frac{G_F V_{ud}^*}{\sqrt{2}}  \Big\{  
   (\bar{u}_{\nu_\tau} \gamma_{\rho} (1-\gamma^5) u_{\tau}) (\bar{u}_{\tau}\gamma_{\mu} (1+\gamma^5) v_{{\bar{N}}} )i f_{\pi} p_{\pi}^{\rho}\Big\}\,.
    \\
\end{aligned}    
\end{align}
With the definitions,
\begin{align}\label{eq:bdlnuampdefs}
P(\lambda_{D^*}) &\equiv \eds^{\alpha} (\lds)(p_D)_{\alpha} = \frac{1}{2 g_{D^*D\pi}} \langle D \pi | D^*(\lds)  \rangle\,, \\
\eds^{*\beta}(\lds) T^{V_L}_{\beta \mu} &\equiv \langle D^*(\lds) | \bar{c} \gamma_{\mu} (1-\gamma^5) b | \bar{B} \rangle\,,  \\
R_{VA}^\mu  &\equiv (\bar{u}_{\nu_\tau} \gamma_{\rho} (1-\gamma^5) u_{\tau}) (\bar{u}_{\tau}\gamma^{\mu} (1+\gamma^5) v_{{\bar{N}}} ) p_{\pi}^{\rho} \,,
\end{align}
and writing the $D^*$ and $\tau$ propagators explicitly,
\begin{align}
\sum_{\lds} \eds^{\alpha}(\lds) \eds^{* \beta}(\lds) \rightarrow \frac{\sum_{\lds} \eds^{\alpha}(\lds) \eds^{* \beta}(\lds)}{p^2_{D^*} - m^2_{D^*} + i m_{D^*} \Gamma_{D^*}}, \quad\quad u_{\tau} \bar{u}_{\tau} \rightarrow \frac{\slashed{p}_\tau + m_\tau}{p^2_\tau - m^2_\tau + i m_\tau \Gamma_\tau}\,,
\end{align}
the amplitude becomes
\begin{align}
\begin{aligned}
\Amp^{\rm NP} &=  iG_F^2 V_{cb} V_{ud}^* f_{\pi} g_{D^*D\pi}  \sum_{\lds} P_D(\lds) \eds^{*\beta}(\lds) T_{\beta \mu}^{V_L} R ^{\mu}_{VA}  \\
&\times \frac{1}{p^2_\tau - m^2_\tau + i m_\tau \Gamma_\tau} \, \frac{1}{p^2_{D^*} - m^2_{D^*} + i m_{D^*} \Gamma_{D^*}}\,.
\end{aligned}
\end{align}
The product of the hadronic matrix element $\eds^{\beta}(\lds)T^{V_L}_{\beta\mu}$ and the leptonic factor $R^{\mu}_{VA}$ can be written as a product of Lorentz-invariant objects by utilizing the completeness relation for the $W$ polarization vectors,
\begin{align}\label{eq:completenessrelation}
\sum_{mn} \ew^{*\mu}(m) \ew^{\mu'}(n)g_{mn} = g^{\mu \mu'},
\end{align}
where $g_{mn}$ is the metric tensor and $m,n$ label the $W$ polarizations. For a $W$ moving in the $-z$ direction in the $B$ rest frame we define,
\begin{align} \label{eq:W_polarization_vectors}
\begin{aligned}
\ew^{\mu} (\pm) &= \frac{1}{\sqrt{2}} (0,\pm 1, -i,0)\,, \\
\ew^{\mu} (0) &= \frac{1}{\sqrt{q^2}} (|\vec{q}|,0,0,-q_0)\,, \\
\ew^{\mu} (t) &= \frac{1}{\sqrt{q^2}} (q_0,0,0,-|\vec{q}|)\,,
\end{aligned}
\end{align}
where $q_0$ and $\vec q$ are the energy and momentum of the $W$ in the $B$ rest frame. The hadronic and leptonic parts then separate as
\begin{align}
\begin{aligned}
\eds^{*\beta}(\lds) T_{\beta}^{\mu, V_L} R_{\mu, VA} &= \eds^{*\beta}(\lds) T_{\beta \mu'}^{V_L}g^{\mu'\mu} R_{\mu,VA} \\
&=\sum_{m}  \underbrace{\eds^{*\beta}(\lds) T_{\beta \mu'}^{V_L} \ew^{*\mu'}(m)}_{H_{ V_L,m}^{\lambda_{D^*}}} g_{mm} \underbrace{\ew^{\mu}(m)R_{\mu, VA}}_{{L}^{V}_m}\,,
\end{aligned}
\end{align}
where $H$ and $L$ denote the hadronic and leptonic helicity amplitudes, defined as projections in the $W$ polarization basis. Since both are Lorentz invariant, they can be computed in any frame. An advantage of this formalism is that helicity amplitudes for any NP operator in Eq.~\eqref{eq:effective_hamiltonian} can be obtained directly by replacing the operator in the matrix element as follows:
\begin{align}\label{eq:lephelamp}
\begin{aligned}
{L}^{V}_m &\equiv 2\ew^{\mu}(m) (\bar{u}_\tau \slashed{p}_\pi m_{\tau} \gamma_{\mu}   (1+\gamma^5) v_{\bar{N}})\,,  \\
{L}^{S} &\equiv   2(\bar{u}_\tau \slashed{p}_{\pi} \slashed{p}_{\tau}  (1+\gamma^5) v_{\bar{N}})\,,  \\
{L}^{T}_{mn} &\equiv - 2 i  \ew^{\mu}(m) \ew^{\nu}(n) 
(\bar{u}_\tau \slashed{p}_{\pi} \slashed{p}_{\tau} \sigma_{\mu\nu}  (1+\gamma^5) v_{\bar{N}} ) \,,
\end{aligned}
\end{align}
\begin{align}\label{eq:hadhelamp}
\begin{aligned}
 H_{V_L,m}^{\lds}(q^2)
 &\equiv  \ew^{*\mu}(m) \langle D^* (p_{D^*},\eds \left(\lds)\right)| \bar c \gamma_\mu(1-\gamma^5) b| \bar B (p_B)\rangle\,,\\
 H_{V_R,m}^{\lds}(q^2)
 &\equiv  \ew^{*\mu}(m) \langle D^* (p_{D^*},\eds \left(\lds)\right)| \bar c \gamma_\mu(1+\gamma^5) b| \bar B (p_B)\rangle\,,\\
 H_{S_L}^{\lds}(q^2)
 &\equiv \langle D^* (p_{D^*},\eds \left(\lds)\right)| \bar c (1 - 
 \gamma^5)  b| \bar B (p_B)\rangle\,,\\
 H_{S_R}^{\lds}(q^2)
 &\equiv \langle D^* (p_{D^*},\eds \left(\lds)\right)| \bar c (1 + 
 \gamma^5) b| \bar B (p_B)\rangle\,,\\
 H_{ T,mn}^{\lds}(q^2)
 &\equiv i \ew^{*\mu}(m) \ew^{*\nu}(n) \langle D^* (p_{D^*},\eds \left(\lds)\right)| \bar c \sigma_{\mu\nu}(1-\gamma^5) b| \bar B (p_B)\rangle\,.
\end{aligned}
\end{align}
The factor $i$ in the hadronic tensor helicity amplitude is a conventional choice that ensures that the amplitude is real. To compensate this, a factor $-i$ is introduced in the leptonic tensor amplitude. Expressions for the hadronic helicity amplitudes in terms of form factors are provided in Appendix~\ref{app:tradformfactors}. The scalar parts of $H_{S_L}^{\lds}$ and $H_{S_R}^{\lds}$ vanish~\cite{tensorff} so that $H_{S_L}^{\lds} = -H_{S_R}^{\lds}$. 

The squared amplitude in terms of helicity amplitudes is given by
\begin{align}\label{eq:ampsqv1}
\begin{aligned}
|\Amp^{\rm NP}|^2 &=  G_F^4 |V_{cb}|^2|V_{ud}^*|^2 f_{\pi}^2 g_{D^*D\pi}^2   \frac{1}{\Gamma_\tau m_\tau} \pi \delta(p_\tau^2 - m_\tau^2) \frac{1}{\Gamma_{D^*} m_{D^*}} \pi \delta(p_{D^*}^2 - m_{D^*}^2) \\
&\times \Bigg| \sum_{\lds} P(\lds) \bigg[C_{S}^{LR} H_{S_L}^{\lds}  {L}^{S}  + C_{S}^{RR} H_{S_R}^{\lds}  {L}^{S}  + C_{V}^{RR} \sum_{m} g_{mm} H_{ V_R,m}^{\lds}  {L}^{V}_m \\
&  +C_T^{RR} \sum_{m,n}  g_{mm}g_{nn} H_{{T},mn}^{\lds}  {L}^{T}_{mn} \bigg] \Bigg|^2.
\end{aligned}
\end{align}
Since the branching fractions for $\tau\to\pi\nuta$ and
$D^*\to D\pi$ in the SM are
\beq
\cB(\tau\to\pi\nuta) = \frac{G^2_F \, |V_{ud}|^2 \, f^2_\pi}{16\pi\, m_\tau\,
\Ga_\tau}(m_\tau^2 - m_\pi^2)^2\,,~~~~
\cB(D^*\to D\pi) =\frac{g^2|\vp_{D}|^3}{6\pi\,m^2_{D^*}\,\Ga_{D^*}}\,,
\eeq
Eq.~\eqref{eq:ampsqv1} can be expressed as
\begin{align}\label{eq:ampsquared}
\begin{aligned}
\mathcal{|M^{\rm NP}|}^2 &= 
\frac{3 \times 2^5 \, {G_F^2 \,|V_{cb}|^2}\,\pi^4\,m_{D^*}}{ (m_\tau^2 - m_\pi^2)^2 |\vp_{D}|^3 }\, \cB(D^*\to D\pi) 
\cB(\tau\to\pi\nuta) \delta(p_{D^*}^2-m_{D^*}^2)   \delta(p_{\tau}^2-m_{\tau}^2)\\
&\times \Bigg| \sum_{\lds} P(\lds) \bigg[C_{S}^{LR} H_{S_L}^{\lds}  {L}^{S}  + C_{S}^{RR} H_{S_R}^{\lds}  {L}^{S}  + C_{V}^{RR} \sum_{m} g_{mm} H_{ V_R,m}^{\lds}  {L}^{V}_m \\
&  +C_T^{RR} \sum_{m,n}  g_{mm}g_{nn} H_{{T},mn}^{\lds}  {L}^{T}_{mn} \bigg] \Bigg|^2\,.
\end{aligned}
\end{align}

Finally, putting Eq.~\eqref{eq:ampsquared} in Eq.~\eqref{eq:decayratewithphasespace}, the measurable differential distribution can be written as  
\begin{align}\label{eq:rateJtau}
\begin{aligned}
\frac{d \Gamma(\BDstaufull)}{d q^2 d E_{\pi} d\!\cos\theta_D d\!\cos\theta_{\pi} d \chi_{\pi}} 
&=  \frac{  3 G_F^2 |V_{cb}|^2  |\vec{p}_{D^*}| \cB(\tau\to\pi\nuta)  \cB(D^*\to D\pi)}{2^{12} \pi^5 m_B^2 (m_\tau^2 - m_\pi^2)^2|\vp_{D}|^2 \sqrt{q^2}} \\
&\times \Big\{ J_{1s} \sin^2\theta_D+J_{1c}\cos^2\theta_D \\
&+(J_{2s} \sin^2\theta_D+J_{2c}\cos^2\theta_D )\cos 2\theta_\pi   \\
& +J_3 \sin^2\theta_D\sin^2\theta_\pi\cos 2\chi_\pi  \\
&  +J_4\sin 2\theta_D\sin 2\theta_\pi \cos\chi_\pi 
+J_5 \sin 2\theta_D\sin\theta_\pi\cos\chi_\pi \\ 
& +(J_{6s} \sin^2\theta_ D+J_{6c}\cos^2\theta_D)\cos\theta_\pi  \\
& +J_7 \sin 2\theta_D\sin\theta_\pi \sin\chi_\pi+J_8\sin 2\theta_D \sin 2\theta_\pi\sin\chi_\pi  \\
& +J_9 \sin^2\theta_D\sin^2\theta_\pi \sin2\chi_\pi  \Big\}\,,
\end{aligned}
\end{align}
where the $J$ functions are provided in Appendix~\ref{app:J functions} for the SM and different NP cases. Our results are in perfect agreement with those of Ref.~\cite{Bernlochner:2024xiz} for light leptons.\footnote{We have confirmed in a private communication with the authors that the lepton angle definition mentioned in their text is not consistent with their $J$-functions; the latter can be made consistent by replacing $\theta_\ell \to \pi - \theta_\ell$, where $\theta_\ell$ is the angle between the charged-lepton momentum and the direction opposite to the $B$-meson momentum in the virtual $W$ rest-frame.}   

\begin{table}[t]
    \centering
    \begin{tabular}{c|c|c|c|c|c}
    \hline
        & \multicolumn{2}{c|}{$\mathcal{B}(B_c^-\to\tau^-X)<10\,(30)\%$}
        & \multicolumn{2}{c}{$R(D^*)$ (2$\sigma$ bound)} & \multicolumn{1}{|c}{$R(D)$ (2$\sigma$ bound)} \\
        \hline
        $m_N$~(GeV)
        & $|C_V^{RR}|$
        & $|C_S^{LR}|\,, |C_S^{RR}|$
        & $|C_V^{RR}|$
        & $|C_T^{RR}|$
        & $|C_S^{LR}|$,$|C_S^{RR}|$ \\
    \hline
        0.01 & 1.90 (3.58) & 0.47 (0.88) & 0.43 & 0.12 & 0.66 \\
        0.1  & 1.90 (3.58) & 0.47 (0.88) & 0.43 & 0.12 & 0.66 \\
        0.5  & 1.82 (3.43) & 0.47 (0.89) & 0.50 & 0.14 & 0.68 \\
        1.0  & 1.64 (3.09) & 0.48 (0.91) & 0.85 & 0.24 & 0.79 \\
    \hline
    \end{tabular}
\caption{Upper bounds on the Wilson coefficients for several RHN masses obtained from the requirement that $\mathcal{B}(B_c^-\to\tau^-X)<10\,(30)\%$, and from $R(D^*)$ and $R(D)$ at $2\sigma$,   assuming that only a single real WC is nonzero. The tensor coefficient $C_T^{RR}$ is not constrained by $\mathcal{B}(B_c^-\to\tau^-X)$.}
    \label{tab:BcTauNu}
\end{table}

For $m_N < m_{B_c} - m_\tau$, it is possible to have the 
two-body decay $B_c \to \tau \bar{N}$ in addition to the SM $B_c^{-} \to \tau^- \bar\nu_\tau$ decay. In the basis of the Hamiltonian in Eq.~\eqref{eq:effective_hamiltonian}, the $B_c \to \tau \bar{N}$ decay width can be written as

\begin{align}
\label{eq:width-BctauN}
\Gamma(B_c^-\to\tau^-\bar N)
&=
\frac{G_F^2|V_{cb}|^2f_{B_c}^2m_{B_c}^3}{8\pi}
\sqrt{
\lambda\left(
1,
r_\tau,
r_N
\right)
}
\nonumber\\
&\quad\times
\Bigg\{
\left[
\left(
1-r_\tau
-r_N
\right)
\left(
r_\tau+r_N
\right)
+
4 r_\tau r_N
\right]
\left|C_V^{RR}\right|^2
\nonumber\\
&\qquad
+
\left(
1-r_\tau
-r_N
\right)
\left(\frac{m_{B_c}}{m_b+m_c}\right)^2
\left|C_S^{LR}-C_S^{RR}\right|^2
\nn \\
&\qquad
+
2\sqrt{r_\tau}\,
\left(
1-r_\tau
+r_N
\right)
\left(\frac{m_{B_c}}{m_b+m_c}\right)
\mathrm{Re}
\left[
C_V^{RR}
\left(C_S^{LR}-C_S^{RR}\right)^\ast
\right]
\Bigg\}\,,
\end{align}
which does not interfere with the SM contribution,
\beq
\Gamma_{\rm SM}(B_c^-\to \tau^-\bar\nu_\tau)
=
\frac{G_F^2|V_{cb}|^2f_{B_c}^2m_{B_c}^3 r_\tau}{8\pi}
\left(1-r_\tau\right)^2\,.
\eeq 
Here, $r_\tau = m_\tau^2/m_{B_c}^2$ and $r_N = m_N^2/m_{B_c}^2$.  Using $f_{B_c} = 0.427(6)$~GeV~\cite{Colquhoun:2015oha}, $|V_{cb}| = (41.1\pm 1.2) \times 10^{-3}$~\cite{pdg24}, $m_{B_c} = 6.27447$~GeV~\cite{pdg24} and $\tau_{B_c} = 0.510(9)~\rm{ps}$~\cite{pdg24}, we obtain $\mathcal{B}_{SM}(B_c^- \to \tau^- \bar\nu_\tau) = 2.16\%$. The tensor operator does not contribute since the matrix element is exactly zero. In Table~\ref{tab:BcTauNu}, we show the upper limit on each of the WCs, taken one at a time. We require that the total branching fraction of $B_c^- \to \tau^- \bar X$ does not exceed 10~(30)\%.
For the $R(D^*)$ constraints we take the HFLAV averaged value of $R(D^*)$~\cite{hflav} and the $\bar B^0 \to D^* \ell \bar \nu_{\ell}$ branching fraction from 
Ref.~\cite{pdg24}, and 
ensure that the Wilson coefficients do not increase the value of the $\bar B^0 \to D^* \tau \bar X$  branching fraction by more than $2\sigma$. Similarly, we use the HFLAV averaged value of $R(D)$ and the PDG average of the branching fraction of $B \to D \ell \bar \nu_{\ell}$ to constrain the scalar WCs.
Note that the bounds for $m_N=0.01$~GeV and $m_N=0.1$~GeV are the same. For scalar operators, the $B_c^- \to \tau^- \bar X$ constraint applies to $|C_S^{LR}-C_S^{RR}|$. As expected, $B_c^-$ decay constrains the scalar WCs more tightly than the vector WC.

\section{Univariate distributions}
\label{univ}
\begin{figure}[t]
\centering


\begin{subfigure}{0.325\textwidth}
    \centering
    \includegraphics[width=\linewidth]{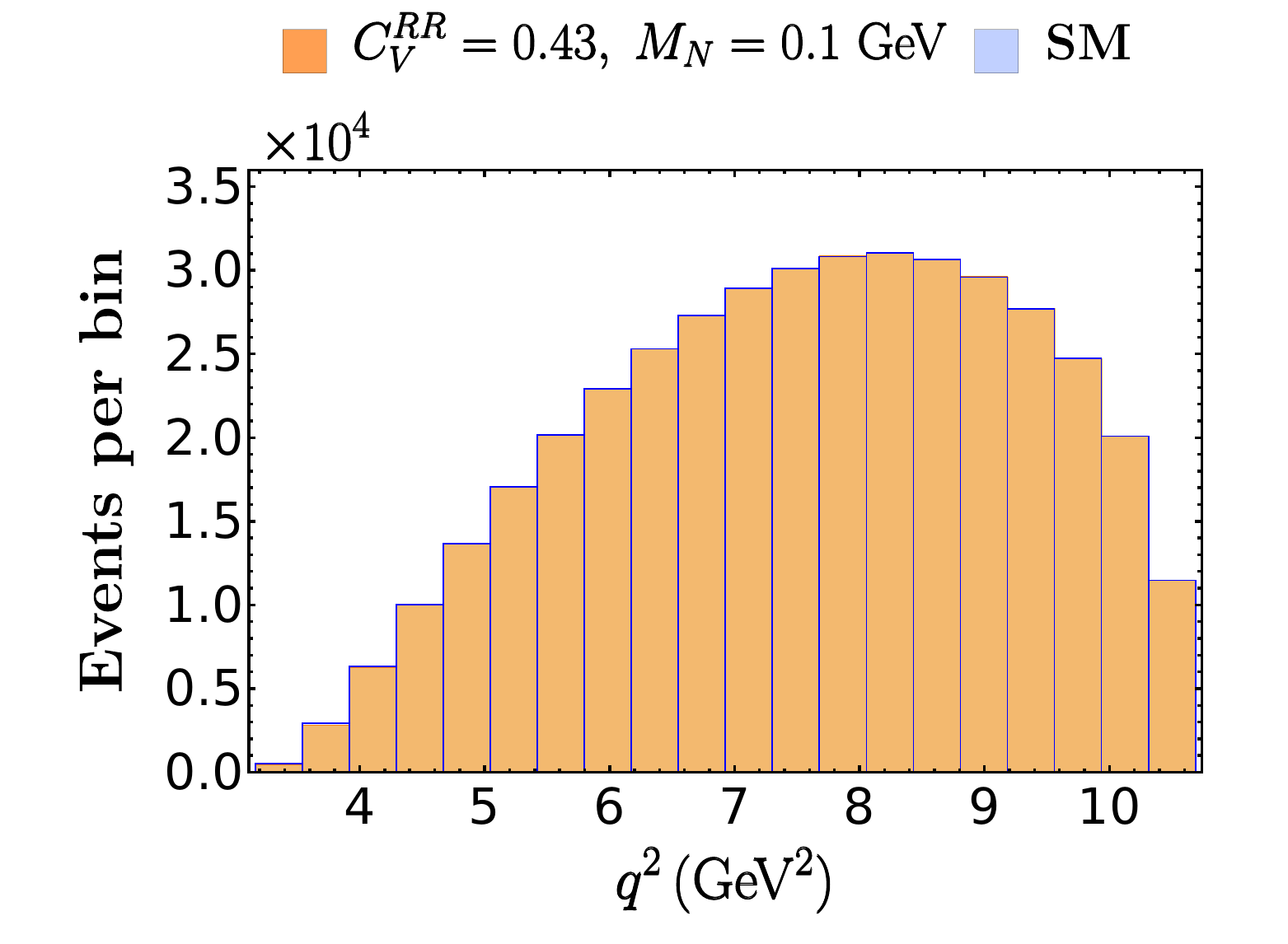}
\end{subfigure}
\hfill
\begin{subfigure}{0.325\textwidth}
    \centering
    \includegraphics[width=\linewidth]{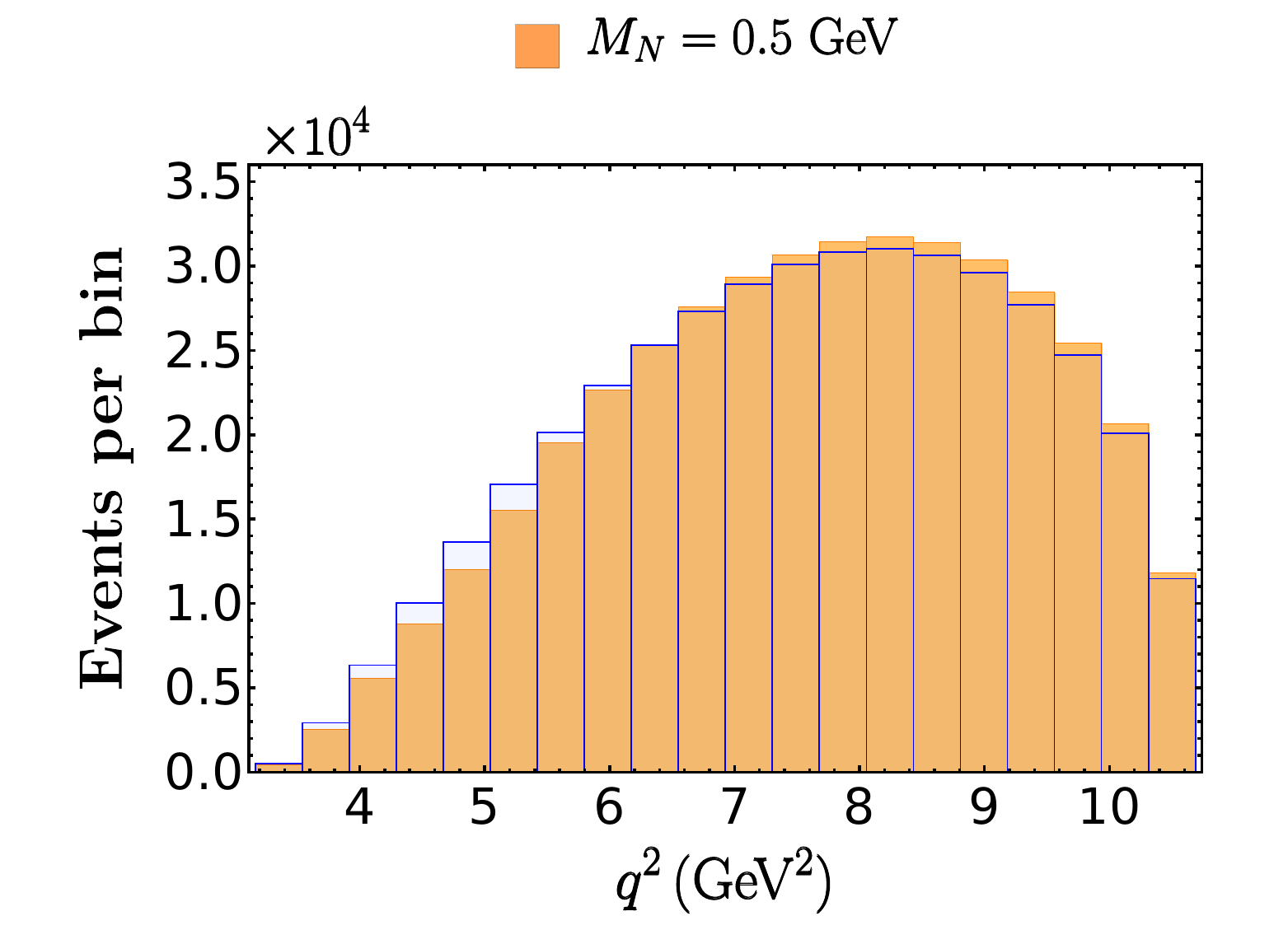}
\end{subfigure}
\hfill
\begin{subfigure}{0.325\textwidth}
    \centering
    \includegraphics[width=\linewidth]{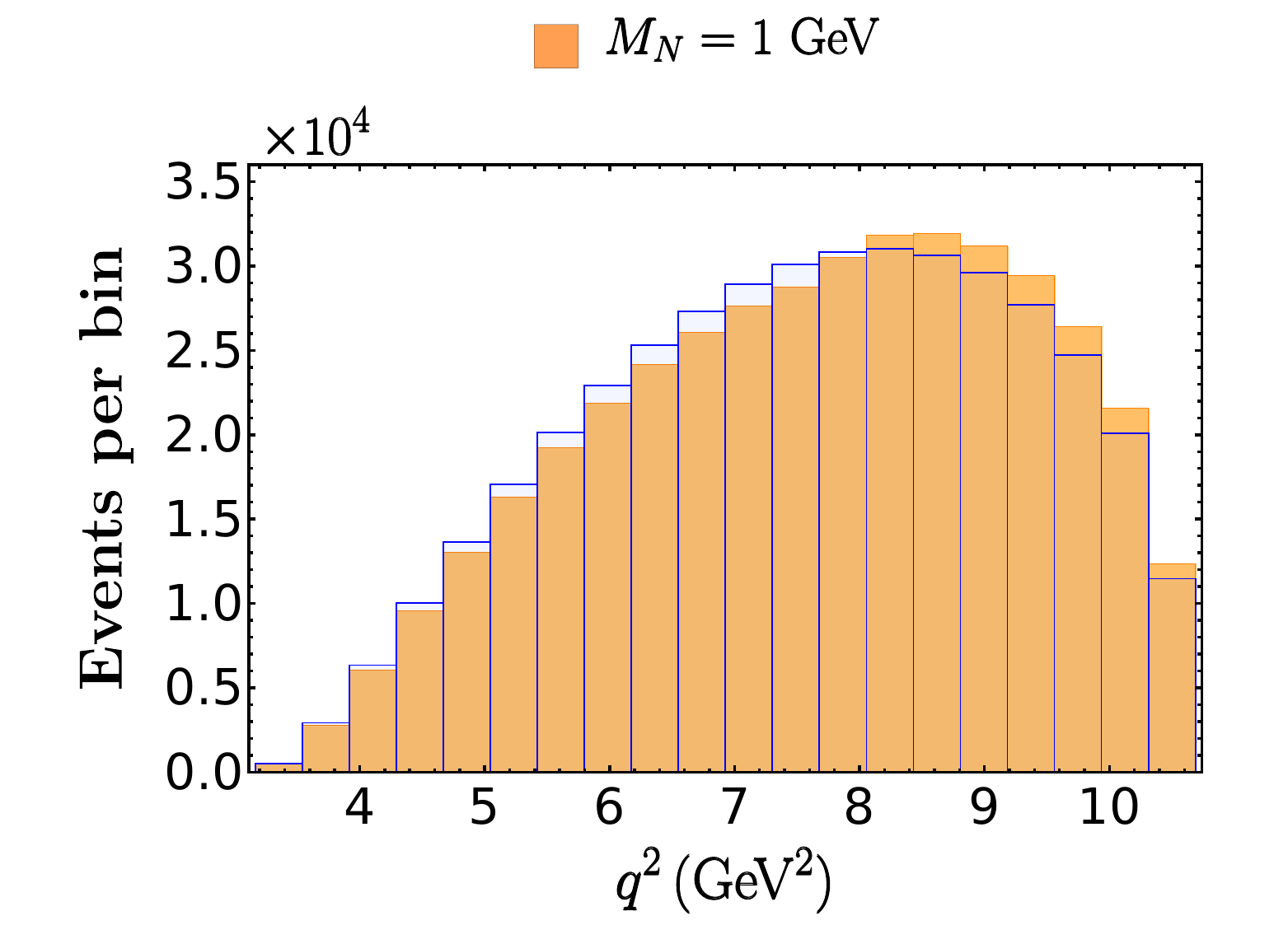}
\end{subfigure}

\vspace{0.5cm}


\begin{subfigure}{0.325\textwidth}
    \centering
    \includegraphics[width=\linewidth]{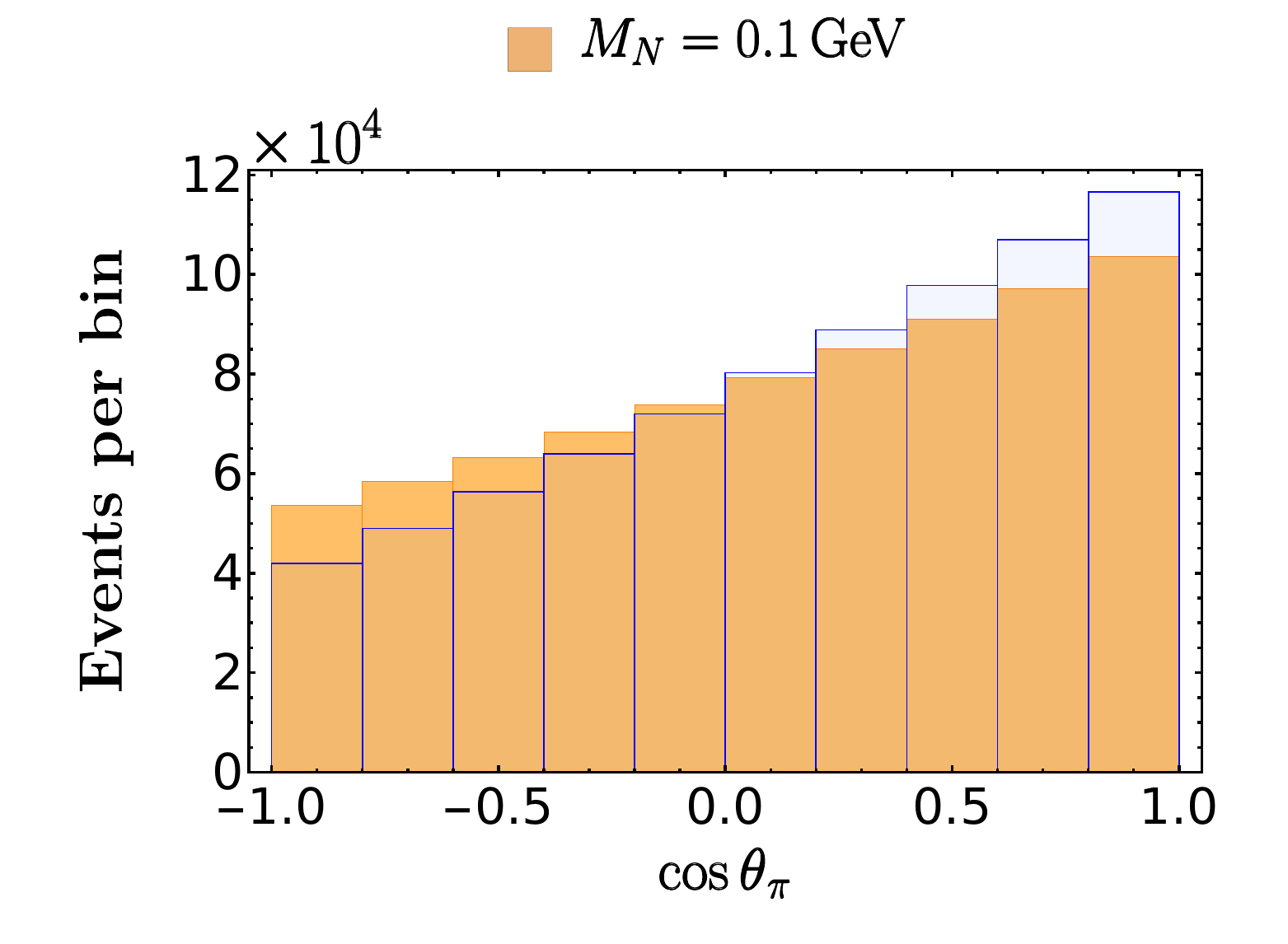}
\end{subfigure}
\hfill
\begin{subfigure}{0.325\textwidth}
    \centering
    \includegraphics[width=\linewidth]{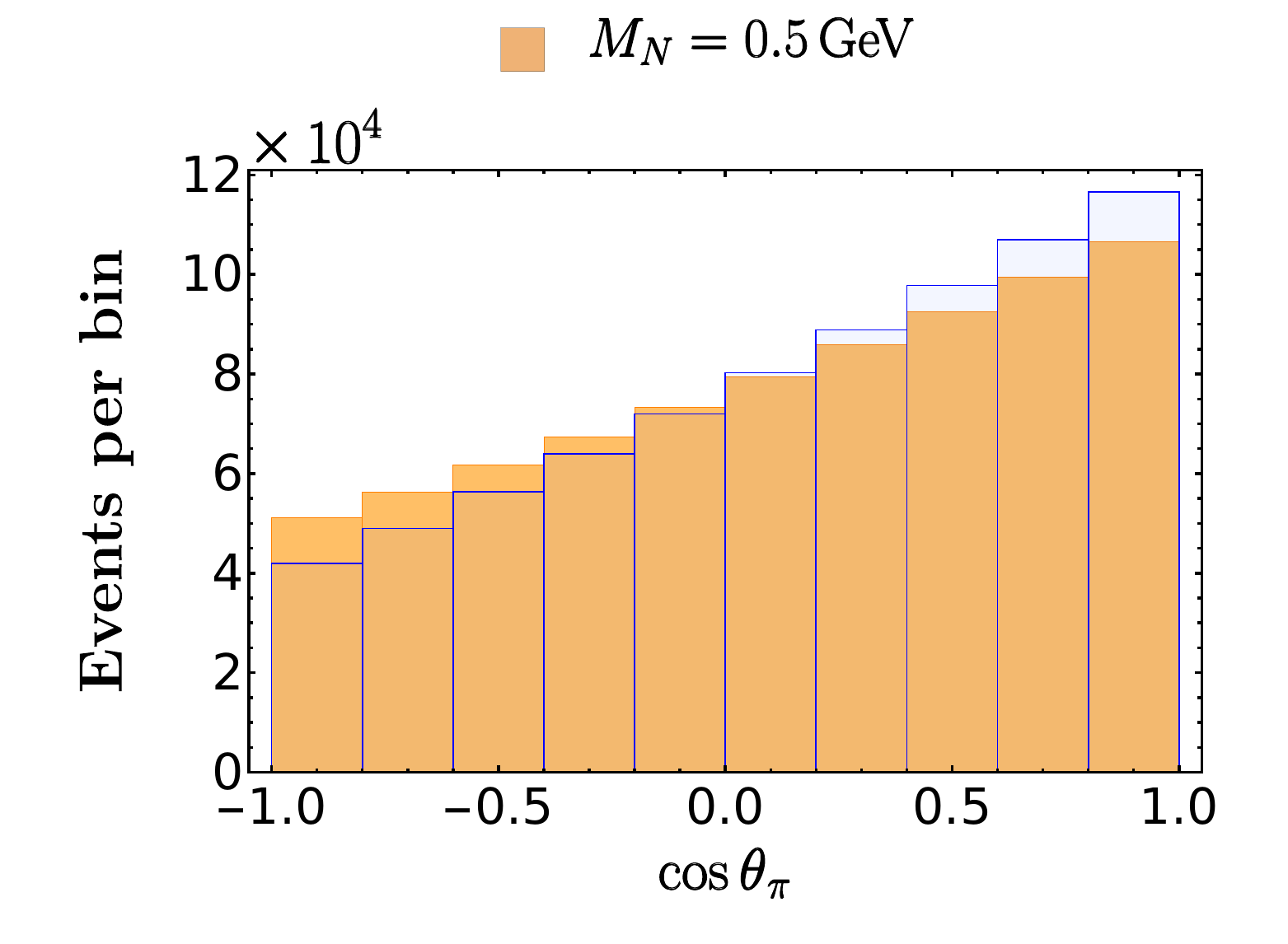}
\end{subfigure}
\hfill
\begin{subfigure}{0.325\textwidth}
    \centering
    \includegraphics[width=\linewidth]{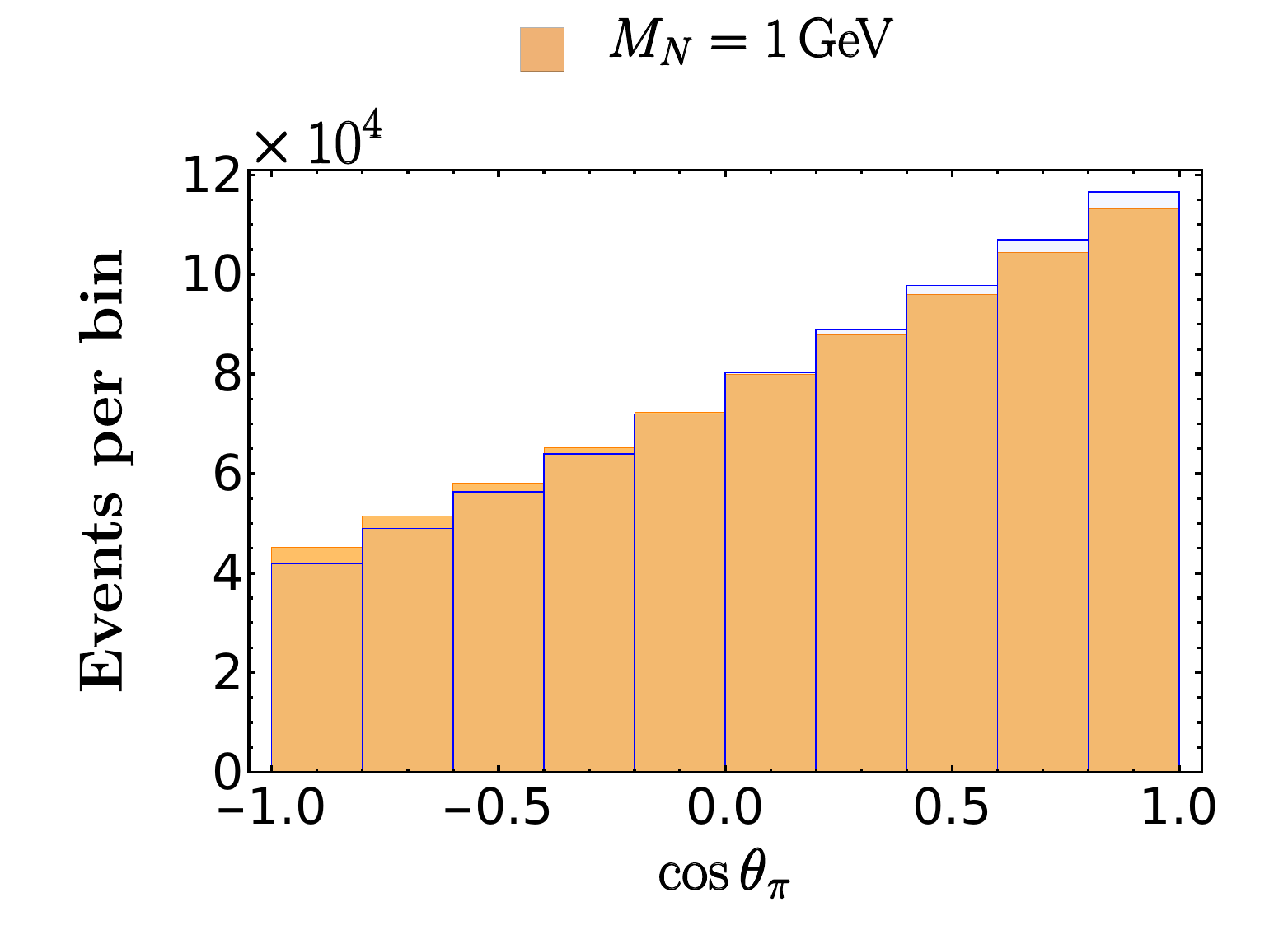}
\end{subfigure}

\vspace{0.5cm}


\begin{subfigure}{0.325\textwidth}
    \centering
    \includegraphics[width=\linewidth]{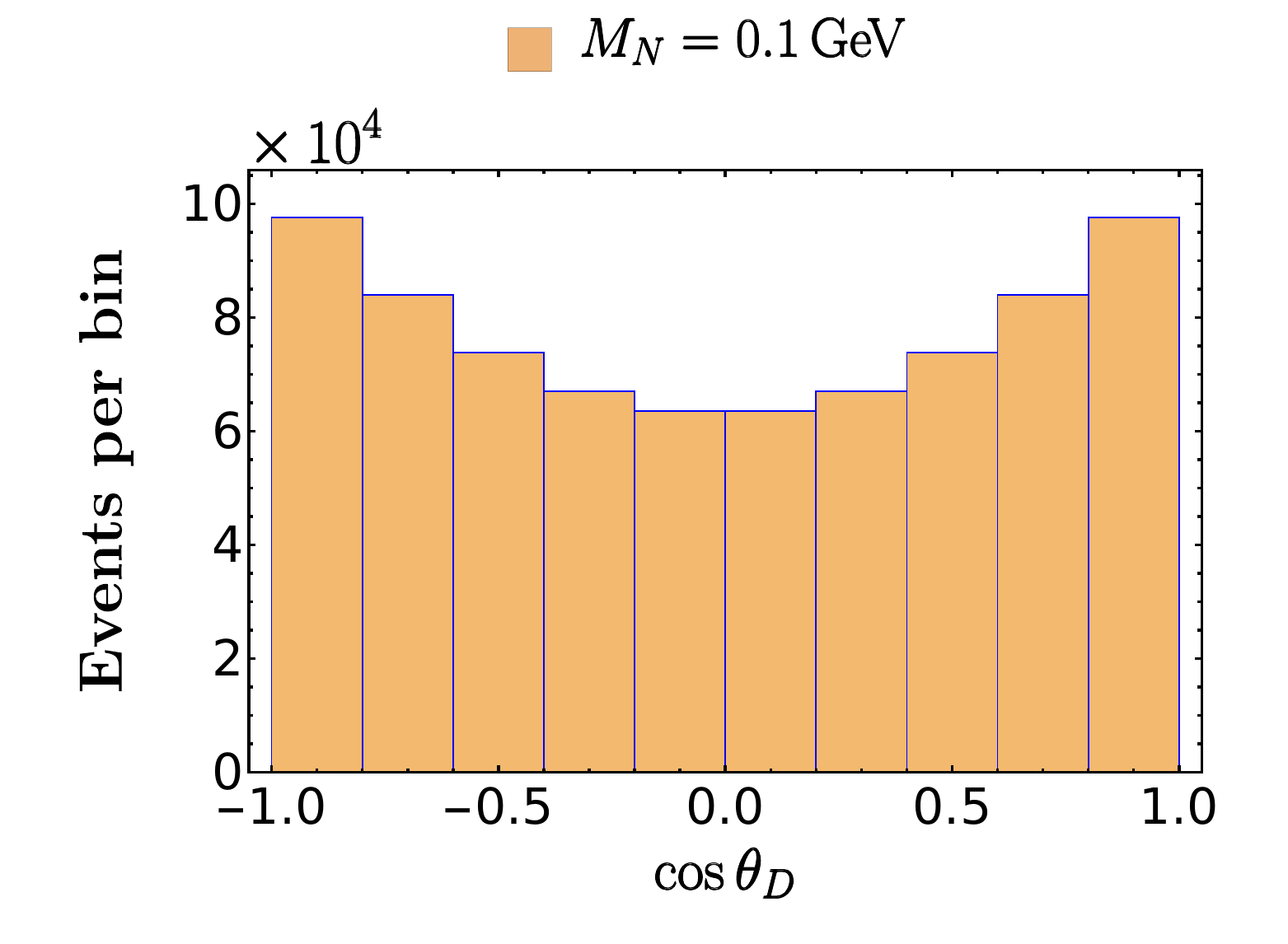}
\end{subfigure}
\hfill
\begin{subfigure}{0.325\textwidth}
    \centering
    \includegraphics[width=\linewidth]{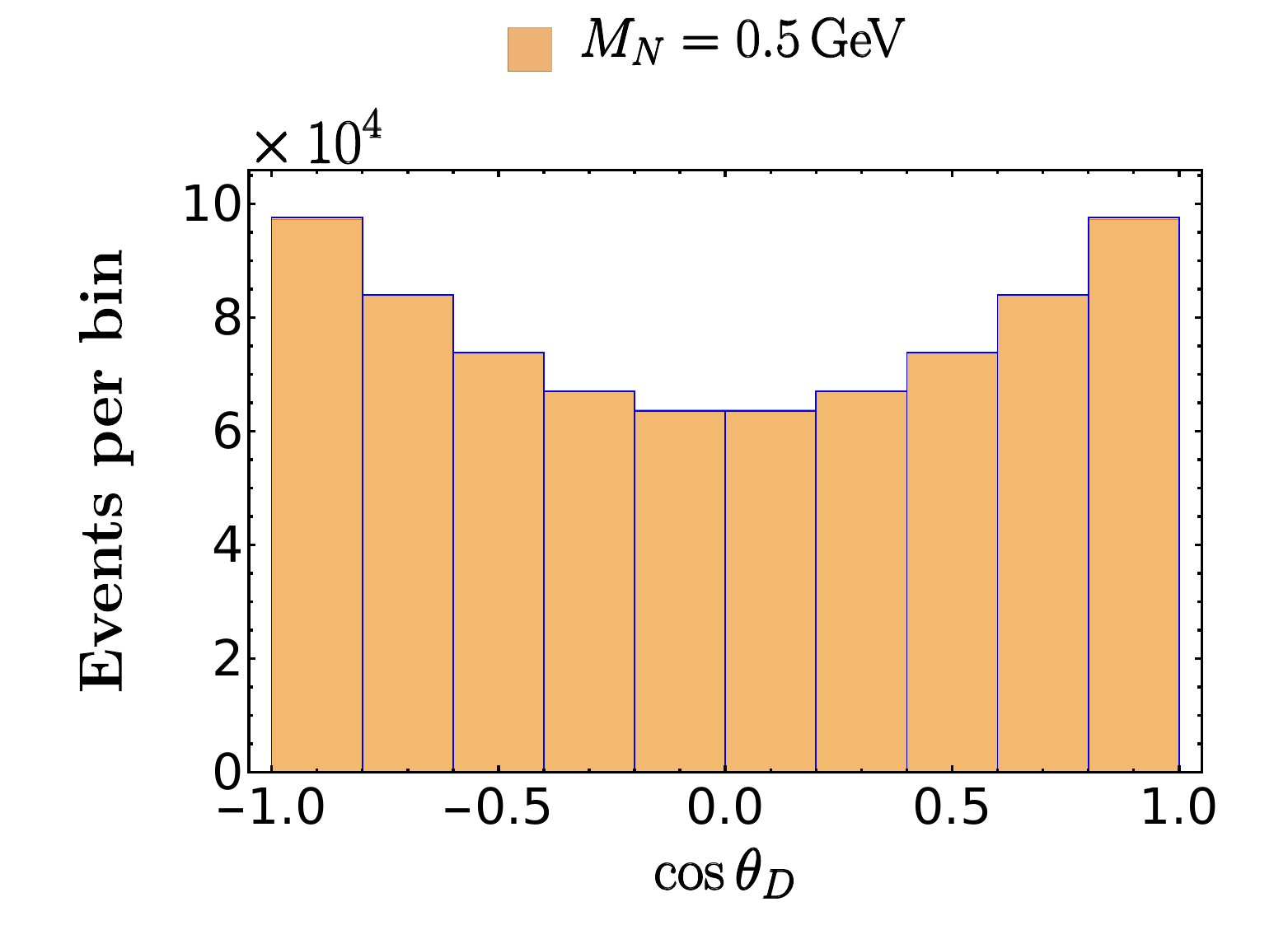}
\end{subfigure}
\hfill
\begin{subfigure}{0.325\textwidth}
    \centering
    \includegraphics[width=\linewidth]{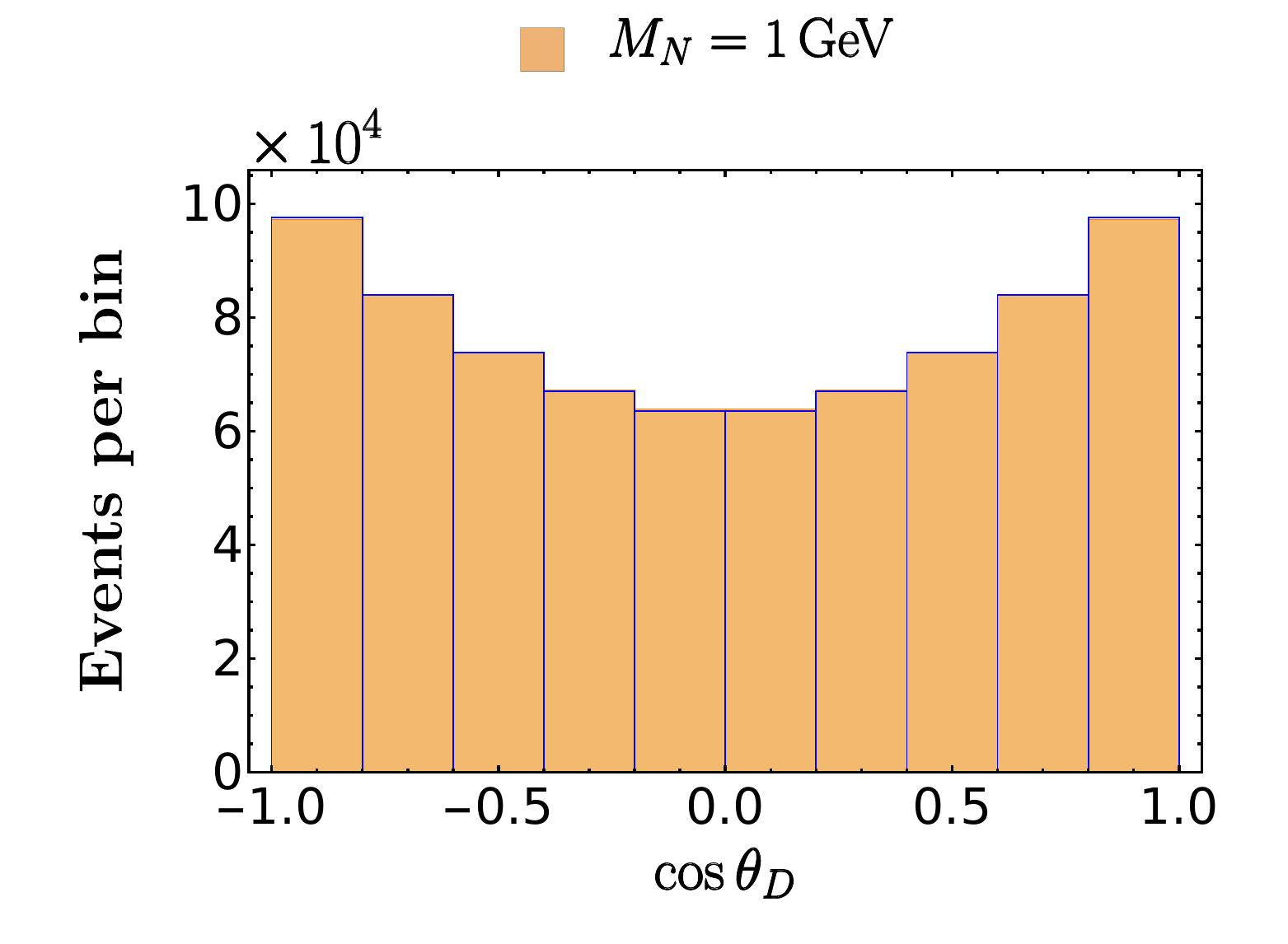}
\end{subfigure}

\vspace{0.5cm}


\begin{subfigure}{0.325\textwidth}
    \centering
    \includegraphics[width=\linewidth]{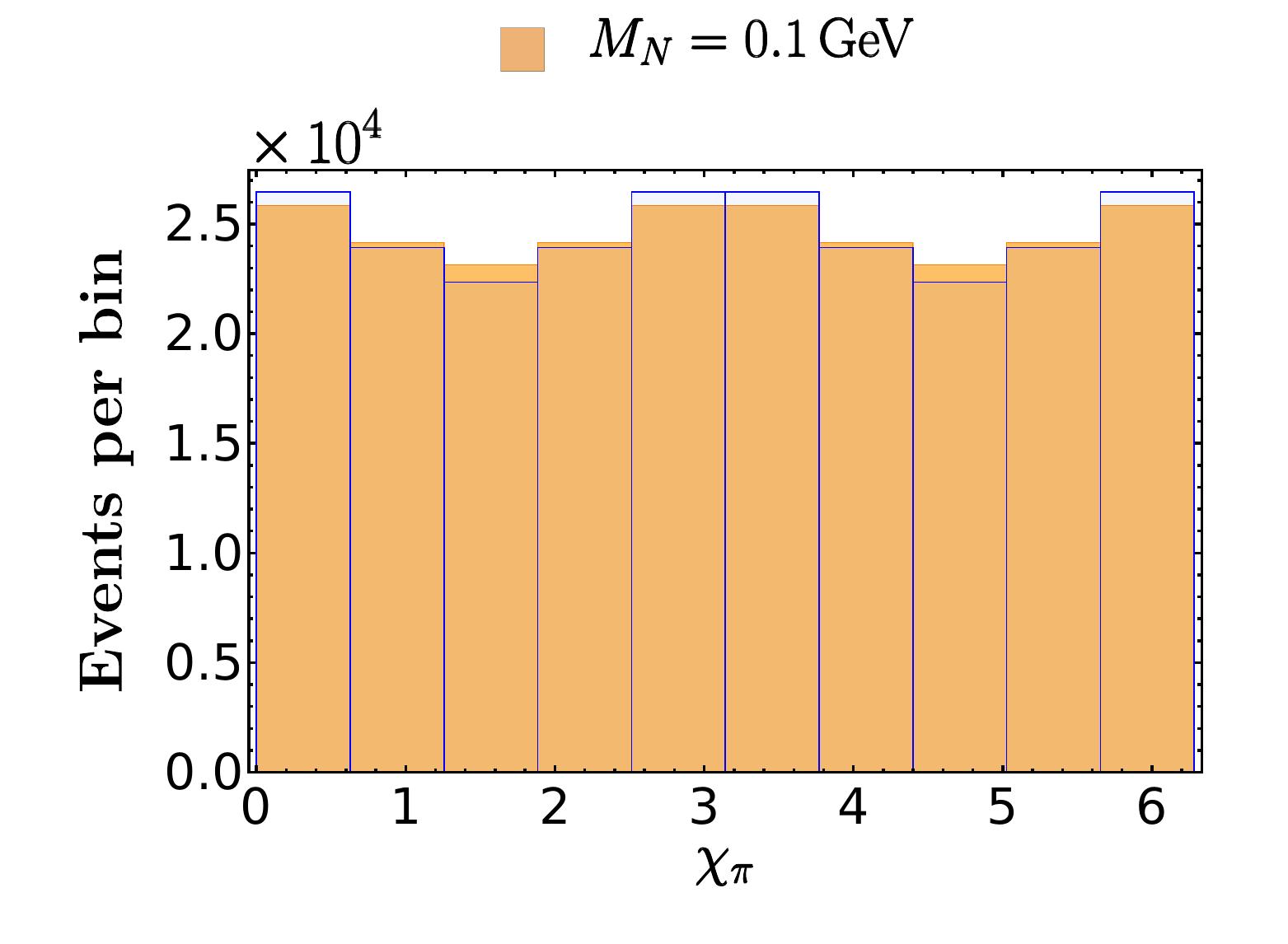}
\end{subfigure}
\hfill
\begin{subfigure}{0.325\textwidth}
    \centering
    \includegraphics[width=\linewidth]{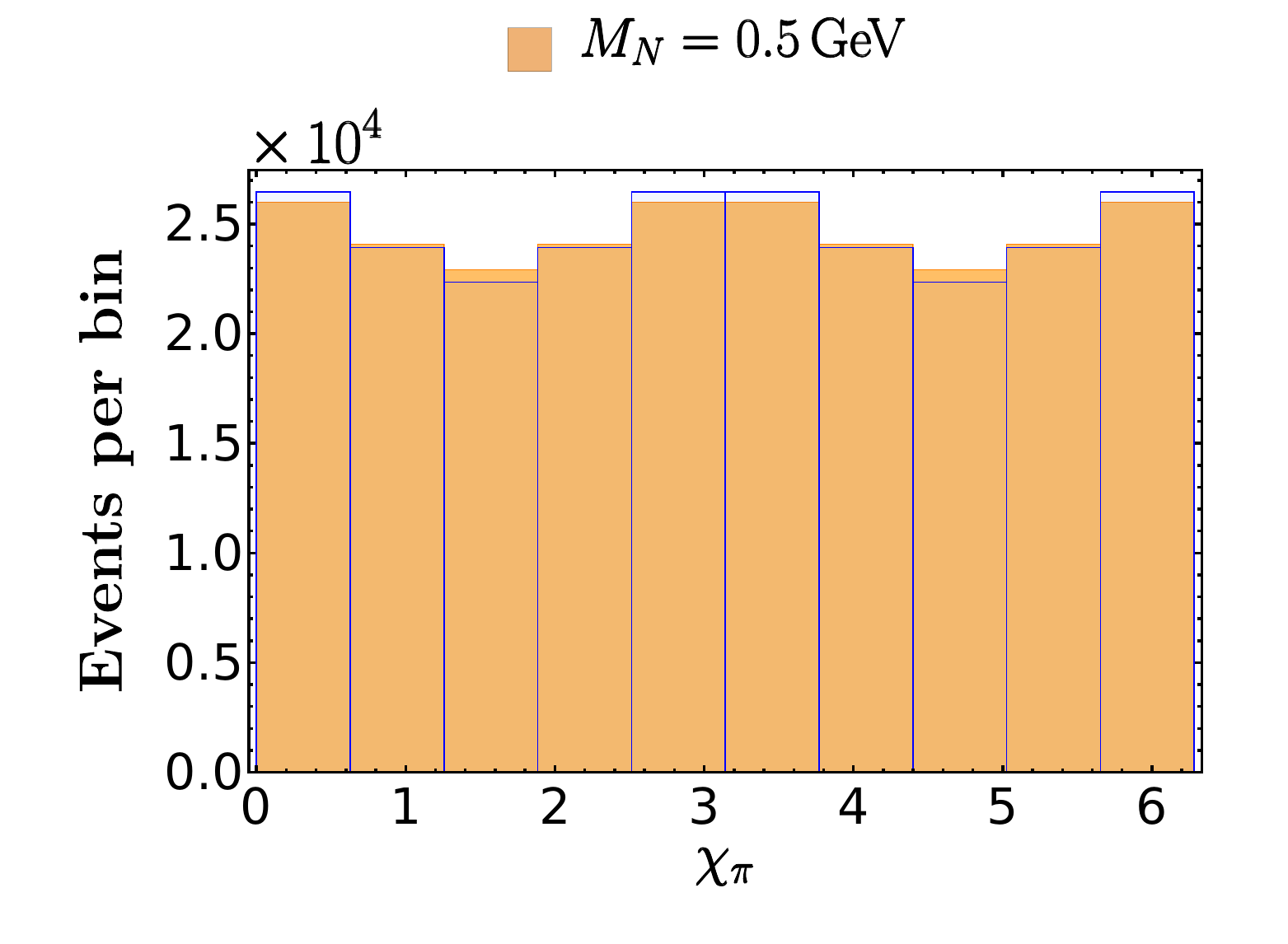}
\end{subfigure}
\hfill
\begin{subfigure}{0.325\textwidth}
    \centering
    \includegraphics[width=\linewidth]{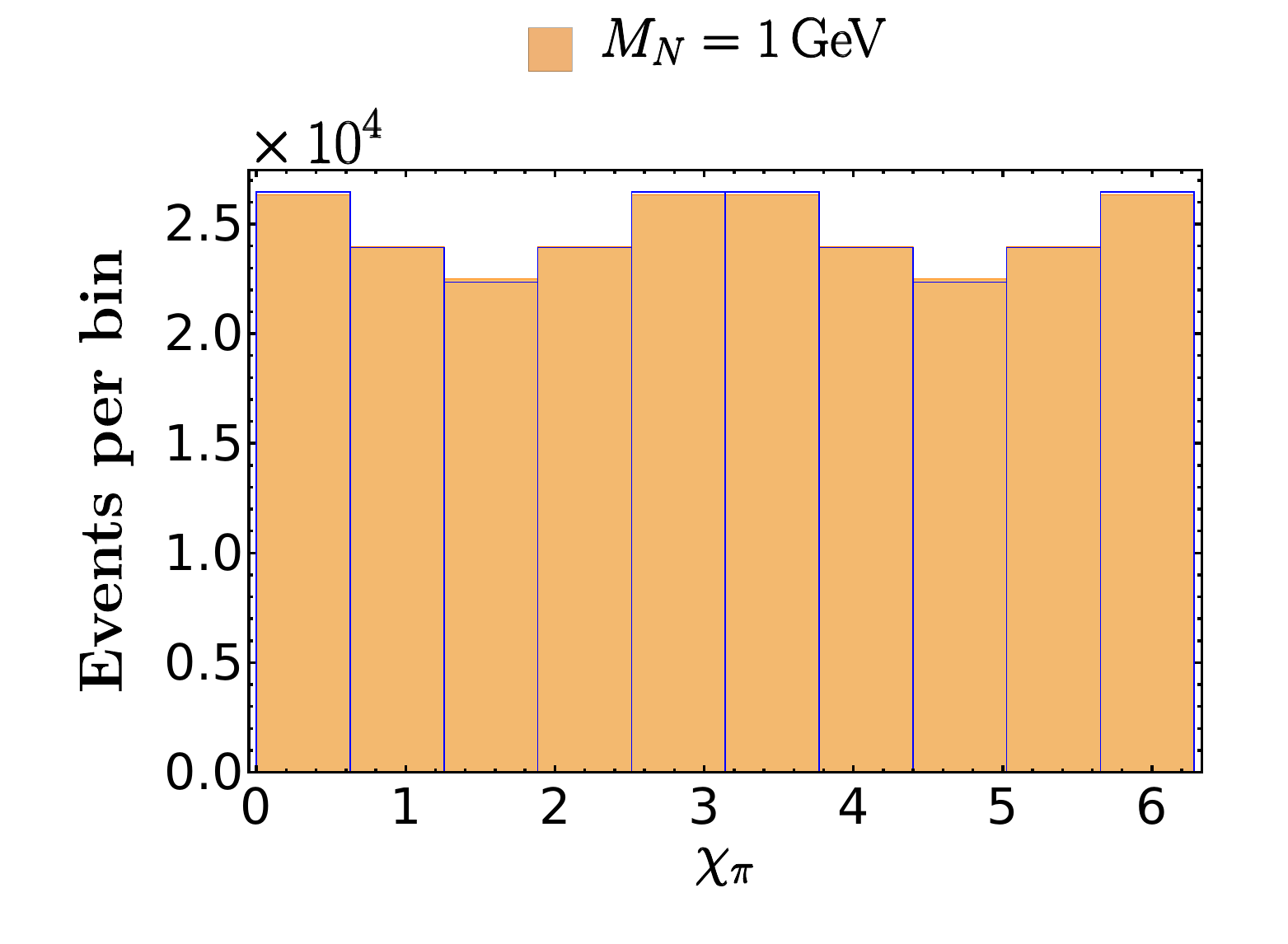}
\end{subfigure}

\caption{Differential distributions for the vector operator with $|C_{V}^{RR}|=0.43$ for $m_N = 0.1~\rm{GeV},~0.5$~GeV and 1~GeV are shown in yellow. The SM distributions are overlaid in blue for comparison. From top to bottom the following univariate distributions are plotted: $q^2$, $\cos\theta_{\pi}$, $\cos\theta_{D}$, and $\chi_\pi$.}
\label{fig:cvrrp5_combined}
\end{figure}

\begin{figure}[t]
\centering


\begin{subfigure}{0.325\textwidth}
    \centering
    \includegraphics[width=\linewidth]{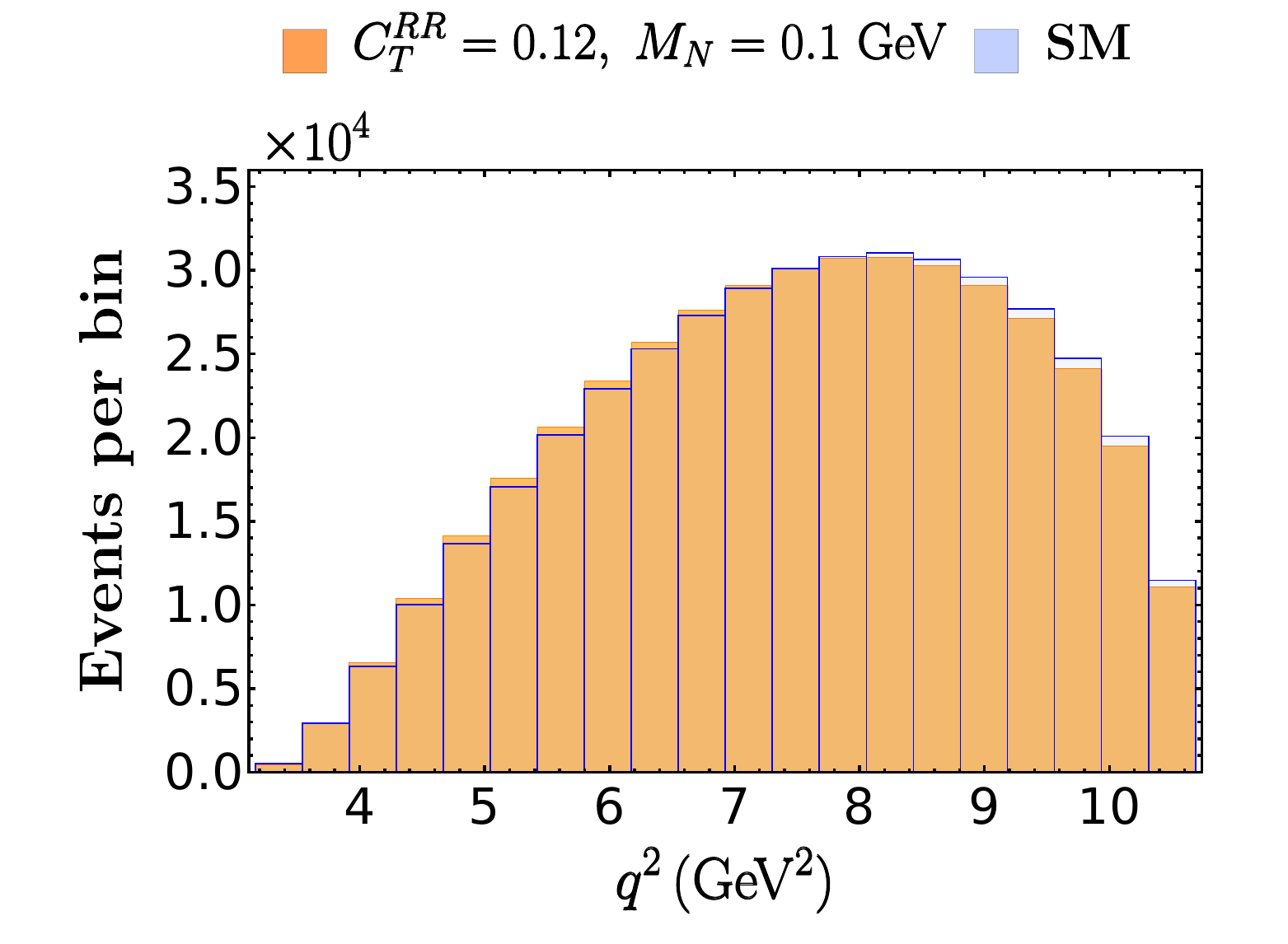}
\end{subfigure}
\hfill
\begin{subfigure}{0.325\textwidth}
    \centering
    \includegraphics[width=\linewidth]{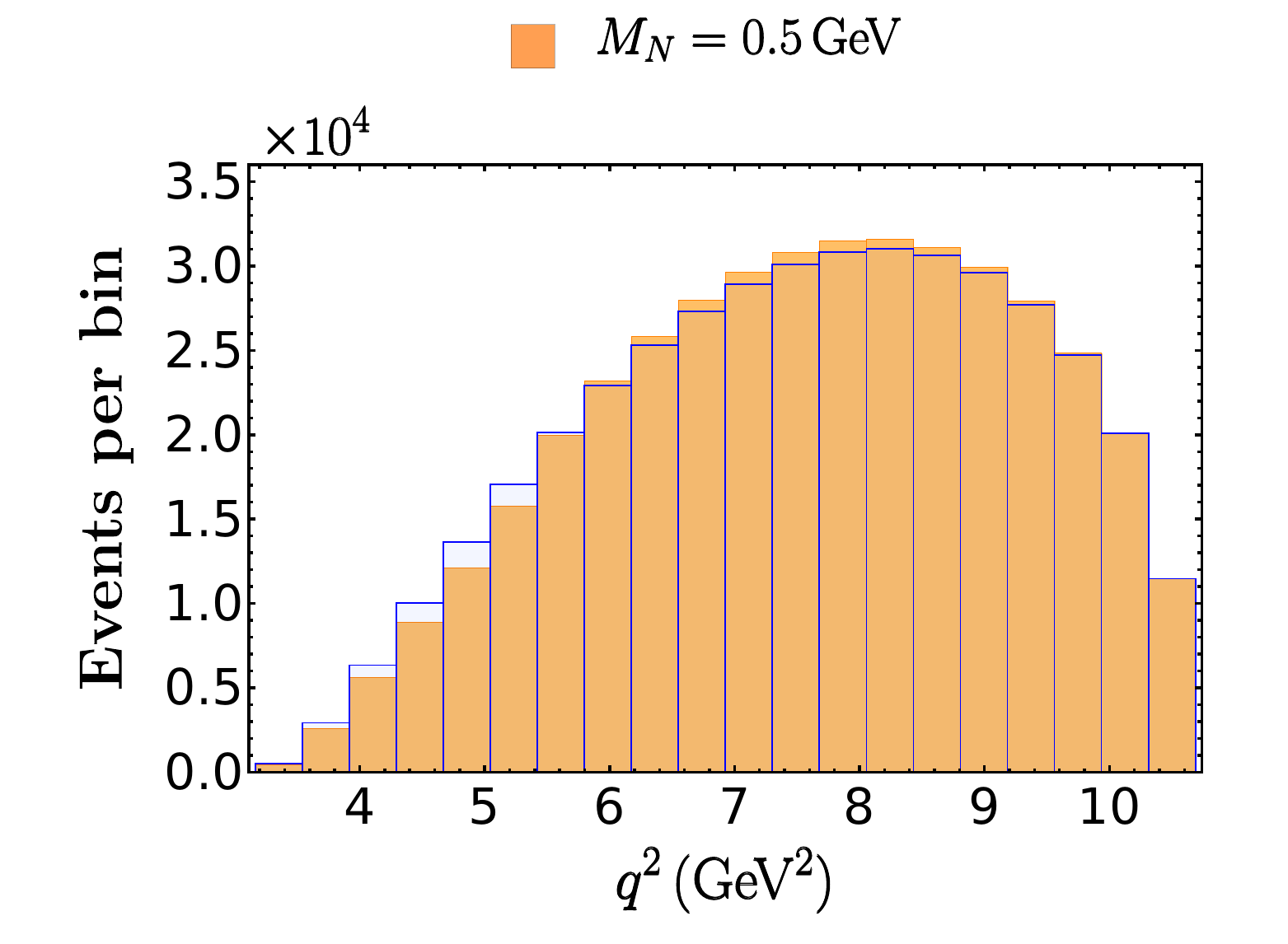}
\end{subfigure}
\hfill
\begin{subfigure}{0.325\textwidth}
    \centering
    \includegraphics[width=\linewidth]{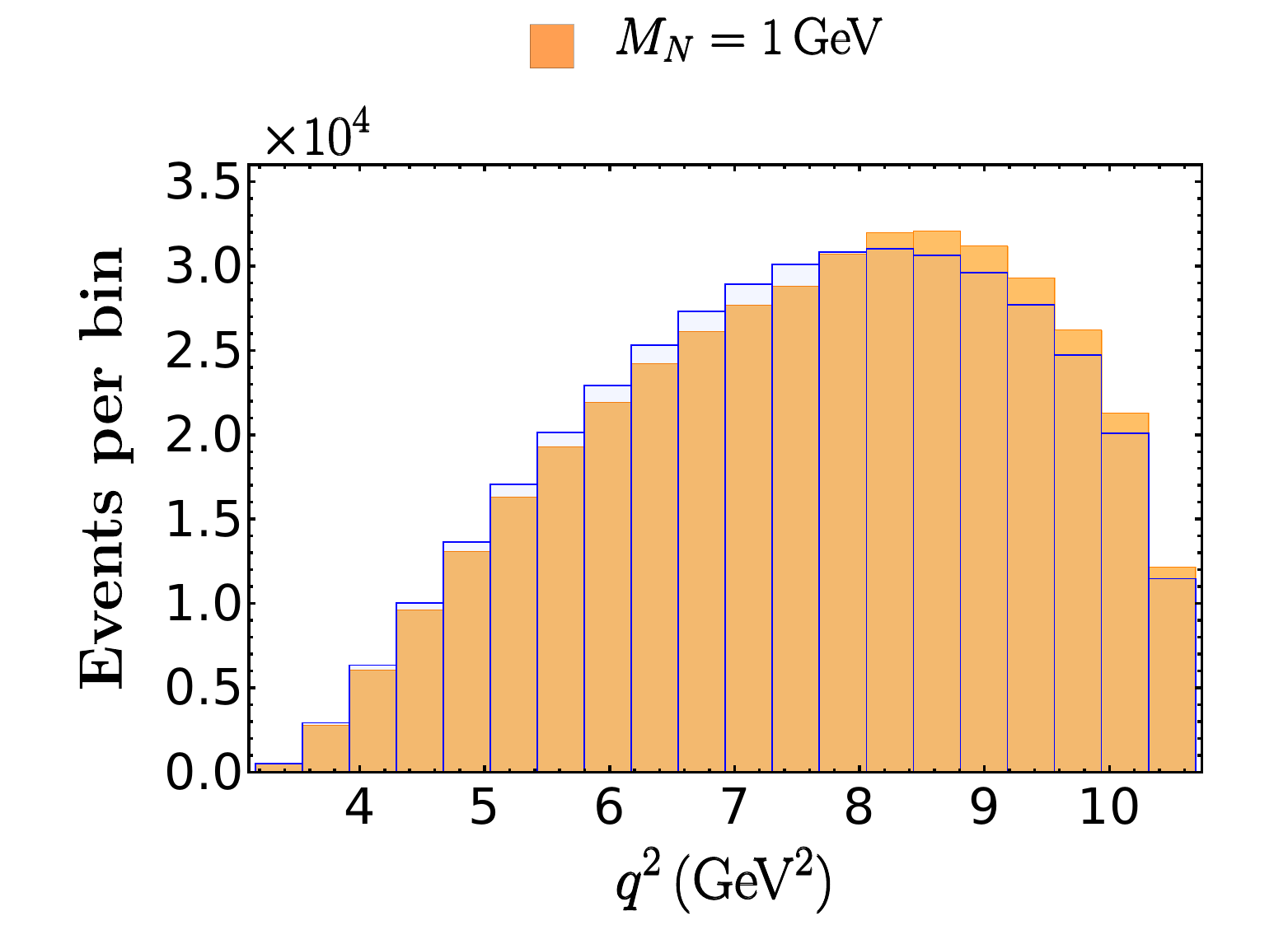}
\end{subfigure}

\vspace{0.5cm}


\begin{subfigure}{0.325\textwidth}
    \centering
    \includegraphics[width=\linewidth]{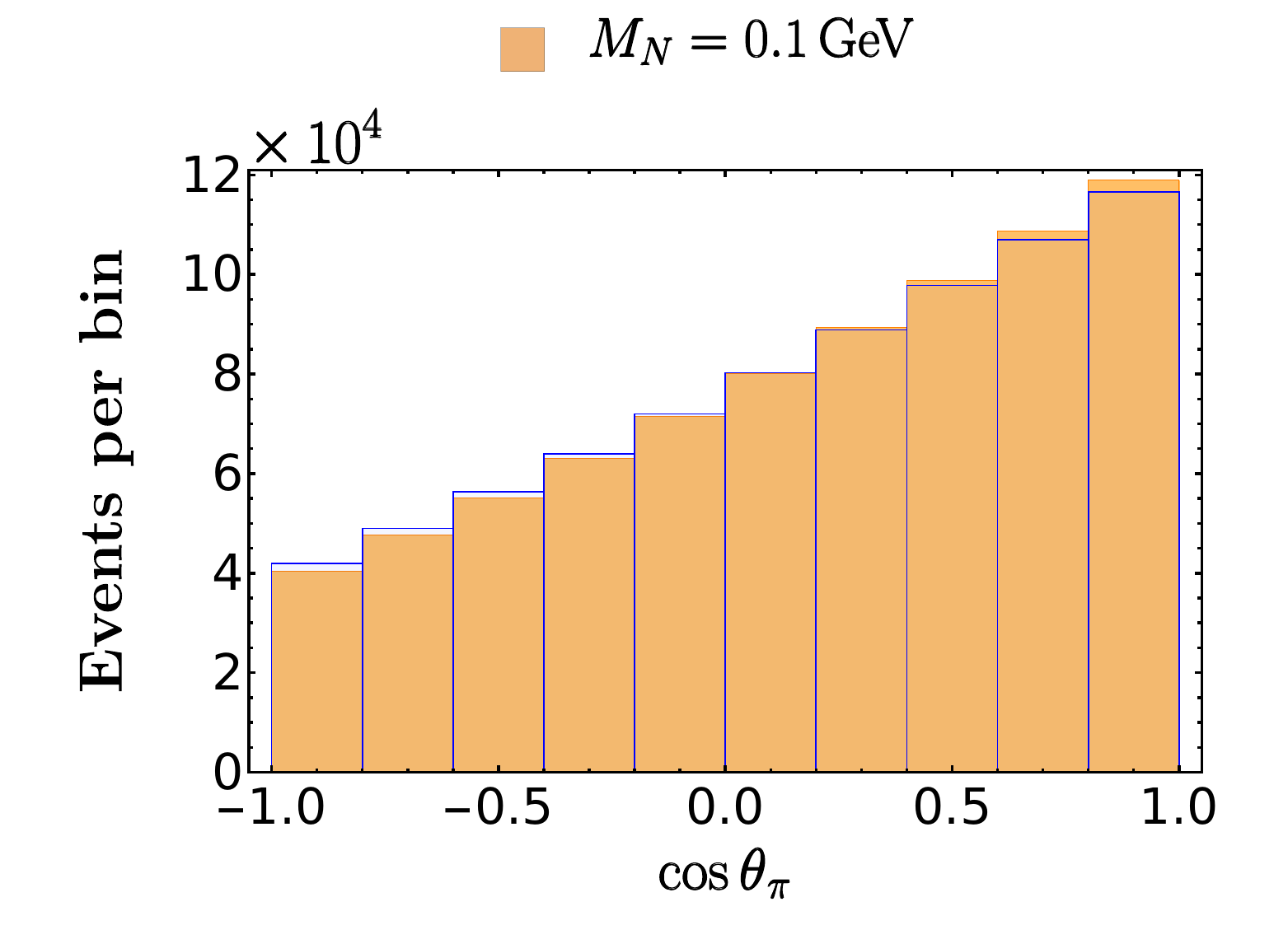}
\end{subfigure}
\hfill
\begin{subfigure}{0.325\textwidth}
    \centering
    \includegraphics[width=\linewidth]{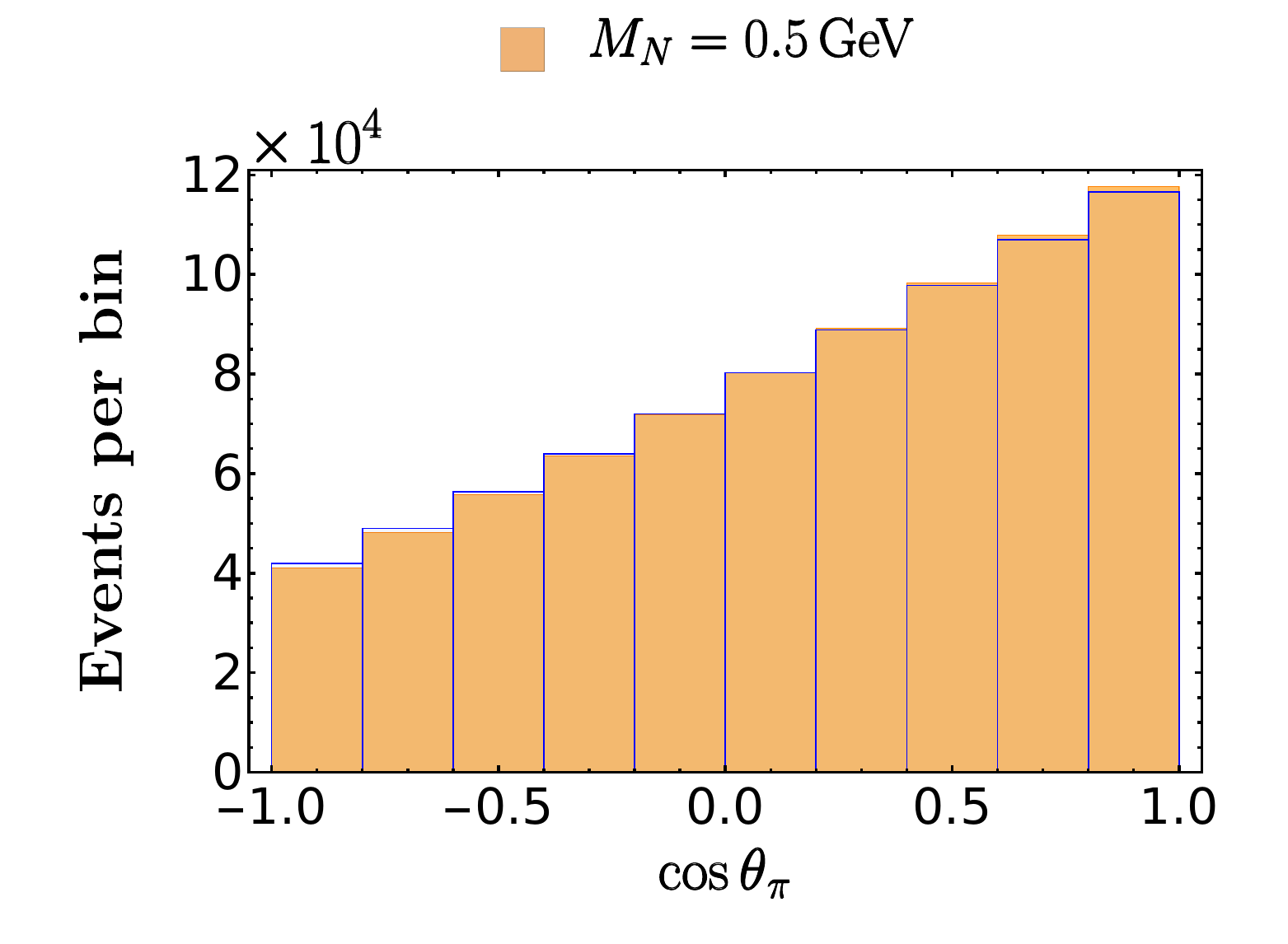}
\end{subfigure}
\hfill
\begin{subfigure}{0.325\textwidth}
    \centering
    \includegraphics[width=\linewidth]{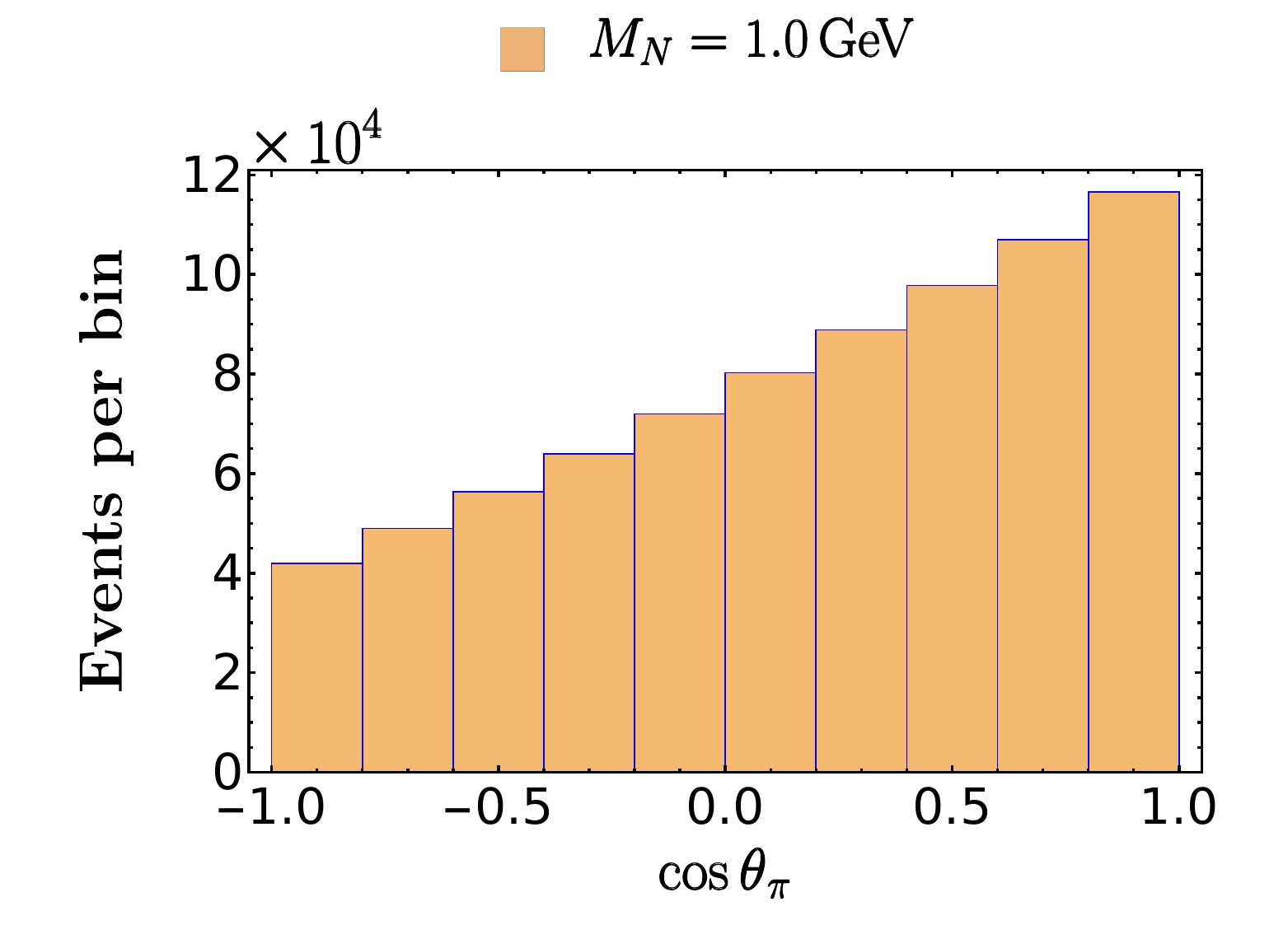}
\end{subfigure}

\vspace{0.5cm}


\begin{subfigure}{0.325\textwidth}
    \centering
    \includegraphics[width=\linewidth]{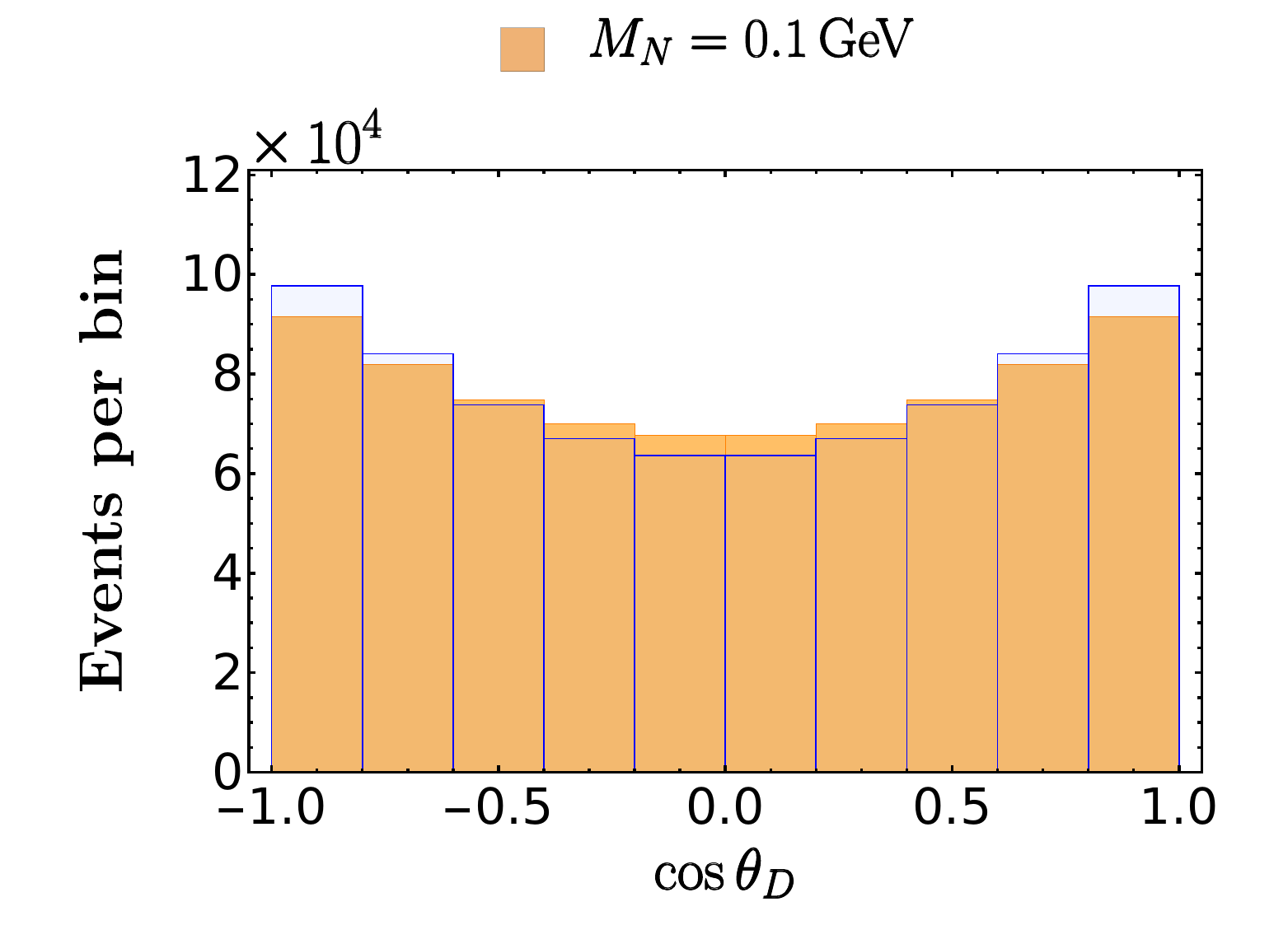}
\end{subfigure}
\hfill
\begin{subfigure}{0.325\textwidth}
    \centering
    \includegraphics[width=\linewidth]{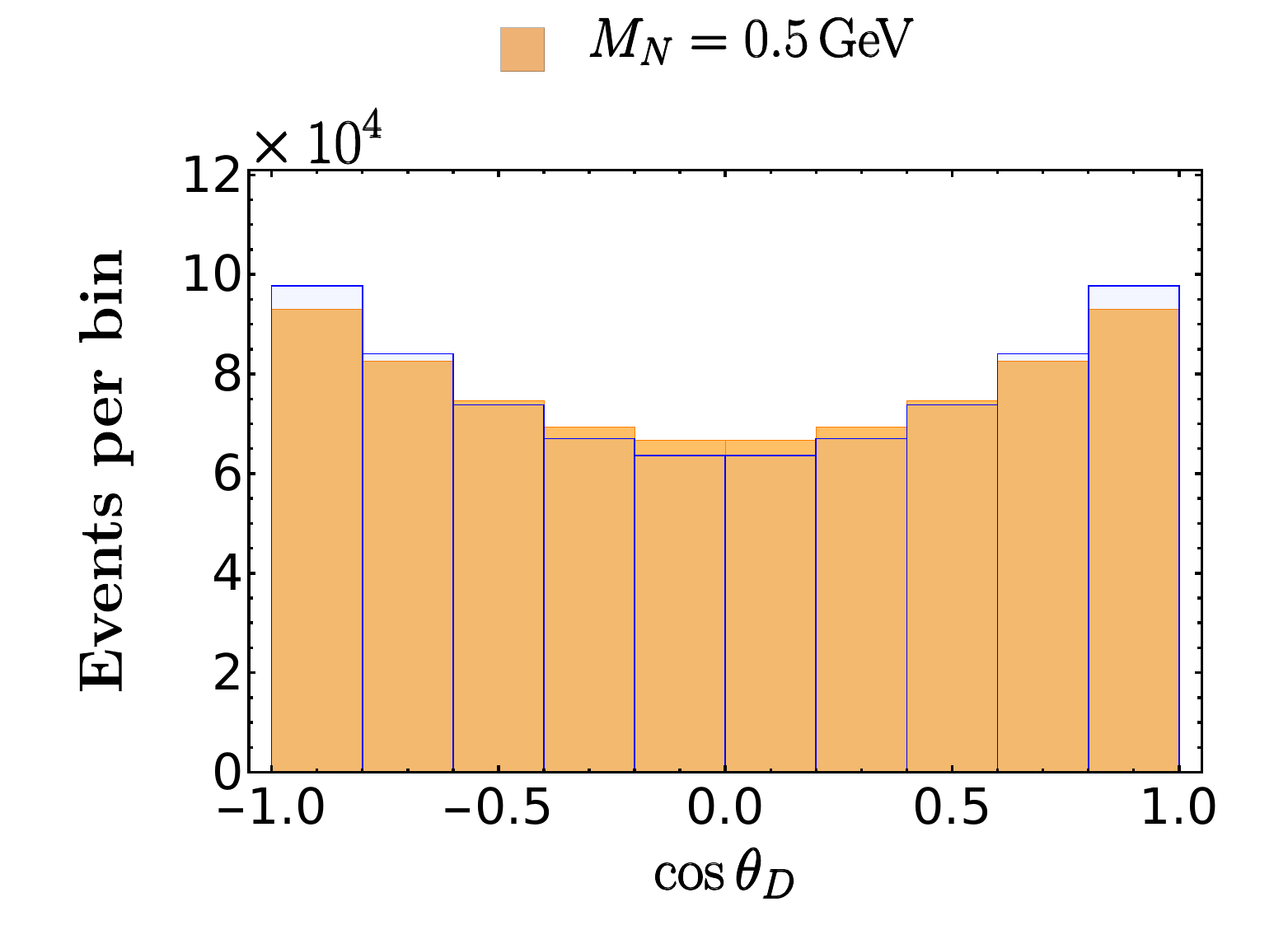}
\end{subfigure}
\hfill
\begin{subfigure}{0.325\textwidth}
    \centering
    \includegraphics[width=\linewidth]{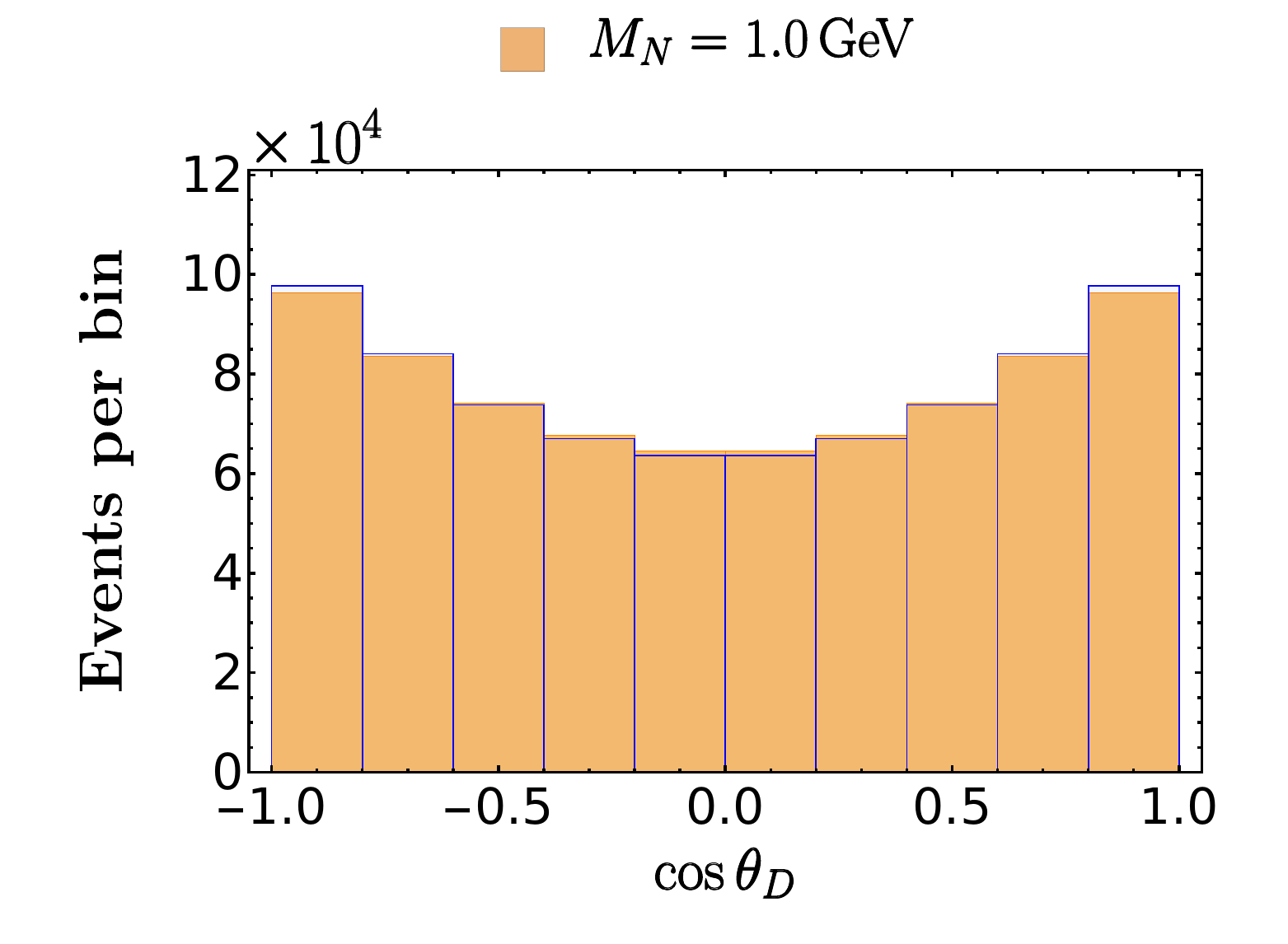}
\end{subfigure}

\vspace{0.5cm}


\begin{subfigure}{0.325\textwidth}
    \centering
    \includegraphics[width=\linewidth]{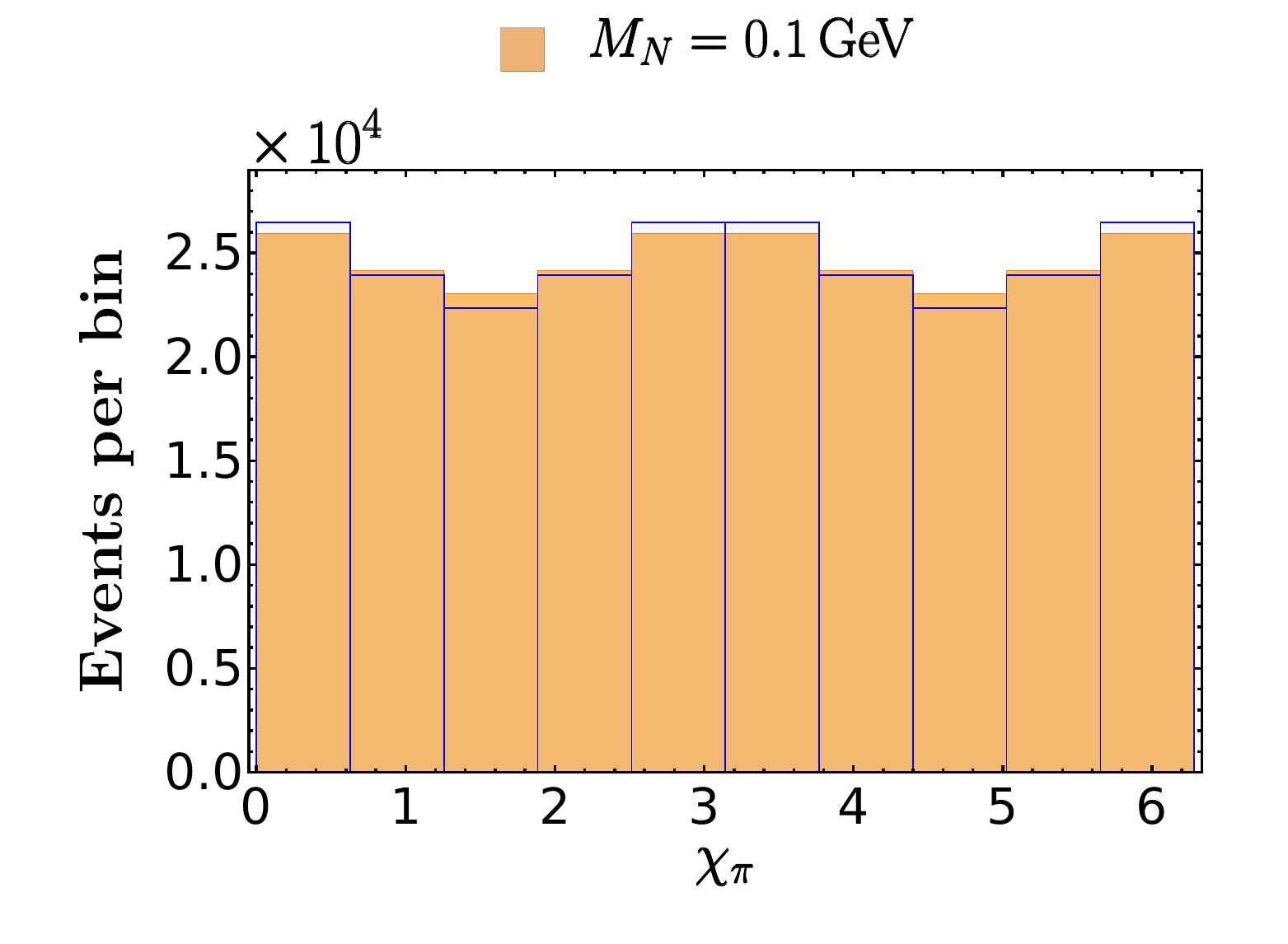}
\end{subfigure}
\hfill
\begin{subfigure}{0.325\textwidth}
    \centering
    \includegraphics[width=\linewidth]{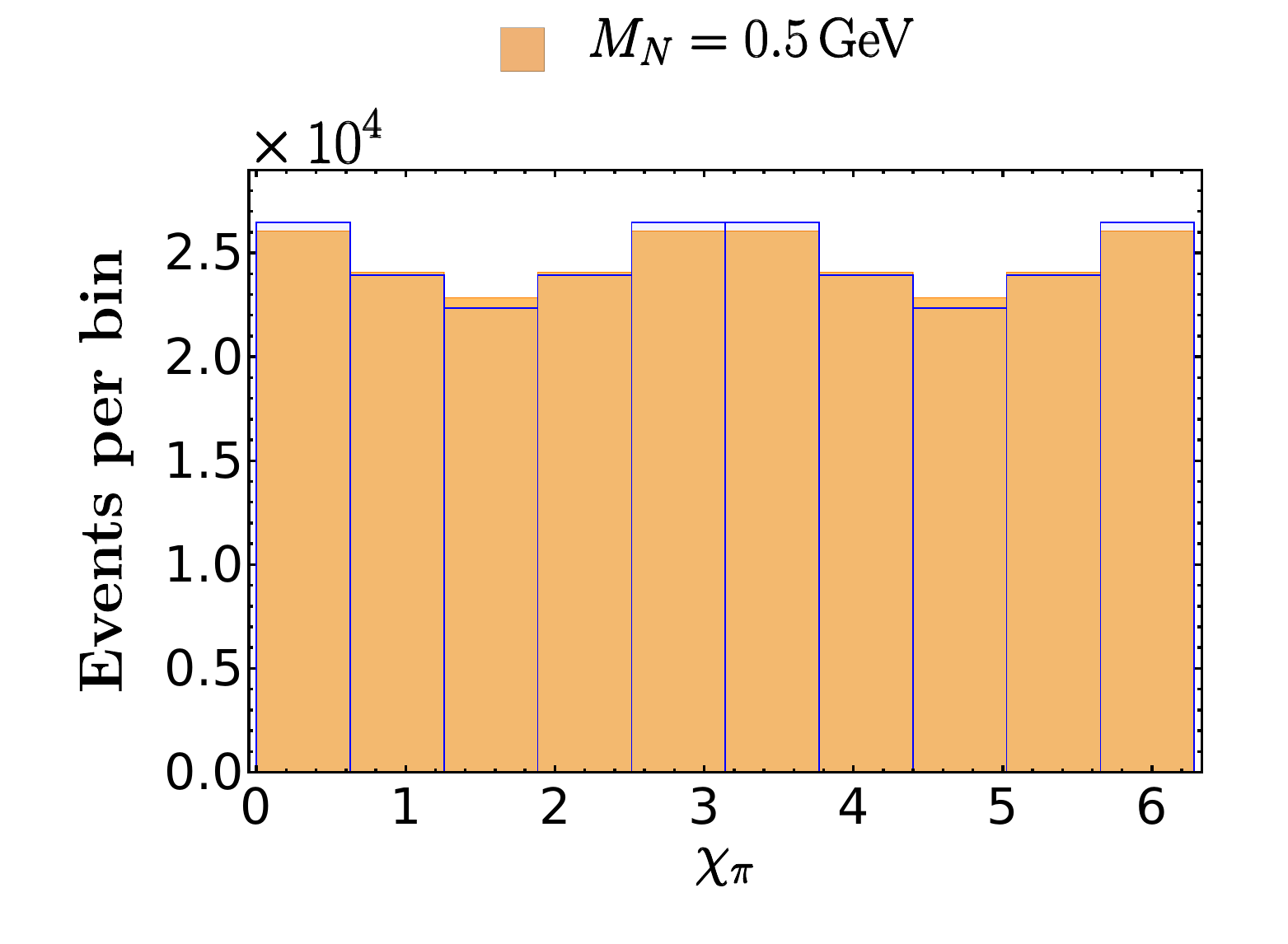}
\end{subfigure}
\hfill
\begin{subfigure}{0.325\textwidth}
    \centering
    \includegraphics[width=\linewidth]{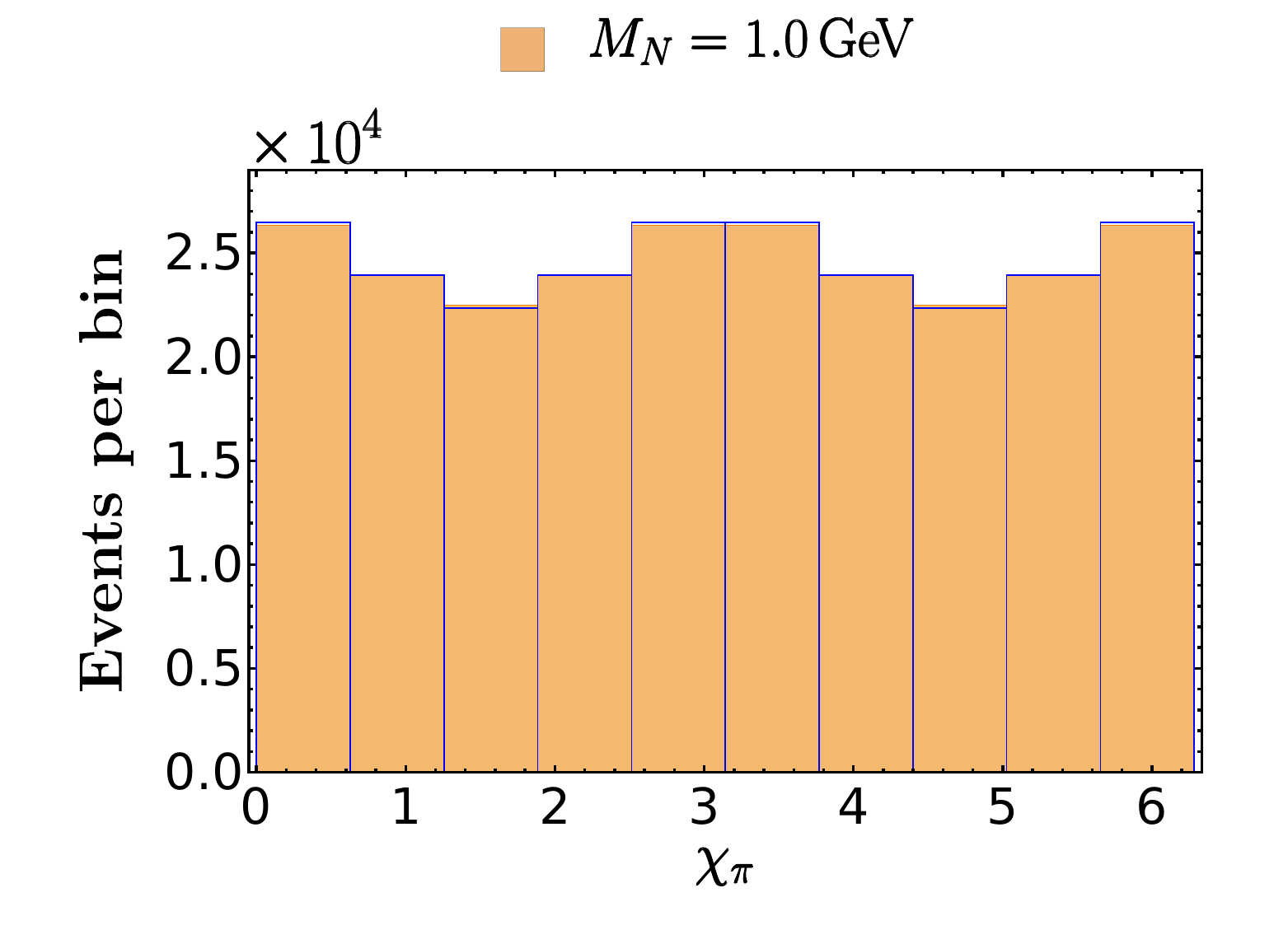}
\end{subfigure}

\caption{Similar to Fig.~\ref{fig:cvrrp5_combined} for the tensor operator with $|C_{T}^{RR}|=0.12$. }
\label{fig:4}
\end{figure}

We present the $q^2$, $\cos\theta_D$, $\cos\theta_{\pi}$, and $\chi_{\pi}$ distributions at Belle~II. The relationship between the differential decay rate and the differential number of events is 
\begin{equation}
    \frac{dN_{\bar B \to D^* \tau \bar\nu_\tau}}{dq^2} = N_{B\bar{B}} \cdot \epsilon(q^2) \cdot \tau_B \cdot \frac{d\Gamma}{dq^2}\,,
\end{equation}
where $\tau_B = 1.58 {\rm\ ps}$ is the $B$ lifetime and
$N_{B\bar{B}}$ is the total number of $B\bar{B}$ pairs produced, which is related to the integrated luminosity by 
$N_{B\bar{B}} = \mathcal{L}_{\text{int}} \times \sigma(\Upsilon(4S)) $. Adopting the target Belle~II integrated luminosity of $\mathcal{L}_{\text{int}} =50 {\rm\ ab^{-1}}$, and taking $\sigma(\Upsilon(4S)) = 1.1 {\rm\ nb}$ \cite{BaBar:2014omp}, we get $N_{B\bar{B}} = 5.5\times 10^{10}  $ . 
 We assume a constant net efficiency of 0.02\% by approximately accounting for improvements of Belle~II relative to Belle~\cite{belleanomaly1}, and arrive at 
\begin{equation}
    \frac{dN_{\bar B \to D^* \tau \bar\nu_\tau}}{dq^2} \approx 2.64 \times 10^{19} \ \text{GeV}^{-1} \cdot \frac{d\Gamma}{dq^2}\,,
\end{equation}
 which corresponds to $1.55 \times 10^5$ $\bar B^0 \to D^* \tau \bar\nu_\tau$ events. 
In Figs.~\ref{fig:cvrrp5_combined} and~\ref{fig:4} we show distributions for three values of $m_N$, and $|C_V^{RR}| = 0.43$ and $|C_T^{RR}| = 0.12$, the smallest values in Table~\ref{tab:BcTauNu} for the respective WCs. 
Following Belle~II, we choose 10 equal-sized bins for the angular distributions~\cite{Belle:2018ezy}.
 Although the projected $q^2$ resolution expected at Belle~II~\cite{Adachi:2018qme} is
\begin{equation}
\sigma_{q^2} \simeq q^2 \sqrt{(0.0011 \sqrt{q^2})^2 + (0.0025)^2} \,,
\end{equation}
we conservatively choose only 20 equal-sized bins for the $q^2$ distribution.

To calculate these distributions, we use the FNAL/MILC determination of the $B \to D^*$ form factors~\cite{fermilab}. Note that the JLQCD~\cite{jlqcd} and HPQCD~\cite{hpqcd} form factors deviate from the FNAL/MILC~\cite{fermilab} form factors. An analysis with these alternative form factors may shift the predicted distributions and the inferred sensitivity to the WCs. Qualitatively, it has been shown that $C_V^{RR}$ has a strong correlation with the $a^g_0$ form factor~\cite{Kapoor:2024ufg}. Therefore, using the JLQCD or HPQCD form factors can change $a^g_0$, and consequently, $C_V^{RR}$. More details on the FNAL/MILC form factor are provided in Appendix~\ref{app:bglformfactors}.

The NP and SM distributions in Figs.~\ref{fig:cvrrp5_combined} and~\ref{fig:4} are normalized to have the same number of events. The NP contribution in yellow is superimposed on the SM contribution in blue to enable comparison of their shapes. Because we expect little to no correlation between the different Wilson coefficients (as the corresponding operators are sufficiently orthogonal to each other), we assume a single nonzero NP WC for each fit. In addition, since there is no interference between the SM and NP operators, the angular coefficients only contain the squared modulus of the WCs. Therefore, only the magnitude of the WCs can be constrained. 

\subsection{Vector operator}
 In Fig.~\ref{fig:cvrrp5_combined}, a noticeable step at $q^2 = (m_\tau + m_N)^2$ is evident in the  $q^2$ distribution from NP. In addition, the $\cos\theta_{\pi}$ distribution is flatter than for the SM. These features can be exploited to identify a RHN  in experimental data, as we investigate with a $\chi^2$ analysis in Section~\ref{sec:statisticalanalysis}. The other distributions are only slightly affected and are thus not very discriminating.


\subsection{Scalar operators}
As mentioned in Section~\ref{sec:amplitudecalculation}, $H_{S_R}^{\lds} = -H_{S_L}^{\lds}$. In addition, the $J$ functions for $O_S^{RR}$ and $O_S^{LR}$ are the same, except for $J^{(C_S^{RR})} \propto |H_{S_R}^{\lds}|^2$ and $J^{(C_S^{LR})} \propto |H_{S_L}^{\lds}|^2$; see Appendices~\ref{app:csrrjfunc} and~\ref{app:cslrjfunc}, respectively. Consequently, the distributions for the two scalar operators are exactly the same, and negligibly different from those for the SM because each operator affects only one of the twelve $J$ functions. We confirm this with a $\chi^2$ analysis in Section~\ref{sec:statisticalanalysis}, and do not present their distributions.

\subsection{Tensor operator}
Similarly to the scalar operator, there is a noticeable but larger step in the $q^2$ distribution at $q^2 = (m_\tau + m_N)^2$ for the tensor operator. 
Also, its $\cos\theta_{D}$ distribution is significantly different from that of the SM. The fact that the scalar and tensor operators yield different spectral modifications from the SM can help discriminate between scalar and tensor operators; see 
Section~\ref{sec:statisticalanalysis}.

\section{Statistical analysis}\label{sec:statisticalanalysis}
We now estimate the size of the Wilson coefficients needed to discover NP at Belle~II
via a shape analysis of the various distributions. 
We interpret the total number of events at Belle~II as arising from the SM or from the SM plus NP, and apply the $\chi^2$ function defined in~\cite{Han:2022uho}:
\begin{align}
\chi^2 =
\sum_{i=1}^{N_{\text{bin}}}
\left[
2 \left(
(1+\alpha_i) N_i^{\text{th}} - N_i^{\text{obs}}
+ N_i^{\text{obs}} \ln
\frac{N_i^{\text{obs}}}{(1+\alpha_i) N_i^{\text{th}}}
\right)
+ \frac{\alpha_i^2}{\sigma_s^2}
\right],
\end{align}
where $N_i^{\rm th}$ and $N_i^{\rm obs}$ are the SM and SM+NP number of events, respectively, in the $i^{\rm th}$ bin. 
We take $N_{\rm bin}=20$ for the $q^2$ spectra and $N_{\rm bin}=10$ for the angular distributions. 
Correspondingly, we introduce 20 and 10 nuisance parameters $\alpha_i$, respectively, to allow for per-bin systematic scaling with a penalty $\sigma_s=10\%$ that includes hadronic form-factor uncertainties and accounts for systematic uncertainties.
We marginalize over $\alpha_i$ to obtain the minimum $\chi^2$, with $\chi^2_{\rm min} = 9$ for $3\sigma$ evidence for NP (for effectively one degree of freedom).

\begin{figure}[t]
\centering

 \includegraphics[width=0.480\textwidth]{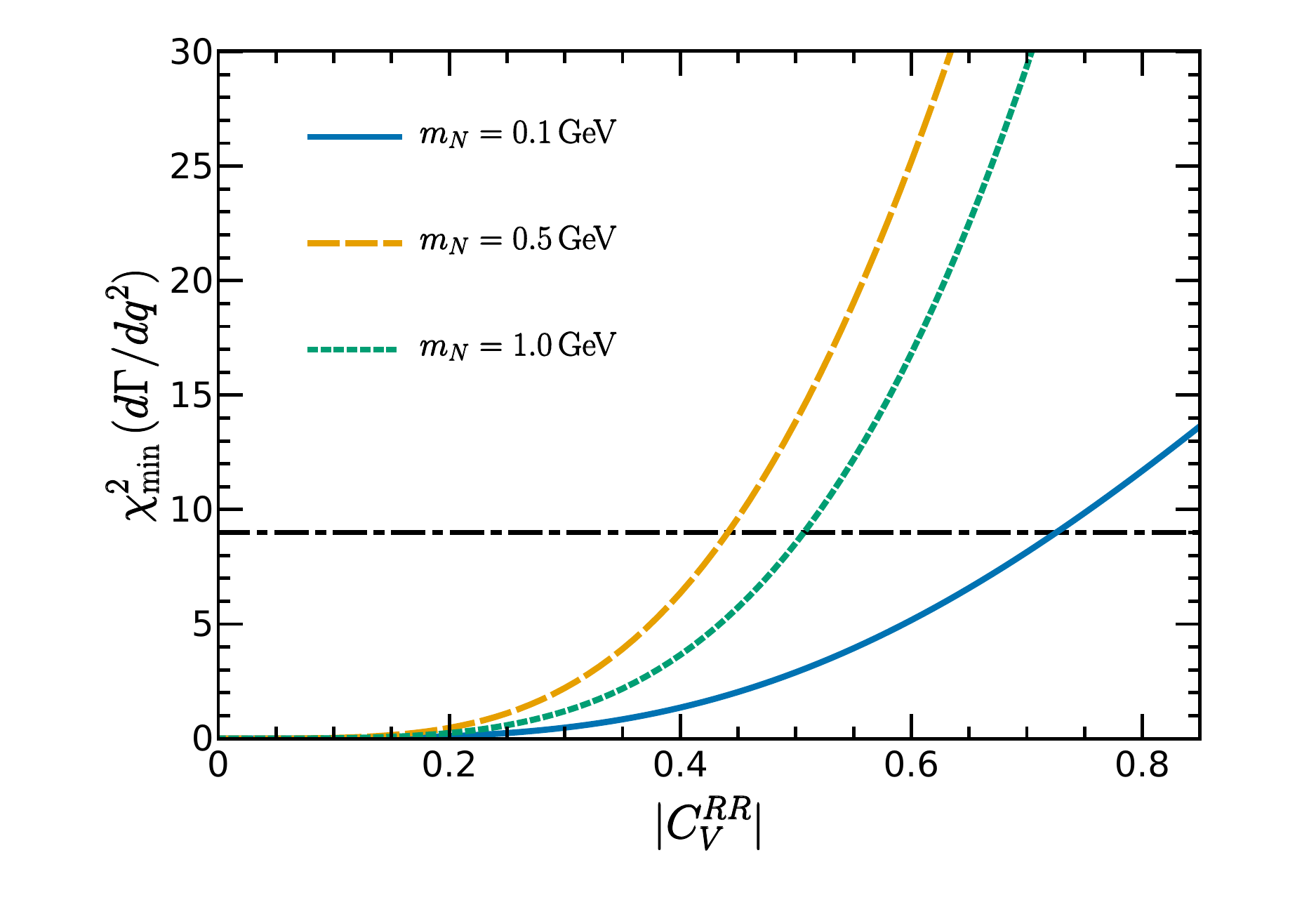}
\includegraphics[width=0.480\textwidth]{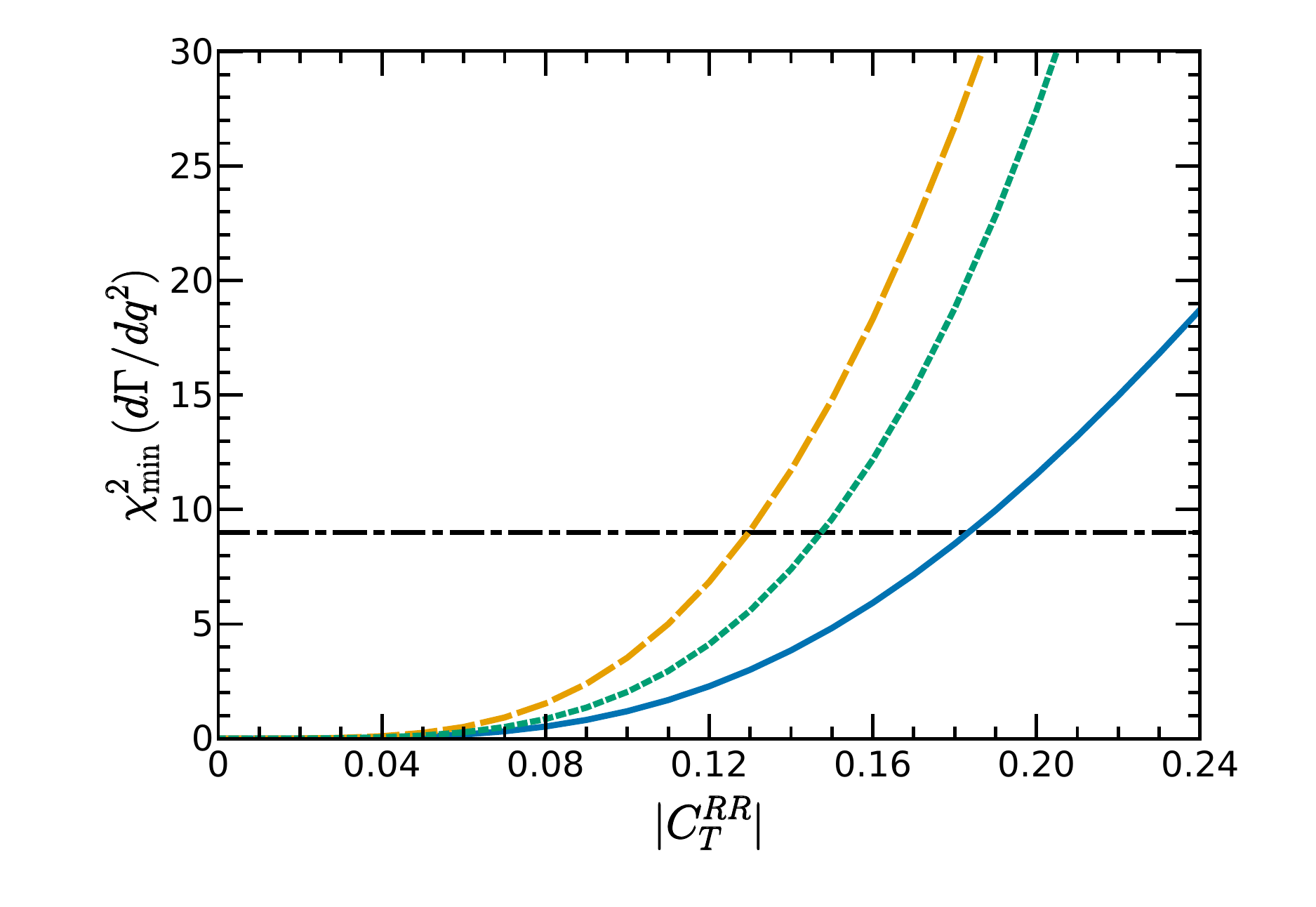}

\vspace{0.3cm}

\includegraphics[width=0.480\textwidth]{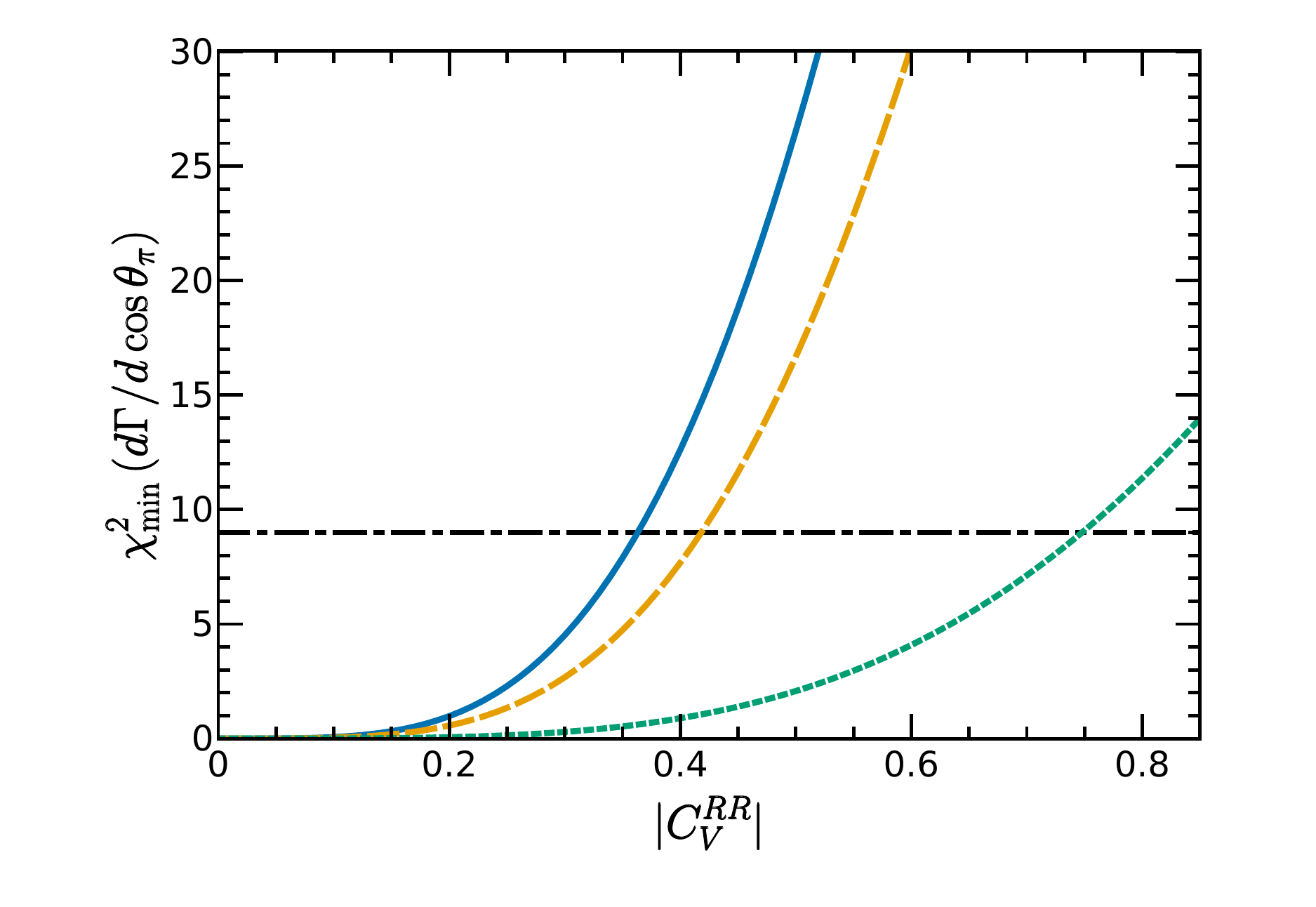}
\includegraphics[width=0.480\textwidth]{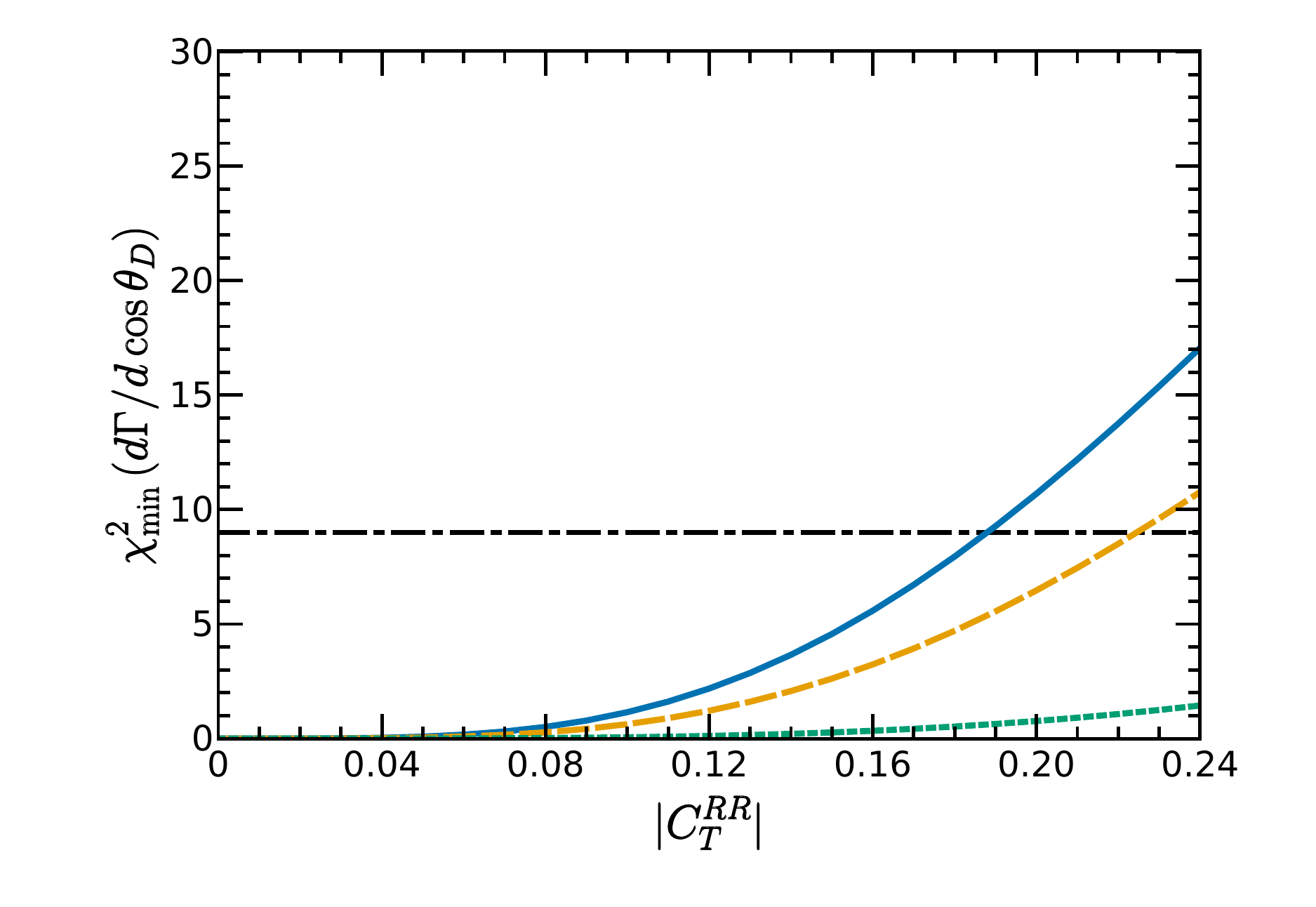}

\vspace{0.3cm}

\caption{$\chi^2_{\rm min}$ as a function of $|C_V^{RR}|$ and $|C_T^{RR}|$ from an analysis of $q^2$ distributions (top panels), and $\cos \theta_\pi$ and $\cos \theta_D$ distributions (bottom panels) for $m_N = 0.1$~GeV (solid blue curves), $m_N =0.5$~GeV (dashed yellow curves) and $m_N =1$~GeV (dotted green curves). The horizontal dot-dashed lines mark the threshold for $3\sigma$ evidence. Note the $2\sigma$ upper bounds on the WCs from $R(D^*)$ in Table~\ref{tab:BcTauNu}.
}
\label{fig:chi2_all_operators}
\end{figure}

In Fig.~\ref{fig:chi2_all_operators} we show $\chi^2_{\rm min}$ as a function of $|C_V^{RR}|$ and $|C_T^{RR}|$, obtained from analyses of the $q^2$, $\cos \theta_\pi$ and $\cos \theta_D$ distributions for $m_N=0.1$~GeV, 0.5~GeV and 1~GeV.
Clearly, the sensitivity of Belle-II data to NP effects depends strongly on the  RHN mass and on the observable. Although the $q^2$ distribution has $\lesssim 1.3\sigma$ sensitivity to $m_N=0.1$~GeV for $|C_V^{RR}| < 0.43$ (the $2\sigma$ bound from $R(D^*)$), the $\cos \theta_\pi$ distribution provides more than $3\sigma$ sensitivity for $|C_V^{RR}| \gtrsim 0.35$.  In contrast, the $q^2$ distribution is much more sensitive to $m_N=1$~GeV than the $\cos \theta_\pi$ distribution. It is also evident that the $q^2$ distribution has much greater sensitivity to $|C_T^{RR}|$ than the 
$\cos \theta_D$ distribution. In fact, for values of $|C_T^{RR}|$ compatible with the $R(D^*)$ bounds in Table~\ref{tab:BcTauNu}, the  $\cos \theta_D$ distribution has $\lesssim 1.5\sigma$ sensitivity for all RHN masses. The qualitative aspects of these results could have been anticipated from Figs.~\ref{fig:cvrrp5_combined} and~\ref{fig:4}.
Note that heavier RHNs may also produce large effects in missing mass distributions. 
For the values of $|C_S^{LR}|$ and $|C_S^{RR}|$ allowed by $R(D)$ at $2\sigma$, the highest statistical significance is $0.33\sigma$ for 
$m_N = 0.1$~GeV and $0.17\sigma$ for 
$m_N = 1$~GeV, which justifies why we do not present their distributions.

Our analysis includes only the SM and SM+NP signal distributions and does not explicitly model experimental backgrounds. At Belle~II, other one-prong decays may feed into the $\tau^- \to \pi^- \nu_\tau$ sample. The most relevant channel is $\tau^-\to \rho^- (\to \pi^-\pi^0)\nu_\tau $ with the $\pi^0$  not reconstructed. Feed down can also arise from $\tau^- \to a_1^- \nu_\tau$, particularly $a_1^- \to \pi^-\pi^0 \pi^0$, when one or both neutral pions are not reconstructed, and from leptonic tau decays when the charged lepton is misidentified as a pion \cite{belleanomaly3}. In addition, fake $D^*$ combinations and imperfect modeling of the SM $B\to D^* \tau^-\nu_\tau$ can further distort the reconstructed $q^2$ and angular distributions and reduce the sensitivity in Fig.~\ref{fig:chi2_all_operators}. A quantitative assessment requires the Belle~II event selection and full detector simulation, which is beyond the scope of this work.

\section{Conclusions}
\label{sec:conclusions}
The decays $\BDtaunu$ and $\BDstartaunu$ have been of great interest over the past decade as their measurements deviate from the SM, hinting at possible lepton universality-violating new physics. A right-handed neutrino is well-motivated by the observation of neutrino masses and mixing. We showed that RHNs can be probed in various kinematic distributions of $\BDstaufull$. We employed an effective theory framework to parameterize the effects of a RHN in terms of dimension-6 operators. We calculated the differential distributions in terms of visible final states, including the decay $\tau \to \pi \nu_\tau$ under the assumption that $\tau$ decay is not affected by NP. Our method can be easily adapted to other decays of the $\tau$ lepton. Finally, we performed a $\chi^2$ analysis to study the sensitivity of Belle~II data to a RHN and found that vector and tensor operators can produce significant deviations from the SM in the $q^2$ and $\cos \theta_\pi$ distributions.

\acknowledgments
We thank B.~Bhattacharya, T.~Kretz and M.~Prim for discussions during the initial stages of this work. N.D. thanks K.~Pandey for help with running the code and acknowledges support from the
SERB grant SPG/2022/001238. A.D. is supported in part by the U.S. National Science Foundation under Grant No.~PHY-2309937. D.M. is supported in part by the U.S. Department of Energy under Grant No.~de-sc0010504. L.M. acknowledges ANRF, Government of India, for financial support through the Project
ANRF/CRG/2025/000133.

\section*{Appendices}
\appendix

\section{Canonical form factor parameterization and helicity amplitudes}\label{app:tradformfactors}
The helicity amplitudes are projections of the $\bar{B} \to D^*$ matrix elements in the basis of $W$ boson polarization vectors. To write expressions for the helicity amplitudes, we first express the matrix elements in terms of the form factors.
\par
The vector and axial vector operator matrix elements are given by 
\begin{align}
\langle D^* (p_{D^*},\eds) | \bar{c} \gamma_\mu b | \bar{B}(p_B) \rangle &= -i \epsilon_{\mu\nu\rho\sigma} \eds^{*\nu}p_B^{\rho}p_{D^*}^{\sigma} \frac{2 V(q^2)}{m_B + m_{D^*}}\,, \label{eq:vectormatrixcanonical} \\
\langle D^* (p_{D^*},\eds) | \bar{c} \gamma_\mu \gamma_5 b | \bar{B}(p_B) \rangle &= \epsilon^{*}_{D^*,\mu} (m_B + m_{D^*}) A_1(q^2) \nonumber \\
&- (p_B+p_{D^*})_{\mu}(\eds^*.q) \frac{A_2(q^2)}{m_B + m_{D^*}} \nonumber \\
&- q_\mu (\eds^*.q) \frac{2 m_{D^*}}{q^2} [A_3(q^2) - A_0(q^2)] \,,\label{eq:axialvectormatrixcanonical}
\end{align}
where
\begin{align}
A_3(q^2) = \frac{m_B + m_{D^*}}{2m_{D^*}} A_1(q^2) - \frac{m_B - m_{D^*}}{2m_{D^*}} A_2(q^2)\,.
\end{align}
Here, $q^\mu = p_B^\mu - p_{D^*}^\mu$, $\eds$ is the polarization vector of $D^*$, and $p_{B(D^*)}$ is the momentum of $B(D^*)$. $V$ is the vector form factor, while $A_{0,1,2,3}$ are the axial-vector form factors. Since $A_3(0) = A_0(0)$, only four form factors are independent.
\par
The tensor matrix element can be parameterized as
\begin{align}\label{eq:tensormatrixcanonical}
\langle D^* (p_{D^*},\eds) | \bar{c} \sigma_{\mu\nu}  b | \bar{B}(p_B) \rangle &= \epsilon_{\mu\nu\rho\sigma} \Bigg\{ -\eds^{*\rho} (p_B + p_{D^*})^{\sigma} T_1(q^2) \\ \nonumber
&+ \eds^{*\rho}q^{\sigma} \frac{m_B^2 - m_{D^*}^2}{q^2} [T_1(q^2) - T_2(q^2)] \\
&+ 2 \frac{(\eds^* . q)}{q^2} p_B^{\rho} p_{D^*}^{\sigma} \left[T_1(q^2) - T_2(q^2) - \frac{q^2}{m_B^2 - m_{D^*}^2}T_3(q^2) \right] \Bigg\}\,, \nonumber
\end{align}
where $T_{1,2,3}$ are the tensor form factors. The pseudotensor matrix elements can be related to the tensor matrix elements by the relation $\bar{c}\sigma_{\mu\nu} \gamma^5 b = -\frac{i}{2} \epsilon_{\mu\nu\alpha\beta} \bar{c}\sigma^{\alpha\beta}b$, with the convention $\epsilon_{0123} = 1$.

The pseudoscalar form factor can be expressed in terms of $A_0$ as
\begin{align}\label{eq:pseudoscalarmatrixcanonical}
\langle D^* (p_{D^*},\eds) | \bar{c} \gamma_5 b | \bar{B} (p_B) \rangle = -(\eds^*.q) \frac{2 m_{D^*}}{m_b + m_c} A_0(q^2)\,,
\end{align}
where $m_{b(c)}$ is the $b(c)$-quark mass. 
\par 
Having defined the scalar, vector and tensor matrix elements,  using Eq.~\eqref{eq:hadhelamp}, we define the hadronic helicity amplitudes as follows:
\begin{align}\label{eq:helamp}
\begin{aligned}
H_{ V}^\pm(q^2)  &\equiv H_{ V_L,\pm}^{\pm}(q^2) = -H_{ V_R,\mp}^{\mp}(q^2) \\
&= (m_B + m_{D^*}) A_1(q^2) \mp \frac{\sqrt{\lambda_{D^*}(q^2)}}{m_B + m_{D^*}}V(q^2)\,, \\
H_{ V}^0(q^2) &\equiv H_{ V_L,0}^{0}(q^2) = - H_{ V_R,0}^{0}(q^2) \\
&= \frac{m_B + m_{D^*}}{2 m_{D^*}\sqrt{q^2}} \left[ -(m_B^2 - m_{D^*}^2 - q^2) A_1(q^2) + \frac{\lambda_{D^*}(q^2)}{(m_B + m_{D^*})^2}A_2(q^2) \right]\,, \\
H_{P}(q^2) &\equiv H_{P}^0(q^2) = H_{S_R}^0 = - H_{S_L}^0 \\
&= -\frac{\sqrt{\lambda_{D^*}(q^2)}}{m_b + m_c} A_0(q^2)\,, \\
H_{T}^\pm(q^2) &\equiv \pm H_{\rm{T},\pm t}^{\pm}(q^2) \\
&= \frac{1}{\sqrt{q^2}} \bigg[ \sqrt{\lambda_{D^*}(q^2)} T_1(q^2) \pm (m_B^2 - m_{D^*}^2) T_2(q^2) \bigg]\,, \\
H_{T}^0(q^2) &\equiv H^0_{T,+-}(q^2) = H^0_{T,0t}(q^2) \\
&= \frac{1}{2 m_{D^*}} \bigg[ (m_B^2 + 3 m_{D^*}^2 - q^2) T_2(q^2) - \frac{\lambda_{D^*}(q^2) }{(m_B^2 - m_{D^*}^2)}T_3(q^2) \bigg]\,. \\
\end{aligned}
\end{align}
These expressions agree with the results in Refs.~\cite{london} 
and~\cite{helampref2}, and up to an overall sign difference in $H_{T}^0$ with respect to Ref.~\cite{tensorff}.

\section{BGL form factor parameterization}\label{app:bglformfactors}

The BGL parameterization of hadronic form factors uses a series expansion in terms of a small parameter $z$ derived from the conformal mapping of the kinematical variable $w$~\cite{bgl}. This mapping allows $z$ to map the kinematic range of $q^2$ (or $w$) to a small interval, improving the convergence properties of the series.  In addition, with more parameters, the BGL parameterization can accommodate a wider range of possible form-factor behaviors, potentially leading to better fits to experimental data.  The canonical form factors are related to them as follows: 
\begin{align}\label{eq:bgltostdff}
\begin{aligned}
g&=\frac{2}{m_B+m_{D^*}}V\,, \\
f&=(m_B+m_{D^*})A_1\,, \\
\mathcal{F}_1 &= \frac{1}{2m_{D^*}}\left[ (m_B^2-m_{D^*}^2-q^2)(m_B+m_{D^*}) A_1  -  
\frac{4m_B^2|{ \vec{p}}_{D^*}|^2}{ m_B+m_{D^*}} A_2 \right]\,,  \\
\mathcal{F}_2 &= 2A_0\,,
\end{aligned}
\end{align}
where the $q^2$ dependence of the form factors is implicit. 
With these definitions, the helicity amplitudes in Eq.~\eqref{eq:helamp} can be written in a very simple manner: 
\begin{align} \label{eq:helampbgl}
\begin{aligned}
H_{V}^\pm(w) &= f\mp g m_B |{\vec{p}_{D^*}}|\,, \\
H_{V}^0(w) &=\frac{\mathcal{F}_1 }{m_B\sqrt{1-2wr+r^2}}\,, \\
H_{P}(w)&=  -\frac{\mathcal{F}_2 m_B |{ \vec{p}_{D^*}}|}{m_b + m_c}\,, \\
H_{T}^\pm(w)&=\frac{\pm f (m_b - m_c) + g m_B |{ \vec{p}_{D^*}}|(m_b + m_c)}{ m_B\sqrt{1-2wr+r^2} }\,, \\
H_{T}^0(w) &= \frac{-(m_b-m_c)(-\mathcal{F}_1 (m_B^2 - m_{D^*}^2) + 2 \mathcal{F}_2 m_B^2 |{ \vec{p}_{D^*}}|^2 ) }{(m_B^2 - m_{D^*}^2) {(m_B^2 + m_{D^*}^2 - 2m_B m_{D^*} w  )}}\,,
\end{aligned}
\end{align}
where  we have made a change of variable from $q^2$ to the dimensionless variable $w = \frac{m_B^2 + m_{D^*}^2 - q^2}{2 m_B m_{D^*}}$.  Note that the pseudoscalar and tensor form factors are reduced to the above four form factors via the relations given in Appendix~A of Ref.~\cite{bdlnumypaper}.
\par
The momentum dependence of these form factors is given by a 
$z$-expansion with
\begin{equation}
z \equiv \frac{\sqrt{w+1}-\sqrt{2}}{\sqrt{w+1}+\sqrt{2}}\,.
\end{equation}
The form factors are
\begin{align} \label{eq:bglff}
\begin{aligned}
g(z) &= \frac{1}{P_{1^-}(z)\phi_g(z)} \sum_{n=0}^{\infty} a^g_n z_n\,,  \quad \quad \;
f(z) = \frac{1}{P_{1^+}(z)\phi_f(z)} \sum_{n=0}^{\infty} a^f_n z^n\,, \\
\mathcal{F}_1(z) &= \frac{1}{P_{1^+}(z)\phi_{\mathcal{F}_1}(z)} \sum_{n=0}^{\infty} a^{\mathcal{F}_1}_n z_n\,,  \quad
\mathcal{F}_2(z) = \frac{1}{P_{0^-}(z)\phi_{\mathcal{F}_2}(z)} \sum_{n=0}^{\infty} a^{\mathcal{F}_2}_n z_n\,.
\end{aligned}
\end{align}
Note that the expansion coefficients satisfy the unitarity conditions,
\begin{align}\label{unitarity}
\sum_{n=0}^\infty  (a^g_n)^2<1\,, \quad 
\sum_{n=0}^\infty  (a^f_n)^2+(a^{\mathcal{F}_1}_n)^2<1\,, \quad 
\sum_{n=0}^\infty  (a^{\mathcal{F}_2}_n)^2<1\,. 
\end{align}
From Eq.~\eqref{eq:bgltostdff}, we find that the form factors are not completely independent. 

In the limit of zero-recoil, i.e., $|{ \vec{p}_{D^*}}|=0$ $(w_{\rm min}=1)$, we find that both $f$ and $\mathcal{F}_1$ depend only on the $A_1$ form factor and are related by 
\begin{align}\label{eq:relF1}
\mathcal{F}_1(0)=(m_B-m_{D^*})f(0) \,.
\end{align}
On the other hand, for maximum recoil, $q^2=0$ $(w_{\rm max}=\frac{m_B^2+m_{D^*}^2}{2m_Bm_{D^*}})$, we require  $A_0=\frac{m_B+m_{D^*}}{2m_{D^*}}A_1-\frac{m_B-m_{D^*}}{2m_{D^*}}A_2$ to avoid the $q^2$ pole in the axial-vector operator matrix element~\cite{tensorff}. This leads to 
\begin{align}\label{eq:relF2}
\mathcal{F}_1(z(w_{\rm max}))=\frac{(m_B^2-m_{D^*}^2)}{2}\mathcal{F}_2(z(w_{\rm max})) \,.
\end{align}

With these relations two expansion parameters can be eliminated. We choose to eliminate the lowest order $\mathcal{F}_1$ constant $a^{\mathcal{F}_1}_0$ and the highest order $\mathcal{F}_2$ constant $a^{\mathcal{F}_2}_{j_{\rm max}}$.

The functions $P_{1^-, 1^+, 0^-}$ and $\phi_{g, f, \mathcal{F}_1, \mathcal{F}_2}$, whose expressions can be found in Ref.~\cite{bgl}, contain several input parameters. We employ the values in Ref.~\cite{fermilab} (originally obtained in Ref.~\cite{gambino}) as input parameters,\footnote{These values are very different from the ones of Ref.~\cite{bgl}, which are used in Ref.~\cite{belle19}. The choice of these values affects the relationship between the form factors and their expansion coefficients (see Eq.~\ref{eq:bglff}). } and consider the number of spectator quarks including SU(3) breaking to be 2.6,  $m_B = 5.280$~GeV and $m_{D^*} = 2.010$~GeV; see Table~\ref{table:bglinput}. 

\begin{table}[t]
\begin{center}
\renewcommand{\arraystretch}{1.4}
\begin{tabular}{|c||c|c|}
\hline 
Type & $B_c^{(*)}$ mass (GeV) & $\chi^{T, L}_{1^{\pm}}$ \\ \hline\hline
$f$, $\mathcal{F}_1$ & 6.739, 6.750, 7.145, 7.150  & $3.894\times 10^{-4}\  {\rm GeV}^{-2}$\\ \hline
$g$ & 6.329, 6.920, 7.020, 7.280 &$5.131\times 10^{-4}\  {\rm GeV}^{-2} $ \\ \hline
$\mathcal{F}_2$ & 6.275, 6.842, 7.250 &$1.9421\times 10^{-2}$  \\\hline
\end{tabular}
\end{center}
\caption{\small Input parameters for the $P_{1^-, 1^+, 0^-}$ and $\phi_{g, f, \mathcal{F}_1, \mathcal{F}_2}$ functions used in the BGL form factors in Eq.~\eqref{eq:bglff}~\cite{fermilab}.}
\label{table:bglinput}
\end{table}

\section{$J$ functions}\label{app:J functions}
We present the $J$ functions necessary to write the angular distribution of $\BDstaufull$ in Eq.~\eqref{eq:rateJtau}.

\subsection{SM}
The angular coefficients for the SM are
\begin{align}
J_{1s}^{(\mathrm{SM})}
&= \frac{\pi\, \, m_\tau^4\, |p_D|^2}{8 q^2 (E_\pi - m_\pi)(E_\pi + m_\pi)}
\Biggl[ \bigl( |H_{V-}|^2 + |H_{V+}|^2 \bigr) \nonumber\\[2pt]
&\quad \times \Bigl( 4 E_\pi^3 \sqrt{q^2} (m_\tau^2 - 3 q^2)
+ E_\pi^2 \bigl( 4 q^2 (m_\pi^2 + m_\tau^2) - 2 m_\tau^4 + 6 q^4 \bigr) \nonumber\\[2pt]
&\quad - 2 E_\pi \sqrt{q^2} \bigl( m_\pi^2 (3 m_\tau^2 - 5 q^2)
+ m_\tau^2 (m_\tau^2 + q^2) \bigr) \nonumber\\[2pt]
&\quad - 3 m_\pi^4 q^2 + m_\pi^2 (3 m_\tau^4 - 5 q^4) + m_\tau^4 q^2 \Bigr)
\Biggr]\,,\\[1.5ex]
J_{1c}^{(\mathrm{SM})}
&= -\frac{\pi\, \, m_\tau^4\, |p_D|^2}{2 q^2 (E_\pi^2 - m_\pi^2)}
\Biggl[ H_V^0 H_V^{0*} \Bigl( 4 E_\pi^3 \sqrt{q^2} (q^2 - m_\tau^2)
+ 2 E_\pi^2 (m_\tau^4 + 2 m_\tau^2 q^2 - q^4) \nonumber\\[2pt]
&\quad + 2 E_\pi \sqrt{q^2} \bigl( m_\pi^2 (m_\tau^2 - 3 q^2)
- m_\tau^2 (m_\tau^2 + q^2) \bigr)
+ m_\pi^4 q^2 - m_\pi^2 (m_\tau^4 - 3 q^4) + m_\tau^4 q^2 \Bigr) \nonumber\\[2pt]
&\quad - 4 H_V^t H_V^{t*} (E_\pi^2 - m_\pi^2)
\Bigl( 2 E_\pi m_\tau^2 \sqrt{q^2} - m_\pi^2 q^2 - m_\tau^4 \Bigr)
\Biggr]\,, \\[1.5ex]
J_{2s}^{(\mathrm{SM})}
&= \frac{\pi\, \, m_\tau^4\, |p_D|^2}{8 q^2 (E_\pi^2 - m_\pi^2)}
\Biggl[ \bigl( H_{V-} H_{V-}^* + H_{V+} H_{V+}^* \bigr) \nonumber\\[2pt]
&\quad \times \Bigl( -4 E_\pi^3 \sqrt{q^2} (m_\tau^2 + q^2)
+ 2 E_\pi^2 \bigl( q^2 (2 m_\pi^2 + q^2) + m_\tau^4 + 6 m_\tau^2 q^2 \bigr) \nonumber\\[2pt]
&\quad - 2 E_\pi \sqrt{q^2} (m_\pi^2 + 3 m_\tau^2)(m_\tau^2 + q^2)
- m_\pi^4 q^2 + m_\pi^2 (m_\tau^4 + q^4) + 3 m_\tau^4 q^2 \Bigr)
\Biggr]\,, \\[1.5ex]
J_{2c}^{(\mathrm{SM})}
&= \frac{\pi\, \, m_\tau^4\, |p_D|^2}{2 q^2 (E_\pi^2 - m_\pi^2)}
\Biggl[ H_V^0 H_V^{0*} \Bigl( 4 E_\pi^3 \sqrt{q^2} (m_\tau^2 + q^2)
- 2 E_\pi^2 \bigl( q^2 (2 m_\pi^2 + q^2) + m_\tau^4 + 6 m_\tau^2 q^2 \bigr) \nonumber\\[2pt]
&\quad + 2 E_\pi \sqrt{q^2} (m_\pi^2 + 3 m_\tau^2)(m_\tau^2 + q^2)
+ m_\pi^4 q^2 - m_\pi^2 (m_\tau^4 + q^4) - 3 m_\tau^4 q^2 \Bigr)
\Biggr]\,,  \\[1.5ex]
J_{3}^{(\mathrm{SM})}
&= \frac{\pi\, \, m_\tau^4\, |p_D|^2}{4 q^2 (E_\pi^2 - m_\pi^2)}
\Biggl[ \bigl( H_{V+} H_{V-}^* + H_{V-} H_{V+}^* \bigr) \nonumber\\[2pt]
&\quad \times \Bigl( 4 E_\pi^3 \sqrt{q^2} (m_\tau^2 + q^2)
- 2 E_\pi^2 \bigl( q^2 (2 m_\pi^2 + q^2) + m_\tau^4 + 6 m_\tau^2 q^2 \bigr) \nonumber\\[2pt]
&\quad + 2 E_\pi \sqrt{q^2} (m_\pi^2 + 3 m_\tau^2)(m_\tau^2 + q^2)
+ m_\pi^4 q^2 - m_\pi^2 (m_\tau^4 + q^4) - 3 m_\tau^4 q^2 \Bigr)
\Biggr]\,, \\[1.5ex]
J_{4}^{(\mathrm{SM})}
&= \frac{\pi\, \, m_\tau^4\, |p_D|^2}{8 q^2 (E_\pi^2 - m_\pi^2)}
\Biggl[ \bigl( H_V^{0*} (H_{V-} + H_{V+}) + H_V^0 H_{V-}^* + H_V^0 H_{V+}^* \bigr) \nonumber\\[2pt]
&\quad \times \Bigl( 4 E_\pi^3 \sqrt{q^2} (m_\tau^2 + q^2)
- 2 E_\pi^2 \bigl( q^2 (2 m_\pi^2 + q^2) + m_\tau^4 + 6 m_\tau^2 q^2 \bigr) \nonumber\\[2pt]
&\quad + 2 E_\pi \sqrt{q^2} (m_\pi^2 + 3 m_\tau^2)(m_\tau^2 + q^2)
+ m_\pi^4 q^2 - m_\pi^2 (m_\tau^4 + q^4) - 3 m_\tau^4 q^2 \Bigr)
\Biggr]\,, \\[1.5ex]
J_{5}^{(\mathrm{SM})}
&= \frac{\pi\, \, m_\tau^4\, |p_D|^2}{2 q^2 \sqrt{E_\pi^2 - m_\pi^2}}
\Biggl[ H_{V-}^* \biggl( -2 E_\pi^2 \sqrt{q^2} \bigl(H_V^0 q^2 + H_V^t m_\tau^2\bigr)
+ E_\pi \Bigl[ H_V^0 q^2 \bigl(m_\pi^2 + 2 m_\tau^2 + q^2\bigr) \nonumber\\[2pt]
&\quad + H_V^t \bigl(m_\pi^2 q^2 + m_\tau^4 + 2 m_\tau^2 q^2\bigr) \Bigr]
- \sqrt{q^2} \Bigl[ H_V^0 m_\tau^2 \bigl(m_\pi^2 + q^2\bigr)
+ H_V^t \bigl(m_\pi^2 q^2 + m_\tau^4\bigr) \Bigr] \biggr) \nonumber\\[2pt]
&\quad - 2 E_\pi^2 H_{V-} m_\tau^2 \sqrt{q^2} H_V^{t*}
+ 2 E_\pi^2 H_V^0 q^{3/2} H_{V+}^*
- 2 E_\pi^2 H_V^t m_\tau^2 \sqrt{q^2} H_{V+}^* \nonumber\\[2pt]
&\quad - 2 E_\pi^2 H_{V+} m_\tau^2 \sqrt{q^2} H_V^{t*}
+ \sqrt{q^2} H_V^{0*} (H_{V-} - H_{V+})
(-2 E_\pi \sqrt{q^2} + m_\pi^2 + q^2) \nonumber\\[2pt]
&\quad \times (E_\pi \sqrt{q^2} - m_\tau^2)
+ E_\pi H_{V-} (m_\pi^2 q^2 + m_\tau^4 + 2 m_\tau^2 q^2) H_V^{t*} \nonumber\\[2pt]
&\quad - E_\pi H_V^0 (m_\pi^2 q^2 + 2 m_\tau^2 q^2 + q^4) H_{V+}^*
+ E_\pi H_V^t (m_\pi^2 q^2 + m_\tau^4 + 2 m_\tau^2 q^2) H_{V+}^* \nonumber\\[2pt]
&\quad + E_\pi H_{V+} (m_\pi^2 q^2 + m_\tau^4 + 2 m_\tau^2 q^2) H_V^{t*}
- H_{V-} (m_\pi^2 q^{3/2} + m_\tau^4 \sqrt{q^2}) H_V^{t*} \nonumber\\[2pt]
&\quad + H_V^0 (m_\pi^2 m_\tau^2 \sqrt{q^2} + m_\tau^2 q^{3/2}) H_{V+}^*
- H_V^t (m_\pi^2 q^{3/2} + m_\tau^4 \sqrt{q^2}) H_{V+}^* \nonumber\\[2pt]
&\quad - H_{V+} (m_\pi^2 q^{3/2} + m_\tau^4 \sqrt{q^2}) H_V^{t*}
\Biggr]\,,\\[1.5ex]
J_{6s}^{(\mathrm{SM})}
&= \frac{\pi\, \, m_\tau^4\, |p_D|^2}{\sqrt{q^2}\, \sqrt{E_\pi^2 - m_\pi^2}}
\Biggl[ \bigl( H_{V-} H_{V-}^* - H_{V+} H_{V+}^* \bigr)
\bigl(-2 E_\pi \sqrt{q^2} + m_\pi^2 + q^2\bigr)
\bigl(m_\tau^2 - E_\pi \sqrt{q^2}\bigr) \Biggr]\,, \\[1.5ex]
J_{6c}^{(\mathrm{SM})}
&= \frac{2\pi\, \, m_\tau^4\, |p_D|^2}{q^2 \sqrt{E_\pi^2 - m_\pi^2}}
\Biggl[ (E_\pi - \sqrt{q^2}) \bigl( H_V^t H_V^{0*} + H_V^0 H_V^{t*} \bigr) \nonumber\\[2pt]
&\quad \times \bigl( -2 E_\pi m_\tau^2 \sqrt{q^2} + m_\pi^2 q^2 + m_\tau^4 \bigr)
\Biggr]\,,\\[1.5ex]
J_{7}^{(\mathrm{SM})}
&= \frac{i\pi\, \, m_\tau^4\, |p_D|^2}{2 q^2 \sqrt{E_\pi^2 - m_\pi^2}}
\Biggl[ H_{V-}^* \biggl( 2 E_\pi^2 \sqrt{q^2} \bigl(H_V^0 q^2 + H_V^t m_\tau^2\bigr)
- E_\pi \Bigl[ H_V^0 q^2 \bigl(m_\pi^2 + 2 m_\tau^2 + q^2\bigr) \nonumber\\[2pt]
&\quad + H_V^t \bigl(m_\pi^2 q^2 + m_\tau^4 + 2 m_\tau^2 q^2\bigr) \Bigr]
+ \sqrt{q^2} \Bigl[ H_V^0 m_\tau^2 \bigl(m_\pi^2 + q^2\bigr)
+ H_V^t \bigl(m_\pi^2 q^2 + m_\tau^4\bigr) \Bigr] \biggr) \nonumber\\[2pt]
&\quad - 2 E_\pi^2\, H_{V-}\, m_\tau^2 \sqrt{q^2}\, H_V^{t*}
+ 2 E_\pi^2\, H_V^0\, q^{3/2}\, H_{V+}^*
- 2 E_\pi^2\, H_V^t\, m_\tau^2 \sqrt{q^2}\, H_{V+}^* \nonumber\\[2pt]
&\quad + 2 E_\pi^2\, H_{V+}\, m_\tau^2 \sqrt{q^2}\, H_V^{t*}
+ \sqrt{q^2}\, H_V^{0*} \bigl(H_{V-} + H_{V+}\bigr)
\bigl(-2 E_\pi \sqrt{q^2} + m_\pi^2 + q^2\bigr) \nonumber\\[2pt]
&\quad \times \bigl(E_\pi \sqrt{q^2} - m_\tau^2\bigr)
+ E_\pi\, H_{V-} \bigl(m_\pi^2 q^2 + m_\tau^4 + 2 m_\tau^2 q^2\bigr) H_V^{t*} \nonumber\\[2pt]
&\quad - E_\pi\, H_V^0 \bigl(m_\pi^2 q^2 + 2 m_\tau^2 q^2 + q^4\bigr) H_{V+}^*
+ E_\pi\, H_V^t \bigl(m_\pi^2 q^2 + m_\tau^4 + 2 m_\tau^2 q^2\bigr) H_{V+}^* \nonumber\\[2pt]
&\quad - E_\pi\, H_{V+} \bigl(m_\pi^2 q^2 + m_\tau^4 + 2 m_\tau^2 q^2\bigr) H_V^{t*}
- H_{V-} \bigl(m_\pi^2 q^{3/2} + m_\tau^4 \sqrt{q^2}\bigr) H_V^{t*} \nonumber\\[2pt]
&\quad + H_V^0 \bigl(m_\pi^2 m_\tau^2 \sqrt{q^2} + m_\tau^2 q^{3/2}\bigr) H_{V+}^*
- H_V^t \bigl(m_\pi^2 q^{3/2} + m_\tau^4 \sqrt{q^2}\bigr) H_{V+}^* \nonumber\\[2pt]
&\quad + H_{V+} \bigl(m_\pi^2 q^{3/2} + m_\tau^4 \sqrt{q^2}\bigr) H_V^{t*}
\Biggr]\,, \\[1.5ex]
J_{8}^{(\mathrm{SM})}
&= \frac{i\pi\, \, m_\tau^4\, |p_D|^2}{8 q^2 (E_\pi^2 - m_\pi^2)}
\Biggl[ \bigl( H_V^{0*} (H_{V+} - H_{V-}) + H_V^0 H_{V-}^* - H_V^0 H_{V+}^* \bigr) \nonumber\\[2pt]
&\quad \times \Bigl( -4 E_\pi^3 \sqrt{q^2} (m_\tau^2 + q^2)
+ 2 E_\pi^2 \bigl( q^2 (2 m_\pi^2 + q^2) + m_\tau^4 + 6 m_\tau^2 q^2 \bigr) \nonumber\\[2pt]
&\quad - 2 E_\pi \sqrt{q^2} (m_\pi^2 + 3 m_\tau^2) (m_\tau^2 + q^2)
- m_\pi^4 q^2 + m_\pi^2 (m_\tau^4 + q^4) + 3 m_\tau^4 q^2 \Bigr)
\Biggr]\,, \\[1.5ex]
J_{9}^{(\mathrm{SM})}
&= \frac{i\pi\, \, m_\tau^4\, |p_D|^2}{4 q^2 (E_\pi^2 - m_\pi^2)}
\Biggl[ \bigl( H_{V+} H_{V-}^* - H_{V-} H_{V+}^* \bigr) \nonumber\\[2pt]
&\quad \times \Bigl( -4 E_\pi^3 \sqrt{q^2} (m_\tau^2 + q^2)
+ 2 E_\pi^2 \bigl( q^2 (2 m_\pi^2 + q^2) + m_\tau^4 + 6 m_\tau^2 q^2 \bigr) 
\nonumber \\[2pt] 
&\quad - 2 E_\pi \sqrt{q^2} (m_\pi^2 + 3 m_\tau^2) (m_\tau^2 + q^2)
- m_\pi^4 q^2 + m_\pi^2 (m_\tau^4 + q^4) + 3 m_\tau^4 q^2 \Bigr)
\Biggr]\,. 
\end{align}

\subsection{$O_V^{RR}$}
The angular coefficients for $O_V^{RR}$ are   
\begin{align}
J_{1s}^{(C_V^{RR})}
&= -\frac{\pi\, |C_V^{RR}|^2|\, m_\tau^2\, |p_D|^2}{8 (E_\pi^2 - m_\pi^2)}
\Biggl[ \bigl( H_{V-} H_{V-}^* + H_{V+} H_{V+}^* \bigr) \nonumber\\[2pt]
&\quad \times \Bigl( 8 E_\pi^3 m_\tau^2 (E_{\tau \bar{N}} - 2 \sqrt{q^2})
+ 4 E_\pi^2 \bigl( -3 E_{\tau \bar{N}}^2 m_\pi^2 + 4 E_{\tau \bar{N}} m_\pi^2 \sqrt{q^2} \nonumber\\[2pt]
&\quad + m_\pi^2 (m_\tau^2 + p_{\tau \bar{N}}^2) + m_\tau^4 \bigr)
- 4 E_\pi m_\pi^2 \bigl( E_{\tau \bar{N}} (m_\pi^2 + 3 m_\tau^2)
- 4 m_\tau^2 \sqrt{q^2} \bigr) \nonumber\\[2pt]
&\quad + 16 E_{\tau \bar{N}}^2 m_\pi^4 - 16 E_{\tau \bar{N}} m_\pi^4 \sqrt{q^2}
+ m_\pi^6 - 2 m_\pi^4 m_\tau^2 - 4 m_\pi^4 p_{\tau \bar{N}}^2 - 3 m_\pi^2 m_\tau^4 \Bigr)
\Biggr]\,, \\[1.5ex]
J_{1c}^{(C_V^{RR})}
&= \frac{\pi\, |C_V^{RR}|^2|\, m_\tau^2\, |p_D|^2}{2}
\Biggl[ 4 H_V^t H_V^{t*} \Bigl( -2 E_{\tau \bar{N}} (2 E_\pi m_\tau^2 + m_\pi^2 \sqrt{q^2}) \nonumber\\[2pt]
&\quad + 2 E_\pi m_\tau^2 \sqrt{q^2} + 2 E_{\tau \bar{N}}^2 m_\pi^2
+ m_\pi^2 (m_\tau^2 + 2 p_{\tau \bar{N}}^2) + m_\tau^4 \Bigr) \nonumber\\[2pt]
&\quad - \frac{H_V^0 H_V^{0*}}{E_\pi^2 - m_\pi^2}
\Bigl( 8 E_\pi^3 m_\tau^2 (E_{\tau \bar{N}} - \sqrt{q^2}) \nonumber\\[2pt]
&\quad + 4 E_\pi^2 m_\pi^2 (-3 E_{\tau \bar{N}}^2 + 2 E_{\tau \bar{N}} \sqrt{q^2} + p_{\tau \bar{N}}^2) \nonumber\\[2pt]
&\quad + 4 E_\pi m_\pi^2 (E_{\tau \bar{N}} (m_\pi^2 - m_\tau^2)
+ 2 m_\tau^2 \sqrt{q^2}) \nonumber\\[2pt]
&\quad + 8 E_{\tau \bar{N}}^2 m_\pi^4 - 8 E_{\tau \bar{N}} m_\pi^4 \sqrt{q^2}
- m_\pi^2 (m_\pi^4 + 2 m_\pi^2 (m_\tau^2 + 2 p_{\tau \bar{N}}^2) + m_\tau^4) \Bigr)
\Biggr]\,, \\[1.5ex]
J_{2s}^{(C_V^{RR})}
&= \frac{\pi\, |C_V^{RR}|^2|\, m_\tau^2\, |p_D|^2}{8 (E_\pi^2 - m_\pi^2)}
\Biggl[ \bigl( H_{V-} H_{V-}^* + H_{V+} H_{V+}^* \bigr) \nonumber\\[2pt]
&\quad \times \Bigl( 8 E_\pi^3 E_{\tau \bar{N}} m_\tau^2
- 4 E_\pi^2 \bigl( 3 E_{\tau \bar{N}}^2 m_\pi^2
+ m_\pi^2 (m_\tau^2 - p_{\tau \bar{N}}^2) + m_\tau^4 \bigr) \nonumber\\[2pt]
&\quad + 4 E_\pi E_{\tau \bar{N}} m_\pi^2 (3 m_\pi^2 + m_\tau^2)
- 3 m_\pi^6 - 2 m_\pi^4 (m_\tau^2 + 2 p_{\tau \bar{N}}^2)
+ m_\pi^2 m_\tau^4 \Bigr)
\Biggr]\,, \\[1.5ex]
J_{2c}^{(C_V^{RR})}
&= \frac{\pi\, |C_V^{RR}|^2|\, m_\tau^2\, |p_D|^2}{2 (E_\pi^2 - m_\pi^2)}
\Biggl[ H_V^0 H_V^{0*} \Bigl( -8 E_\pi^3 E_{\tau \bar{N}} m_\tau^2
+ 4 E_\pi^2 \bigl( 3 E_{\tau \bar{N}}^2 m_\pi^2 \nonumber\\[2pt]
&\quad + m_\pi^2 (m_\tau^2 - p_{\tau \bar{N}}^2) + m_\tau^4 \bigr)
- 4 E_\pi E_{\tau \bar{N}} m_\pi^2 (3 m_\pi^2 + m_\tau^2) \nonumber\\[2pt]
&\quad + 3 m_\pi^6 + 2 m_\pi^4 (m_\tau^2 + 2 p_{\tau \bar{N}}^2)
- m_\pi^2 m_\tau^4 \Bigr)
\Biggr]\,, \\[1.5ex]
J_{3}^{(C_V^{RR})}
&= -\frac{\pi\, |C_V^{RR}|^2|\, m_\tau^2\, |p_D|^2}{4 (E_\pi^2 - m_\pi^2)}
\Biggl[ \bigl( H_{V+} H_{V-}^* + H_{V-} H_{V+}^* \bigr) \nonumber\\[2pt]
&\quad \times \Bigl( 8 E_\pi^3 E_{\tau \bar{N}} m_\tau^2
- 4 E_\pi^2 \bigl( 3 E_{\tau \bar{N}}^2 m_\pi^2
+ m_\pi^2 (m_\tau^2 - p_{\tau \bar{N}}^2) + m_\tau^4 \bigr) \nonumber\\[2pt]
&\quad + 4 E_\pi E_{\tau \bar{N}} m_\pi^2 (3 m_\pi^2 + m_\tau^2)
- 3 m_\pi^6 - 2 m_\pi^4 (m_\tau^2 + 2 p_{\tau \bar{N}}^2)
+ m_\pi^2 m_\tau^4 \Bigr)
\Biggr]\,, \\[1.5ex]
J_{4}^{(C_V^{RR})}
&= -\frac{\pi\, |C_V^{RR}|^2|\, m_\tau^2\, |p_D|^2}{8 (E_\pi^2 - m_\pi^2)}
\Biggl[ \bigl( H_V^{0*} (H_{V-} + H_{V+}) + H_V^0 H_{V-}^* + H_V^0 H_{V+}^* \bigr) \nonumber\\[2pt]
&\quad \times \Bigl( 8 E_\pi^3 E_{\tau \bar{N}} m_\tau^2
- 4 E_\pi^2 \bigl( 3 E_{\tau \bar{N}}^2 m_\pi^2
+ m_\pi^2 (m_\tau^2 - p_{\tau \bar{N}}^2) + m_\tau^4 \bigr) \nonumber\\[2pt]
&\quad + 4 E_\pi E_{\tau \bar{N}} m_\pi^2 (3 m_\pi^2 + m_\tau^2)
- 3 m_\pi^6 - 2 m_\pi^4 (m_\tau^2 + 2 p_{\tau \bar{N}}^2)
+ m_\pi^2 m_\tau^4 \Bigr)
\Biggr]\,, \\[1.5ex]
J_{5}^{(C_V^{RR})}
&= -\frac{\pi\, |C_V^{RR}|^2|\, m_\tau^2\, |p_D|^2}{2 \sqrt{E_\pi^2 - m_\pi^2}}
\Biggl[ H_{V-}^* \biggl( -2 E_\pi^2 m_\tau^2 \bigl( 2 E_{\tau \bar{N}} H_V^t + \sqrt{q^2} (H_V^0 - H_V^t) \bigr) \nonumber\\[2pt]
&\quad + E_\pi \Bigl( 4 E_{\tau \bar{N}}^2 H_V^t m_\pi^2
+ 2 E_{\tau \bar{N}} m_\pi^2 \sqrt{q^2} (H_V^0 - H_V^t)
+ m_\tau^2 (H_V^0 + H_V^t) (m_\pi^2 + m_\tau^2) \Bigr) \nonumber\\[2pt]
&\quad + m_\pi^2 \Bigl( -2 E_{\tau \bar{N}} (H_V^0 m_\tau^2 + H_V^t m_\pi^2)
+ \sqrt{q^2} (H_V^0 - H_V^t) (m_\pi^2 - m_\tau^2) \Bigr) \biggr) \nonumber\\[2pt]
&\quad - H_V^{0*} (H_{V-} - H_{V+})
\Bigl( 2 E_\pi^2 m_\tau^2 \sqrt{q^2}
- E_\pi \bigl( m_\pi^2 (2 E_{\tau \bar{N}} \sqrt{q^2} + m_\tau^2) + m_\tau^4 \bigr) \nonumber\\[2pt]
&\quad + m_\pi^2 (2 E_{\tau \bar{N}} m_\tau^2 + \sqrt{q^2} (m_\pi^2 - m_\tau^2)) \Bigr) \nonumber\\[2pt]
&\quad - 4 E_\pi^2 E_{\tau \bar{N}} m_\tau^2 \bigl( H_{V-} H_V^{t*} + H_V^t H_{V+}^* + H_{V+} H_V^{t*} \bigr) \nonumber\\[2pt]
&\quad + 2 E_\pi^2 m_\tau^2 \sqrt{q^2} \bigl( H_{V-} H_V^{t*} + H_V^0 H_{V+}^* + H_V^t H_{V+}^* + H_{V+} H_V^{t*} \bigr) \nonumber\\[2pt]
&\quad + 4 E_\pi E_{\tau \bar{N}}^2 m_\pi^2 \bigl( H_{V-} H_V^{t*} + H_V^t H_{V+}^* + H_{V+} H_V^{t*} \bigr) \nonumber\\[2pt]
&\quad - 2 E_\pi E_{\tau \bar{N}} m_\pi^2 \sqrt{q^2} \bigl( H_{V-} H_V^{t*} + H_V^0 H_{V+}^* + H_V^t H_{V+}^* + H_{V+} H_V^{t*} \bigr) \nonumber\\[2pt]
&\quad + E_\pi m_\pi^2 m_\tau^2 \bigl( H_{V-} H_V^{t*} - H_V^0 H_{V+}^* + H_V^t H_{V+}^* + H_{V+} H_V^{t*} \bigr) \nonumber\\[2pt]
&\quad + E_\pi m_\tau^4 \bigl( H_{V-} H_V^{t*} - H_V^0 H_{V+}^* + H_V^t H_{V+}^* + H_{V+} H_V^{t*} \bigr) \nonumber\\[2pt]
&\quad - 2 E_{\tau \bar{N}} m_\pi^4 \bigl( H_{V-} H_V^{t*} + H_V^t H_{V+}^* + H_{V+} H_V^{t*} \bigr) \nonumber\\[2pt]
&\quad + 2 E_{\tau \bar{N}} m_\pi^2 m_\tau^2 H_V^0 H_{V+}^* \nonumber\\[2pt]
&\quad + \sqrt{q^2} m_\pi^4 \bigl( H_{V-} H_V^{t*} + H_V^0 H_{V+}^* + H_V^t H_{V+}^* + H_{V+} H_V^{t*} \bigr) \nonumber\\[2pt]
&\quad - \sqrt{q^2} m_\pi^2 m_\tau^2 \bigl( H_{V-} H_V^{t*} + H_V^0 H_{V+}^* + H_V^t H_{V+}^* + H_{V+} H_V^{t*} \bigr)
\Biggr]\,, \\[1.5ex]
J_{6s}^{(C_V^{RR})}
&= \frac{\pi\, |C_V^{RR}|^2|\, m_\tau^2\, |p_D|^2}{\sqrt{E_\pi^2 - m_\pi^2}}
\Biggl[ \bigl( H_{V-} H_{V-}^* - H_{V+} H_{V+}^* \bigr) \nonumber\\[2pt]
&\quad \times \Bigl( -2 E_\pi^2 m_\tau^2 \sqrt{q^2}
+ E_\pi \bigl( m_\pi^2 (2 E_{\tau \bar{N}} \sqrt{q^2} + m_\tau^2) + m_\tau^4 \bigr) \nonumber\\[2pt]
&\quad + m_\pi^2 \bigl( \sqrt{q^2} (m_\tau^2 - m_\pi^2) - 2 E_{\tau \bar{N}} m_\tau^2 \bigr) \Bigr)
\Biggr]\,, \\[1.5ex]
J_{6c}^{(C_V^{RR})}
&= -\frac{2\pi\, |C_V^{RR}|^2|\, m_\tau^2\, |p_D|^2}{\sqrt{E_\pi^2 - m_\pi^2}}
\Biggl[ \bigl( H_V^t H_V^{0*} + H_V^0 H_V^{t*} \bigr) \nonumber\\[2pt]
&\quad \times \Bigl( 2 E_\pi^2 m_\tau^2 (\sqrt{q^2} - 2 E_{\tau \bar{N}})
+ E_\pi \bigl( 4 E_{\tau \bar{N}}^2 m_\pi^2 - 2 E_{\tau \bar{N}} m_\pi^2 \sqrt{q^2} \nonumber\\[2pt]
&\quad + m_\pi^2 m_\tau^2 + m_\tau^4 \bigr)
- 2 E_{\tau \bar{N}} m_\pi^4
+ m_\pi^2 \sqrt{q^2} (m_\pi^2 - m_\tau^2) \Bigr)
\Biggr]\,, \\[1.5ex]
J_{7}^{(C_V^{RR})}
&= \frac{i\pi\, |C_V^{RR}|^2|\, m_\tau^2\, |p_D|^2}{2 \sqrt{E_\pi^2 - m_\pi^2}}
\Biggl[ H_{V-}^* \biggl( 2 E_\pi^2 m_\tau^2 \bigl( 2 E_{\tau \bar{N}} H_V^t + \sqrt{q^2} (H_V^0 - H_V^t) \bigr) \nonumber\\[2pt]
&\quad - E_\pi \Bigl( 4 E_{\tau \bar{N}}^2 H_V^t m_\pi^2
+ 2 E_{\tau \bar{N}} m_\pi^2 \sqrt{q^2} (H_V^0 - H_V^t)
+ m_\tau^2 (H_V^0 + H_V^t) (m_\pi^2 + m_\tau^2) \Bigr) \nonumber\\[2pt]
&\quad + m_\pi^2 \Bigl( 2 E_{\tau \bar{N}} (H_V^0 m_\tau^2 + H_V^t m_\pi^2)
+ \sqrt{q^2} (H_V^0 - H_V^t) (m_\pi^2 - m_\tau^2) \Bigr) \biggr) \nonumber\\[2pt]
&\quad - H_V^{0*} (H_{V-} + H_{V+})
\Bigl( 2 E_\pi^2 m_\tau^2 \sqrt{q^2}
- E_\pi \bigl( m_\pi^2 (2 E_{\tau \bar{N}} \sqrt{q^2} + m_\tau^2) + m_\tau^4 \bigr) \nonumber\\[2pt]
&\quad + m_\pi^2 (2 E_{\tau \bar{N}} m_\tau^2 + \sqrt{q^2} (m_\pi^2 - m_\tau^2)) \Bigr) \nonumber\\[2pt]
&\quad - 4 E_\pi^2 E_{\tau \bar{N}} m_\tau^2 \bigl( H_{V-} H_V^{t*} + H_V^t H_{V+}^* - H_{V+} H_V^{t*} \bigr) \nonumber\\[2pt]
&\quad + 2 E_\pi^2 m_\tau^2 \sqrt{q^2} \bigl( H_{V-} H_V^{t*} + H_V^0 H_{V+}^* + H_V^t H_{V+}^* - H_{V+} H_V^{t*} \bigr) \nonumber\\[2pt]
&\quad + 4 E_\pi E_{\tau \bar{N}}^2 m_\pi^2 \bigl( H_{V-} H_V^{t*} + H_V^t H_{V+}^* - H_{V+} H_V^{t*} \bigr) \nonumber\\[2pt]
&\quad - 2 E_\pi E_{\tau \bar{N}} m_\pi^2 \sqrt{q^2} \bigl( H_{V-} H_V^{t*} + H_V^0 H_{V+}^* + H_V^t H_{V+}^* - H_{V+} H_V^{t*} \bigr) \nonumber\\[2pt]
&\quad + E_\pi m_\pi^2 m_\tau^2 \bigl( H_{V-} H_V^{t*} - H_V^0 H_{V+}^* + H_V^t H_{V+}^* - H_{V+} H_V^{t*} \bigr) \nonumber\\[2pt]
&\quad + E_\pi m_\tau^4 \bigl( H_{V-} H_V^{t*} - H_V^0 H_{V+}^* + H_V^t H_{V+}^* - H_{V+} H_V^{t*} \bigr) \nonumber\\[2pt]
&\quad - 2 E_{\tau \bar{N}} m_\pi^4 \bigl( H_{V-} H_V^{t*} + H_V^t H_{V+}^* - H_{V+} H_V^{t*} \bigr) \nonumber\\[2pt]
&\quad + 2 E_{\tau \bar{N}} m_\pi^2 m_\tau^2 H_V^0 H_{V+}^* \nonumber\\[2pt]
&\quad + \sqrt{q^2} m_\pi^4 \bigl( H_{V-} H_V^{t*} + H_V^0 H_{V+}^* + H_V^t H_{V+}^* - H_{V+} H_V^{t*} \bigr) \nonumber\\[2pt]
&\quad - \sqrt{q^2} m_\pi^2 m_\tau^2 \bigl( H_{V-} H_V^{t*} + H_V^0 H_{V+}^* + H_V^t H_{V+}^* - H_{V+} H_V^{t*} \bigr)
\Biggr]\,, \\[1.5ex]
J_{8}^{(C_V^{RR})}
&= -\frac{i\pi\, |C_V^{RR}|^2|\, m_\tau^2\, |p_D|^2}{8 (E_\pi^2 - m_\pi^2)}
\Biggl[ \bigl( H_V^{0*} (H_{V+} - H_{V-}) + H_V^0 H_{V-}^* - H_V^0 H_{V+}^* \bigr) \nonumber\\[2pt]
&\quad \times \Bigl( 8 E_\pi^3 E_{\tau \bar{N}} m_\tau^2
- 4 E_\pi^2 \bigl( 3 E_{\tau \bar{N}}^2 m_\pi^2
+ m_\pi^2 (m_\tau^2 - p_{\tau \bar{N}}^2) + m_\tau^4 \bigr) \nonumber\\[2pt]
&\quad + 4 E_\pi E_{\tau \bar{N}} m_\pi^2 (3 m_\pi^2 + m_\tau^2)
- 3 m_\pi^6 - 2 m_\pi^4 (m_\tau^2 + 2 p_{\tau \bar{N}}^2)
+ m_\pi^2 m_\tau^4 \Bigr)
\Biggr]\,, \\[1.5ex]
J_{9}^{(C_V^{RR})}
&= -\frac{i\pi\, |C_V^{RR}|^2| \, m_\tau^2\, |p_D|^2}{4 (E_\pi^2 - m_\pi^2)}
\Biggl[ \bigl( H_{V+} H_{V-}^* - H_{V-} H_{V+}^* \bigr) \nonumber\\[2pt]
&\quad \times \Bigl( 8 E_\pi^3 E_{\tau \bar{N}} m_\tau^2
- 4 E_\pi^2 \bigl( 3 E_{\tau \bar{N}}^2 m_\pi^2
+ m_\pi^2 (m_\tau^2 - p_{\tau \bar{N}}^2) + m_\tau^4 \bigr) \nonumber\\[2pt]
&\quad + 4 E_\pi E_{\tau \bar{N}} m_\pi^2 (3 m_\pi^2 + m_\tau^2)
- 3 m_\pi^6 - 2 m_\pi^4 (m_\tau^2 + 2 p_{\tau \bar{N}}^2)
+ m_\pi^2 m_\tau^4 \Bigr)
\Biggr]\,,
\end{align}
where $|p_{\tau \bar{N}}| = \frac{\sqrt{\lambda(q^2,m_\tau^2,m_N^2)}}{2\sqrt{q^2}}$ and $E_{\tau \bar{N}} = \sqrt{p_{\tau\bar{N}}^2+m_\tau^2}$.

\subsection{$O_S^{RR}$}
\label{app:csrrjfunc}
The angular coefficients for $O_S^{RR}$ are
\begin{align}
J_{1c}^{(C_{S}^{RR})}
&= 2\pi\, |C_{S}^{RR}|^2\, H_{SP}\, p_D^2\, H_{SP}^*
\Biggl[ - E_{\tau \bar{N}}^2 \Bigl( 2 E_\pi m_\tau^2 \sqrt{q^2}
+ m_\pi^2 (m_\tau^2 + 4 p_{\tau \bar{N}}^2) + m_\tau^4 \Bigr) \nonumber\\[2pt]
&\quad + p_{\tau \bar{N}}^2 \Bigl( 2 E_\pi m_\tau^2 \sqrt{q^2}
+ m_\pi^2 (m_\tau^2 + 2 p_{\tau \bar{N}}^2) + m_\tau^4 \Bigr) \nonumber\\[2pt]
&\quad + 2 E_{\tau \bar{N}}^4 m_\pi^2
- 2 E_{\tau \bar{N}}^3 m_\pi^2 \sqrt{q^2} \nonumber\\[2pt]
&\quad + 2 E_{\tau \bar{N}} \sqrt{q^2}
\Bigl( m_\pi^2 (m_\tau^2 + p_{\tau \bar{N}}^2) + m_\tau^4 \Bigr)
\Biggr] \,.
\end{align}
The rest of the $J^{(C_S^{RR})}$ functions are zero.

\subsection{$O_S^{LR}$}\label{app:cslrjfunc}
The angular coefficients for $O_S^{LR}$ are
\begin{align}
J_{1c}^{(C_{S}^{LR})}
&= 2\pi\, |C_{S}^{LR}|^2 \, H_{SP}\, p_D^2\, H_{SP}^*
\Biggl[ - E_{\tau \bar{N}}^2 \Bigl( 2 E_\pi m_\tau^2 \sqrt{q^2}
+ m_\pi^2 (m_\tau^2 + 4 p_{\tau \bar{N}}^2) + m_\tau^4 \Bigr) \nonumber\\[2pt]
&\quad + p_{\tau \bar{N}}^2 \Bigl( 2 E_\pi m_\tau^2 \sqrt{q^2}
+ m_\pi^2 (m_\tau^2 + 2 p_{\tau \bar{N}}^2) + m_\tau^4 \Bigr) \nonumber\\[2pt]
&\quad + 2 E_{\tau \bar{N}}^4 m_\pi^2
- 2 E_{\tau \bar{N}}^3 m_\pi^2 \sqrt{q^2} \nonumber\\[2pt]
&\quad + 2 E_{\tau \bar{N}} \sqrt{q^2}
\Bigl( m_\pi^2 (m_\tau^2 + p_{\tau \bar{N}}^2) + m_\tau^4 \Bigr)
\Biggr] \,.
\end{align}
The rest of the $J^{(C_S^{LR})}$ functions are zero. 

\subsection{$O_T^{RR}$}
The angular coefficients for $O_T^{RR}$ are
\begin{align}
J_{1s}^{(C_T^{RR})}
&= \frac{2\pi\, |C_T^{RR}|^2\, p_D^2}{(E_\pi - m_\pi)(E_\pi + m_\pi)}
\Biggl[ \bigl( |H_{T-}|^2 + |H_{T+}|^2 \bigr) \nonumber\\[2pt]
&\quad \times \Bigl( 24 E_\pi^3 E_{\tau} m_\tau^2 (E_{\tau}-p_{\tau}) (E_{\tau}+p_{\tau}) \nonumber\\[2pt]
&\quad + 20 E_\pi^2 E_{\tau}^4 m_\pi^2
- 8 E_\pi^2 E_{\tau}^2 \bigl( m_\pi^2 (3 m_\tau^2+2 p_{\tau}^2)+3 m_\tau^4 \bigr) \nonumber\\[2pt]
&\quad + 16 \sqrt{q^2} (m_\pi^2-E_\pi^2)
\Bigl( E_\pi E_{\tau}^2 m_\tau^2 - E_\pi m_\tau^2 p_{\tau}^2
+ E_{\tau}^3 m_\pi^2 \nonumber\\[2pt]
&\quad - E_{\tau} \bigl( m_\pi^2 (m_\tau^2+p_{\tau}^2)+m_\tau^4 \bigr) \Bigr)
- 4 E_\pi^2 m_\pi^2 p_{\tau}^4 \nonumber\\[2pt]
&\quad - 4 E_\pi E_{\tau}^3 m_\pi^2 (m_\pi^2+7 m_\tau^2)
+ 4 E_\pi E_{\tau}
\Bigl( m_\pi^2 p_{\tau}^2 (m_\pi^2+7 m_\tau^2) \nonumber\\[2pt]
&\quad + m_\tau^2 (m_\pi^2+m_\tau^2)^2 \Bigr)
- 16 E_{\tau}^4 m_\pi^4 \nonumber\\[2pt]
&\quad + E_{\tau}^2 m_\pi^2 \bigl( m_\pi^4
+ 2 m_\pi^2 (11 m_\tau^2+6 p_{\tau}^2)+21 m_\tau^4 \bigr) \nonumber\\[2pt]
&\quad - m_\pi^6 m_\tau^2 - m_\pi^6 p_{\tau}^2
- 3 m_\pi^4 m_\tau^4 - 2 m_\pi^4 m_\tau^2 p_{\tau}^2 \nonumber\\[2pt]
&\quad + 4 m_\pi^4 p_{\tau}^4
- 3 m_\pi^2 m_\tau^6 - m_\pi^2 m_\tau^4 p_{\tau}^2
- m_\tau^8 \Bigr)
\Biggr]\,,  \\[1.5ex]
J_{1c}^{(C_T^{RR})}
&= \frac{8\pi\, |C_T^{RR}|^2\, H_T^0\, p_D^2\, H_T^{0*}}{E_\pi^2 - m_\pi^2}
\Biggl[ 8 E_\pi^3 m_\tau^2 (E_{\tau}^2-p_{\tau}^2) (E_{\tau}-\sqrt{q^2}) \nonumber\\[2pt]
&\quad + 4 E_\pi^2 (E_{\tau}^2-2 E_{\tau} \sqrt{q^2}+p_{\tau}^2)
\bigl( E_{\tau}^2 m_\pi^2 - m_\pi^2 (m_\tau^2+p_{\tau}^2)-m_\tau^4 \bigr) \nonumber\\[2pt]
&\quad + 4 E_\pi \Bigl( E_{\tau}^3 (m_\pi^4-m_\pi^2 m_\tau^2)
+ 2 E_{\tau}^2 m_\pi^2 m_\tau^2 \sqrt{q^2} \nonumber\\[2pt]
&\quad - E_{\tau} \bigl( m_\pi^4 (m_\tau^2+p_{\tau}^2)
+ m_\pi^2 (2 m_\tau^4-m_\tau^2 p_{\tau}^2)+m_\tau^6 \bigr) \nonumber\\[2pt]
&\quad - 2 m_\pi^2 m_\tau^2 p_{\tau}^2 \sqrt{q^2} \Bigr) \nonumber\\[2pt]
&\quad - \bigl( E_{\tau}^2 m_\pi^2 - m_\pi^2 (m_\tau^2+p_{\tau}^2)-m_\tau^4 \bigr) \nonumber\\[2pt]
&\quad \times \Bigl( 8 E_{\tau}^2 m_\pi^2 - 8 E_{\tau} m_\pi^2 \sqrt{q^2}
+ m_\pi^4 + 2 m_\pi^2 (m_\tau^2+2 p_{\tau}^2)+m_\tau^4 \Bigr)
\Biggr]\,,  \\[1.5ex]
J_{2s}^{(C_T^{RR})}
&= -\frac{2\pi\, |C_T^{RR}|^2\, p_D^2}{E_\pi^2 - m_\pi^2}
\Biggl[ \bigl( H_{T-} H_{T-}^* + H_{T+} H_{T+}^* \bigr) \nonumber\\[2pt]
&\quad \times \Bigl( 8 E_\pi^3 E_{\tau} m_\tau^2 (p_{\tau}^2 - E_{\tau}^2)
- 4 E_\pi^2 \Bigl( 3 E_{\tau}^4 m_\pi^2 \nonumber\\[2pt]
&\quad - 4 E_{\tau}^2 \bigl( m_\pi^2 (m_\tau^2 + p_{\tau}^2) + m_\tau^4 \bigr)
+ p_{\tau}^2 \bigl( m_\pi^2 (2 m_\tau^2 + p_{\tau}^2) + 2 m_\tau^4 \bigr) \Bigr) \nonumber\\[2pt]
&\quad + 4 E_\pi E_{\tau} \Bigl( E_{\tau}^2 (3 m_\pi^4 + 5 m_\pi^2 m_\tau^2)
- 3 m_\pi^4 (m_\tau^2 + p_{\tau}^2) \nonumber\\[2pt]
&\quad - m_\pi^2 (6 m_\tau^4 + 5 m_\tau^2 p_{\tau}^2) - 3 m_\tau^6 \Bigr) \nonumber\\[2pt]
&\quad - E_{\tau}^2 \Bigl( 3 m_\pi^6
+ 2 m_\pi^4 (5 m_\tau^2 + 2 p_{\tau}^2)
+ 7 m_\pi^2 m_\tau^4 \Bigr) \nonumber\\[2pt]
&\quad + 3 m_\pi^6 m_\tau^2 + 3 m_\pi^6 p_{\tau}^2
+ 9 m_\pi^4 m_\tau^4 + 14 m_\pi^4 m_\tau^2 p_{\tau}^2 \nonumber\\[2pt]
&\quad + 4 m_\pi^4 p_{\tau}^4
+ 9 m_\pi^2 m_\tau^6 + 11 m_\pi^2 m_\tau^4 p_{\tau}^2
+ 3 m_\tau^8 \Bigr)
\Biggr]\,,  \\[1.5ex]
J_{2c}^{(C_T^{RR})}
&= \frac{8\pi\, |C_T^{RR}|^2\, H_T^0\, p_D^2\,H_T^{0*}}{E_\pi^2 - m_\pi^2}
\Biggl[ 8 E_\pi^3 E_{\tau} m_\tau^2 (p_{\tau}^2 - E_{\tau}^2) \nonumber\\[2pt]
&\quad - 4 E_\pi^2 \Bigl( 3 E_{\tau}^4 m_\pi^2
- 4 E_{\tau}^2 \bigl( m_\pi^2 (m_\tau^2 + p_{\tau}^2) + m_\tau^4 \bigr) \nonumber\\[2pt]
&\quad + p_{\tau}^2 \bigl( m_\pi^2 (2 m_\tau^2 + p_{\tau}^2) + 2 m_\tau^4 \bigr) \Bigr) \nonumber\\[2pt]
&\quad + 4 E_\pi E_{\tau} \Bigl( E_{\tau}^2 (3 m_\pi^4 + 5 m_\pi^2 m_\tau^2)
- 3 m_\pi^4 (m_\tau^2 + p_{\tau}^2) \nonumber\\[2pt]
&\quad - m_\pi^2 (6 m_\tau^4 + 5 m_\tau^2 p_{\tau}^2) - 3 m_\tau^6 \Bigr) \nonumber\\[2pt]
&\quad - E_{\tau}^2 \Bigl( 3 m_\pi^6
+ 2 m_\pi^4 (5 m_\tau^2 + 2 p_{\tau}^2)
+ 7 m_\pi^2 m_\tau^4 \Bigr) \nonumber\\[2pt]
&\quad + 3 m_\pi^6 m_\tau^2 + 3 m_\pi^6 p_{\tau}^2
+ 9 m_\pi^4 m_\tau^4 + 14 m_\pi^4 m_\tau^2 p_{\tau}^2 \nonumber\\[2pt]
&\quad + 4 m_\pi^4 p_{\tau}^4
+ 9 m_\pi^2 m_\tau^6 + 11 m_\pi^2 m_\tau^4 p_{\tau}^2
+ 3 m_\tau^8
\Biggr]\,,  \\[1.5ex]
J_{3}^{(C_T^{RR})}
&= \frac{4\pi\, |C_T^{RR}|^2\, p_D^2}{E_\pi^2 - m_\pi^2}
\Biggl[ \bigl( H_{T+} H_{T-}^* + H_{T-} H_{T+}^* \bigr) \nonumber\\[2pt]
&\quad \times \Bigl( 8 E_\pi^3 E_{\tau} m_\tau^2 (E_{\tau}^2 - p_{\tau}^2)
+ 4 E_\pi^2 \Bigl( 3 E_{\tau}^4 m_\pi^2 \nonumber\\[2pt]
&\quad - 4 E_{\tau}^2 \bigl( m_\pi^2 (m_\tau^2 + p_{\tau}^2) + m_\tau^4 \bigr)
+ p_{\tau}^2 \bigl( m_\pi^2 (2 m_\tau^2 + p_{\tau}^2) + 2 m_\tau^4 \bigr) \Bigr) \nonumber\\[2pt]
&\quad - 4 E_\pi E_{\tau} \Bigl( E_{\tau}^2 (3 m_\pi^4 + 5 m_\pi^2 m_\tau^2)
- 3 m_\pi^4 (m_\tau^2 + p_{\tau}^2) \nonumber\\[2pt]
&\quad - m_\pi^2 (6 m_\tau^4 + 5 m_\tau^2 p_{\tau}^2) - 3 m_\tau^6 \Bigr) \nonumber\\[2pt]
&\quad + E_{\tau}^2 \Bigl( 3 m_\pi^6
+ 2 m_\pi^4 (5 m_\tau^2 + 2 p_{\tau}^2)
+ 7 m_\pi^2 m_\tau^4 \Bigr) \nonumber\\[2pt]
&\quad - 3 m_\pi^6 m_\tau^2 - 3 m_\pi^6 p_{\tau}^2
- 9 m_\pi^4 m_\tau^4 - 14 m_\pi^4 m_\tau^2 p_{\tau}^2 \nonumber\\[2pt]
&\quad - 4 m_\pi^4 p_{\tau}^4
- 9 m_\pi^2 m_\tau^6 - 11 m_\pi^2 m_\tau^4 p_{\tau}^2
- 3 m_\tau^8 \Bigr)
\Biggr]\,,  \\[1.5ex]
J_{4}^{(C_T^{RR})}
&= \frac{2\pi\, |C_T^{RR}|^2\, p_D^2}{E_\pi^2 - m_\pi^2}
\Biggl[ \bigl( H_T^{0*} (H_{T-} - H_{T+}) + H_T^0 H_{T-}^* - H_T^0 H_{T+}^* \bigr) \nonumber\\[2pt]
&\quad \times \Bigl( 8 E_\pi^3 E_{\tau} m_\tau^2 (p_{\tau}^2 - E_{\tau}^2)
- 4 E_\pi^2 \Bigl( 3 E_{\tau}^4 m_\pi^2 \nonumber\\[2pt]
&\quad - 4 E_{\tau}^2 \bigl( m_\pi^2 (m_\tau^2 + p_{\tau}^2) + m_\tau^4 \bigr)
+ p_{\tau}^2 \bigl( m_\pi^2 (2 m_\tau^2 + p_{\tau}^2) + 2 m_\tau^4 \bigr) \Bigr) \nonumber\\[2pt]
&\quad + 4 E_\pi E_{\tau} \Bigl( E_{\tau}^2 (3 m_\pi^4 + 5 m_\pi^2 m_\tau^2)
- 3 m_\pi^4 (m_\tau^2 + p_{\tau}^2) \nonumber\\[2pt]
&\quad - m_\pi^2 (6 m_\tau^4 + 5 m_\tau^2 p_{\tau}^2) - 3 m_\tau^6 \Bigr) \nonumber\\[2pt]
&\quad - E_{\tau}^2 \Bigl( 3 m_\pi^6
+ 2 m_\pi^4 (5 m_\tau^2 + 2 p_{\tau}^2)
+ 7 m_\pi^2 m_\tau^4 \Bigr) \nonumber\\[2pt]
&\quad + 3 m_\pi^6 m_\tau^2 + 3 m_\pi^6 p_{\tau}^2
+ 9 m_\pi^4 m_\tau^4 + 14 m_\pi^4 m_\tau^2 p_{\tau}^2 \nonumber\\[2pt]
&\quad + 4 m_\pi^4 p_{\tau}^4
+ 9 m_\pi^2 m_\tau^6 + 11 m_\pi^2 m_\tau^4 p_{\tau}^2
+ 3 m_\tau^8 \Bigr)
\Biggr] \,, \\[1.5ex]
J_{5}^{(C_T^{RR})}
&= -\frac{8\pi\, |C_T^{RR}|^2\, p_D^2}{\sqrt{E_\pi^2 - m_\pi^2}}
\Biggl[ \bigl( H_T^{0*} (H_{T-} + H_{T+}) + H_T^0 H_{T-}^* + H_T^0 H_{T+}^* \bigr) \nonumber\\[2pt]
&\quad \times \Bigl( 2 E_\pi^2 m_\tau^2 (E_{\tau}^2 - p_{\tau}^2) (2 E_{\tau} - \sqrt{q^2}) \nonumber\\[2pt]
&\quad + E_\pi \Bigl( 4 E_{\tau}^4 m_\pi^2
- 2 E_{\tau}^3 m_\pi^2 \sqrt{q^2} \nonumber\\[2pt]
&\quad - E_{\tau}^2 \bigl( m_\pi^2 (5 m_\tau^2 + 4 p_{\tau}^2) + 5 m_\tau^4 \bigr) \nonumber\\[2pt]
&\quad + 2 E_{\tau} \sqrt{q^2} \bigl( m_\pi^2 (m_\tau^2 + p_{\tau}^2) + m_\tau^4 \bigr)
+ m_\tau^2 p_{\tau}^2 (m_\pi^2 + m_\tau^2) \Bigr) \nonumber\\[2pt]
&\quad - 2 E_{\tau}^3 (m_\pi^4 + 2 m_\pi^2 m_\tau^2)
+ E_{\tau}^2 m_\pi^2 \sqrt{q^2} (m_\pi^2 + 3 m_\tau^2) \nonumber\\[2pt]
&\quad + 2 E_{\tau} \Bigl( m_\pi^4 (m_\tau^2 + p_{\tau}^2)
+ 2 m_\pi^2 m_\tau^2 (m_\tau^2 + p_{\tau}^2)
+ m_\tau^6 \Bigr) \nonumber\\[2pt]
&\quad - \sqrt{q^2} \Bigl( m_\pi^4 (m_\tau^2 + p_{\tau}^2)
+ m_\pi^2 (2 m_\tau^4 + 3 m_\tau^2 p_{\tau}^2)
+ m_\tau^6 \Bigr)
\Bigr)
\Biggr]\,,  \\[1.5ex]
J_{6s}^{(C_T^{RR})}
&= \frac{16\pi\, |C_T^{RR}|^2\, p_D^2}{\sqrt{E_\pi^2 - m_\pi^2}}
\Biggl[ \bigl( H_{T-} H_{T-}^* - H_{T+} H_{T+}^* \bigr) \nonumber\\[2pt]
&\quad \times \Bigl( 2 E_\pi^2 m_\tau^2 (E_{\tau}^2 - p_{\tau}^2) (2 E_{\tau} - \sqrt{q^2}) \nonumber\\[2pt]
&\quad + E_\pi \Bigl( 4 E_{\tau}^4 m_\pi^2
- 2 E_{\tau}^3 m_\pi^2 \sqrt{q^2} \nonumber\\[2pt]
&\quad - E_{\tau}^2 \bigl( m_\pi^2 (5 m_\tau^2 + 4 p_{\tau}^2) + 5 m_\tau^4 \bigr) \nonumber\\[2pt]
&\quad + 2 E_{\tau} \sqrt{q^2} \bigl( m_\pi^2 (m_\tau^2 + p_{\tau}^2) + m_\tau^4 \bigr)
+ m_\tau^2 p_{\tau}^2 (m_\pi^2 + m_\tau^2) \Bigr) \nonumber\\[2pt]
&\quad - 2 E_{\tau}^3 (m_\pi^4 + 2 m_\pi^2 m_\tau^2)
+ E_{\tau}^2 m_\pi^2 \sqrt{q^2} (m_\pi^2 + 3 m_\tau^2) \nonumber\\[2pt]
&\quad + 2 E_{\tau} \Bigl( m_\pi^4 (m_\tau^2 + p_{\tau}^2)
+ 2 m_\pi^2 m_\tau^2 (m_\tau^2 + p_{\tau}^2)
+ m_\tau^6 \Bigr) \nonumber\\[2pt]
&\quad - \sqrt{q^2} \Bigl( m_\pi^4 (m_\tau^2 + p_{\tau}^2)
+ m_\pi^2 (2 m_\tau^4 + 3 m_\tau^2 p_{\tau}^2)
+ m_\tau^6 \Bigr)
\Bigr)
\Biggr]\,,  \\[1.5ex]
J_{6c}^{(C_T^{RR})} &= 0\,, \\[1.5ex]
J_{7}^{({C_T^{RR}})}
&= -\frac{8 i \pi\, |C_T^{RR}|^2\, p_D^2}{\sqrt{E_\pi^2 - m_\pi^2}}
\Biggl[ \bigl( H_T^{0*} (H_{T+} - H_{T-}) + H_T^0 H_{T-}^* - H_T^0 H_{T+}^* \bigr) \nonumber\\[2pt]
&\quad \times \Bigl( 2 E_\pi^2 m_\tau^2 (E_{\tau}^2 - p_{\tau}^2) (2 E_{\tau} - \sqrt{q^2}) \nonumber\\[2pt]
&\quad + E_\pi \Bigl( 4 E_{\tau}^4 m_\pi^2
- 2 E_{\tau}^3 m_\pi^2 \sqrt{q^2} \nonumber\\[2pt]
&\quad - E_{\tau}^2 \bigl( m_\pi^2 (5 m_\tau^2 + 4 p_{\tau}^2) + 5 m_\tau^4 \bigr) \nonumber\\[2pt]
&\quad + 2 E_{\tau} \sqrt{q^2} \bigl( m_\pi^2 (m_\tau^2 + p_{\tau}^2) + m_\tau^4 \bigr)
+ m_\tau^2 p_{\tau}^2 (m_\pi^2 + m_\tau^2) \Bigr) \nonumber\\[2pt]
&\quad - 2 E_{\tau}^3 (m_\pi^4 + 2 m_\pi^2 m_\tau^2)
+ E_{\tau}^2 m_\pi^2 \sqrt{q^2} (m_\pi^2 + 3 m_\tau^2) \nonumber\\[2pt]
&\quad + 2 E_{\tau} \Bigl( m_\pi^4 (m_\tau^2 + p_{\tau}^2)
+ 2 m_\pi^2 m_\tau^2 (m_\tau^2 + p_{\tau}^2)
+ m_\tau^6 \Bigr) \nonumber\\[2pt]
&\quad - \sqrt{q^2} \Bigl( m_\pi^4 (m_\tau^2 + p_{\tau}^2)
+ m_\pi^2 (2 m_\tau^4 + 3 m_\tau^2 p_{\tau}^2)
+ m_\tau^6 \Bigr)
\Bigr)
\Biggr]\,,  \\[1.5ex]
J_{8}^{(C_T^{RR})}
&= \frac{2 i \pi\, |C_T^{RR}|^2\, p_D^2}{E_\pi^2 - m_\pi^2}
\Biggl[ \bigl( - H_T^{0*} (H_{T-} + H_{T+}) + H_T^0 H_{T-}^* + H_T^0 H_{T+}^* \bigr) \nonumber\\[2pt]
&\quad \times \Bigl( 8 E_\pi^3 E_{\tau} m_\tau^2 (p_{\tau}^2 - E_{\tau}^2) \nonumber\\[2pt]
&\quad - 4 E_\pi^2 \Bigl( 3 E_{\tau}^4 m_\pi^2
- 4 E_{\tau}^2 \bigl( m_\pi^2 (m_\tau^2 + p_{\tau}^2) + m_\tau^4 \bigr) \nonumber\\[2pt]
&\quad + p_{\tau}^2 \bigl( m_\pi^2 (2 m_\tau^2 + p_{\tau}^2) + 2 m_\tau^4 \bigr) \Bigr) \nonumber\\[2pt]
&\quad + 4 E_\pi E_{\tau} \Bigl( E_{\tau}^2 (3 m_\pi^4 + 5 m_\pi^2 m_\tau^2)
- 3 m_\pi^4 (m_\tau^2 + p_{\tau}^2) \nonumber\\[2pt]
&\quad - m_\pi^2 (6 m_\tau^4 + 5 m_\tau^2 p_{\tau}^2) - 3 m_\tau^6 \Bigr) \nonumber\\[2pt]
&\quad - E_{\tau}^2 \Bigl( 3 m_\pi^6
+ 2 m_\pi^4 (5 m_\tau^2 + 2 p_{\tau}^2)
+ 7 m_\pi^2 m_\tau^4 \Bigr) \nonumber\\[2pt]
&\quad + 3 m_\pi^6 m_\tau^2 + 3 m_\pi^6 p_{\tau}^2
+ 9 m_\pi^4 m_\tau^4 + 14 m_\pi^4 m_\tau^2 p_{\tau}^2 \nonumber\\[2pt]
&\quad + 4 m_\pi^4 p_{\tau}^4
+ 9 m_\pi^2 m_\tau^6 + 11 m_\pi^2 m_\tau^4 p_{\tau}^2
+ 3 m_\tau^8 \Bigr)
\Biggr]\,,  \\[1.5ex]
J_{9}^{(C_T^{RR})}
&= \frac{4 i \pi\, |C_T^{RR}|^2\, p_D^2}{E_\pi^2 - m_\pi^2}
\Biggl[ \bigl( H_{T+} H_{T-}^* - H_{T-} H_{T+}^* \bigr) \nonumber\\[2pt]
&\quad \times \Bigl( 8 E_\pi^3 E_{\tau} m_\tau^2 (E_{\tau}^2 - p_{\tau}^2)
+ 4 E_\pi^2 \Bigl( 3 E_{\tau}^4 m_\pi^2 \nonumber\\[2pt]
&\quad - 4 E_{\tau}^2 \bigl( m_\pi^2 (m_\tau^2 + p_{\tau}^2) + m_\tau^4 \bigr)
+ p_{\tau}^2 \bigl( m_\pi^2 (2 m_\tau^2 + p_{\tau}^2) + 2 m_\tau^4 \bigr) \Bigr) \nonumber\\[2pt]
&\quad - 4 E_\pi E_{\tau} \Bigl( E_{\tau}^2 (3 m_\pi^4 + 5 m_\pi^2 m_\tau^2)
- 3 m_\pi^4 (m_\tau^2 + p_{\tau}^2) \nonumber\\[2pt]
&\quad - m_\pi^2 (6 m_\tau^4 + 5 m_\tau^2 p_{\tau}^2) - 3 m_\tau^6 \Bigr) \nonumber\\[2pt]
&\quad + E_{\tau}^2 \Bigl( 3 m_\pi^6
+ 2 m_\pi^4 (5 m_\tau^2 + 2 p_{\tau}^2)
+ 7 m_\pi^2 m_\tau^4 \Bigr) \nonumber\\[2pt]
&\quad - 3 m_\pi^6 m_\tau^2 - 3 m_\pi^6 p_{\tau}^2
- 9 m_\pi^4 m_\tau^4 - 14 m_\pi^4 m_\tau^2 p_{\tau}^2 \nonumber\\[2pt]
&\quad - 4 m_\pi^4 p_{\tau}^4
- 9 m_\pi^2 m_\tau^6 - 11 m_\pi^2 m_\tau^4 p_{\tau}^2
- 3 m_\tau^8 \Bigr)
\Biggr] \,.
\end{align}

\bibliographystyle{JHEP}
\bibliography{BDstarRHN.bib}

@article{Abazajian:2012ys,
    author = "Abazajian, K. N. and others",
    title = "{Light Sterile Neutrinos: A White Paper}",
    eprint = "1204.5379",
    archivePrefix = "arXiv",
    primaryClass = "hep-ph",
    reportNumber = "FERMILAB-PUB-12-881-PPD",
    month = "4",
    year = "2012"
}

@article{Liao:2016qyd,
	Archiveprefix = {arXiv},
	Author = {Liao, Yi and Ma, Xiao-Dong},
	Doi = {10.1103/PhysRevD.96.015012},
	Eprint = {1612.04527},
	Journal = {Phys. Rev.},
	Number = {1},
	Pages = {015012},
	Primaryclass = {hep-ph},
	Slaccitation = {%%CITATION = ARXIV:1612.04527;%%},
	Title = {{Operators up to Dimension Seven in Standard Model Effective Field Theory Extended with Sterile Neutrinos}},
	Volume = {D96},
	Year = {2017}}

@article{Bischer:2019ttk,
	author = "Bischer, Ingolf and Rodejohann, Werner",
	title = "{General neutrino interactions from an effective field theory perspective}",
	eprint = "1905.08699",
	archivePrefix = "arXiv",
	primaryClass = "hep-ph",
	doi = "10.1016/j.nuclphysb.2019.114746",
	journal = "Nucl. Phys. B",
	volume = "947",
	pages = "114746",
	year = "2019"
}

@article{Adachi:2018qme,
    author = "Adachi, I. and Browder, T. E. and Kri\v{z}an, P. and Tanaka, S. and Ushiroda, Y.",
    collaboration = "Belle-II",
    title = "{Detectors for extreme luminosity: Belle II}",
    doi = "10.1016/j.nima.2018.03.068",
    journal = "Nucl. Instrum. Meth. A",
    volume = "907",
    pages = "46--59",
    year = "2018"
}

@article{Mandal:2020htr,
    author = "Mandal, Rusa and Murgui, Clara and Pe\~nuelas, Ana and Pich, Antonio",
    title = "{The role of right-handed neutrinos in $b \to c \tau \bar{\nu}$ anomalies}",
    eprint = "2004.06726",
    archivePrefix = "arXiv",
    primaryClass = "hep-ph",
    reportNumber = "IFIC/20-14; FTUV/20-0414; SI-HEP-2020-XX, IFIC/20-14; FTUV/20-0414; SI-HEP-2020-10",
    doi = "10.1007/JHEP08(2020)022",
    journal = "JHEP",
    volume = "08",
    number = "08",
    pages = "022",
    year = "2020"
}

@article{Babu:2018vrl,
    author = "Babu, K. S. and Dutta, Bhaskar and Mohapatra, Rabindra N.",
    title = "{A theory of R(D$^{*}$, D) anomaly with right-handed currents}",
    eprint = "1811.04496",
    archivePrefix = "arXiv",
    primaryClass = "hep-ph",
    reportNumber = "OSU-HEP-18-06, MI-TH-181, UMD-PP-018-08",
    doi = "10.1007/JHEP01(2019)168",
    journal = "JHEP",
    volume = "01",
    pages = "168",
    year = "2019"
}

@article{Greljo:2018ogz,
    author = "Greljo, Admir and Robinson, Dean J. and Shakya, Bibhushan and Zupan, Jure",
    title = "{R(D$^{(?)}$) from W$^{?}$ and right-handed neutrinos}",
    eprint = "1804.04642",
    archivePrefix = "arXiv",
    primaryClass = "hep-ph",
    reportNumber = "LCTP-18-11, MITP-18-028, MITP/18-028",
    doi = "10.1007/JHEP09(2018)169",
    journal = "JHEP",
    volume = "09",
    pages = "169",
    year = "2018"
}

@article{Asadi:2018wea,
    author = "Asadi, Pouya and Buckley, Matthew R. and Shih, David",
    title = "{It\textquoteright{}s all right(-handed neutrinos): a new W$^{?}$ model for the $ {R}_{D^{{\left(\ast \right)}}} $ anomaly}",
    eprint = "1804.04135",
    archivePrefix = "arXiv",
    primaryClass = "hep-ph",
    doi = "10.1007/JHEP09(2018)010",
    journal = "JHEP",
    volume = "09",
    pages = "010",
    year = "2018"
}

@article{He:2012zp,
    author = "He, Xiao-Gang and Valencia, German",
    title = "{$B$ decays with $\tau$ leptons in nonuniversal left-right models}",
    eprint = "1211.0348",
    archivePrefix = "arXiv",
    primaryClass = "hep-ph",
    doi = "10.1103/PhysRevD.87.014014",
    journal = "Phys. Rev. D",
    volume = "87",
    number = "1",
    pages = "014014",
    year = "2013"
}

@article{Cvetic:2017gkt,
    author = "Cveti\v{c}, G. and Halzen, Francis and Kim, C. S. and Oh, Sechul",
    title = "{Anomalies in (semi)-leptonic $B$ decays $B^{\pm} \to \tau^{\pm} \nu$, $B^{\pm} \to D \tau^{\pm} \nu$ and $B^{\pm} \to D^* \tau^{\pm} \nu$, and possible resolution with sterile neutrino}",
    eprint = "1702.04335",
    archivePrefix = "arXiv",
    primaryClass = "hep-ph",
    doi = "10.1088/1674-1137/41/11/113102",
    journal = "Chin. Phys. C",
    volume = "41",
    number = "11",
    pages = "113102",
    year = "2017"
}

@article{Bhattacharya:2024zog,
    author = "Bhattacharya, Bhubanjyoti and Browder, Thomas E. and Datta, Alakabha and Kapoor, Tejhas and Kou, Emi and Mukherjee, Lopamudra",
    title = "{New physics searches via angular distributions of $ \overline{B}\to {D}^{\ast}\left(\to D\pi \right)\tau \left(\to \ell {\nu}_{\tau }{\overline{\nu}}_{\ell}\right){\overline{\nu}}_{\tau } $ decays}",
    eprint = "2411.09414",
    archivePrefix = "arXiv",
    primaryClass = "hep-ph",
    doi = "10.1007/JHEP04(2025)135",
    journal = "JHEP",
    volume = "04",
    pages = "135",
    year = "2025"
}

@article{Bernlochner:2024xiz,
    author = "Bernlochner, Florian U. and Fedele, Marco and Kretz, Tim and Nierste, Ulrich and Prim, Markus T.",
    title = "{Model independent bounds on heavy sterile neutrinos from the angular distribution of B {\textrightarrow} D$^{*}${\ensuremath{\ell}}{\ensuremath{\nu}} decays}",
    eprint = "2410.11945",
    archivePrefix = "arXiv",
    primaryClass = "hep-ph",
    reportNumber = "TTP24-041",
    doi = "10.1007/JHEP01(2025)040",
    journal = "JHEP",
    volume = "01",
    pages = "040",
    year = "2025"
}

@article{Aparici:2009fh,
    author = "Aparici, Alberto and Kim, Kyungwook and Santamaria, Arcadi and Wudka, Jose",
    title = "{Right-handed neutrino magnetic moments}",
    eprint = "0904.3244",
    archivePrefix = "arXiv",
    primaryClass = "hep-ph",
    reportNumber = "FTUV-09-0421, IFIC-09-15, UCRHEP-T466",
    doi = "10.1103/PhysRevD.80.013010",
    journal = "Phys. Rev. D",
    volume = "80",
    pages = "013010",
    year = "2009"
}

@article{Bhattacharya:2020lfm,
    author = "Bhattacharya, Bhubanjyoti and Datta, Alakabha and Kamali, Saeed and London, David",
    title = "{A measurable angular distribution for $ \overline{B}\to {D}^{\ast }{\tau}^{-}{\overline{v}}_{\tau } $ decays}",
    eprint = "2005.03032",
    archivePrefix = "arXiv",
    primaryClass = "hep-ph",
    reportNumber = "UdeM-GPP-TH-20-277; UMISS-HEP-2020-01",
    doi = "10.1007/JHEP07(2020)194",
    journal = "JHEP",
    volume = "07",
    number = "07",
    pages = "194",
    year = "2020"
}

@article{fermilab,
    author = "Bazavov, A. and others",
    collaboration = "Fermilab Lattice, MILC, Fermilab Lattice, MILC",
    title = "{Semileptonic form factors for $B \rightarrow D^* \ell \nu $ at nonzero recoil from $2+1$-flavor lattice QCD: Fermilab Lattice~and~MILC~Collaborations}",
    eprint = "2105.14019",
    archivePrefix = "arXiv",
    primaryClass = "hep-lat",
    reportNumber = "FERMILAB-PUB-21-261-T~, FERMILAB-PUB-21/261-T",
    doi = "10.1140/epjc/s10052-022-10984-9",
    journal = "Eur. Phys. J. C",
    volume = "82",
    number = "12",
    pages = "1141",
    year = "2022",
    note = "[Erratum: Eur.Phys.J.C 83, 21 (2023)]"
}

@article{jlqcd,
    author = "Aoki, Y. and Colquhoun, B. and Fukaya, H. and Hashimoto, S. and Kaneko, T. and Kellermann, R. and Koponen, J. and Kou, E.",
    collaboration = "JLQCD",
    title = {{$B \to D^*\ell \nu_\ell$ semileptonic form factors from lattice QCD with M\"obius domain-wall quarks}},
    eprint = "2306.05657",
    archivePrefix = "arXiv",
    primaryClass = "hep-lat",
    reportNumber = "KEK-CP-393, OU-HET-1186",
    month = "6",
    year = "2023"
}

@article{belle19,
    author = "Waheed, E. and others",
    collaboration = "Belle",
    title = "{Measurement of the CKM matrix element $|V_{cb}|$ from $B^0\to D^{*-}\ell^ {+} \nu_\ell$ at Belle}",
    eprint = "1809.03290",
    archivePrefix = "arXiv",
    primaryClass = "hep-ex",
    doi = "10.1103/PhysRevD.100.052007",
    journal = "Phys. Rev. D",
    volume = "100",
    number = "5",
    pages = "052007",
    year = "2019",
    note = "[Erratum: Phys.Rev.D 103, 079901 (2021)]"
}

@article{tensorff,
    author = "Sakaki, Yasuhito and Tanaka, Minoru and Tayduganov, Andrey and Watanabe, Ryoutaro",
    title = "{Testing leptoquark models in $\bar{B} \rightarrow D^{(*)} \tau \bar{\nu}$ }",
    eprint = "1309.0301",
    archivePrefix = "arXiv",
    primaryClass = "hep-ph",
    reportNumber = "OU-HET-791, KEK-TH-1660, OU-HET 791",
    doi = "10.1103/PhysRevD.88.094012",
    journal = "Phys. Rev. D",
    volume = "88",
    number = "9",
    pages = "094012",
    year = "2013"
}

@article{london,
    author = "Bhattacharya, Bhubanjyoti and Datta, Alakabha and Kamali, Saeed and London, David",
    title = "{CP Violation in ${\bar B}^0\to D^{*+}\mu^-{\bar\nu}_\mu$}",
    eprint = "1903.02567",
    archivePrefix = "arXiv",
    primaryClass = "hep-ph",
    reportNumber = "UdeM-GPP-TH-19-269, MITP/19-016",
    doi = "10.1007/JHEP05(2019)191",
    journal = "JHEP",
    volume = "05",
    pages = "191",
    year = "2019"
}

@article{bgl,
    author = "Boyd, C. Glenn and Grinstein, Benjamin and Lebed, Richard F.",
    title = "{Precision corrections to dispersive bounds on form-factors}",
    eprint = "hep-ph/9705252",
    archivePrefix = "arXiv",
    reportNumber = "CMU-HEP-97-07A, UCSD-PTH-97-12",
    doi = "10.1103/PhysRevD.56.6895",
    journal = "Phys. Rev. D",
    volume = "56",
    pages = "6895--6911",
    year = "1997"
}

@article{gambino,
    author = "Bigi, Dante and Gambino, Paolo and Schacht, Stefan",
    title = "{$R(D^*)$, $|V_{cb}|$, and the Heavy Quark Symmetry relations between form factors}",
    eprint = "1707.09509",
    archivePrefix = "arXiv",
    primaryClass = "hep-ph",
    doi = "10.1007/JHEP11(2017)061",
    journal = "JHEP",
    volume = "11",
    pages = "061",
    year = "2017"
}

@article{pdg24,
  author       = {S. Navas and others},
  title        = {Particle Data Group},
  journal      = {Phys. Rev. D},
  volume       = {110},
  pages        = {030001},
  year         = {2024},
  note         = {to be published}
}

@article{babaranomaly1,
    author = "Lees, J. P. and others",
    collaboration = "BaBar",
    title = "{Evidence for an excess of $\bar{B} \to D^{(*)} \tau^-\bar{\nu}_\tau$ decays}",
    eprint = "1205.5442",
    archivePrefix = "arXiv",
    primaryClass = "hep-ex",
    reportNumber = "BABAR-PUB-12-012, SLAC-PUB-15028",
    doi = "10.1103/PhysRevLett.109.101802",
    journal = "Phys. Rev. Lett.",
    volume = "109",
    pages = "101802",
    year = "2012"
}

@article{babaranomaly2,
    author = "Lees, J. P. and others",
    collaboration = "BaBar",
    title = "{Measurement of an Excess of $\bar{B} \to D^{(*)}\tau^- \bar{\nu}_\tau$ Decays and Implications for Charged Higgs Bosons}",
    eprint = "1303.0571",
    archivePrefix = "arXiv",
    primaryClass = "hep-ex",
    reportNumber = "BABAR-PUB-13-001, SLAC-PUB-15381",
    doi = "10.1103/PhysRevD.88.072012",
    journal = "Phys. Rev. D",
    volume = "88",
    number = "7",
    pages = "072012",
    year = "2013"
}

@article{belleanomaly1,
    author = "Huschle, M. and others",
    collaboration = "Belle",
    title = "{Measurement of the branching ratio of $\bar{B} \to D^{(\ast)} \tau^- \bar{\nu}_\tau$ relative to $\bar{B} \to D^{(\ast)} \ell^- \bar{\nu}_\ell$ decays with hadronic tagging at Belle}",
    eprint = "1507.03233",
    archivePrefix = "arXiv",
    primaryClass = "hep-ex",
    reportNumber = "KEK-REPORT-2015-18",
    doi = "10.1103/PhysRevD.92.072014",
    journal = "Phys. Rev. D",
    volume = "92",
    number = "7",
    pages = "072014",
    year = "2015"
}

@article{belleanomaly2,
    author = "Hirose, S. and others",
    collaboration = "Belle",
    title = "{Measurement of the $\tau$ lepton polarization and $R(D^*)$ in the decay $\bar{B} \to D^* \tau^- \bar{\nu}_\tau$}",
    eprint = "1612.00529",
    archivePrefix = "arXiv",
    primaryClass = "hep-ex",
    reportNumber = "KEK-PREPRINT-2016-53, BELLE-PREPRINT-2016-14",
    doi = "10.1103/PhysRevLett.118.211801",
    journal = "Phys. Rev. Lett.",
    volume = "118",
    number = "21",
    pages = "211801",
    year = "2017"
}

@article{belleanomaly3,
    author = "Hirose, S. and others",
    collaboration = "Belle",
    title = "{Measurement of the $\tau$ lepton polarization and $R(D^*)$ in the decay $\bar{B} \rightarrow D^* \tau^- \bar{\nu}_\tau$ with one-prong hadronic $\tau$ decays at Belle}",
    eprint = "1709.00129",
    archivePrefix = "arXiv",
    primaryClass = "hep-ex",
    reportNumber = "KEK-PREPRINT-2017-26, BELLE-PREPRINT-2017-18",
    doi = "10.1103/PhysRevD.97.012004",
    journal = "Phys. Rev. D",
    volume = "97",
    number = "1",
    pages = "012004",
    year = "2018"
}

@article{belleanomaly4,
    author = "Caria, G. and others",
    collaboration = "Belle",
    title = "{Measurement of $\mathcal{R}(D)$ and $\mathcal{R}(D^*)$ with a semileptonic tagging method}",
    eprint = "1910.05864",
    archivePrefix = "arXiv",
    primaryClass = "hep-ex",
    reportNumber = "Belle-2019-18, KEK-2019-40",
    doi = "10.1103/PhysRevLett.124.161803",
    journal = "Phys. Rev. Lett.",
    volume = "124",
    number = "16",
    pages = "161803",
    year = "2020"
}

@article{lhcbanomaly1,
    collaboration = "LHCb",
    title = "{Measurement of the ratios of branching fractions $\mathcal{R}(D^{*})$ and $\mathcal{R}(D^{0})$}",
    eprint = "2302.02886",
    archivePrefix = "arXiv",
    primaryClass = "hep-ex",
    reportNumber = "LHCb-PAPER-2022-039, CERN-EP-2022-284",
    doi = "10.1103/PhysRevLett.131.111802",
    journal = "Phys. Rev. Lett.",
    volume = "131",
    pages = "111802",
    year = "2023"
}

@article{lhcbanomaly2,
    author = "Aaij, Roel and others",
    collaboration = "LHCb",
    title = "{Test of lepton flavor universality using $B^0 \rightarrow D^{*-} \tau^+ \nu_{\tau}$  decays with hadronic $\tau$ channels}",
    eprint = "2305.01463",
    archivePrefix = "arXiv",
    primaryClass = "hep-ex",
    reportNumber = "LHCb-PAPER-2022-052, CERN-EP-2023-062",
    doi = "10.1103/PhysRevD.108.012018",
    journal = "Phys. Rev. D",
    volume = "108",
    number = "1",
    pages = "012018",
    year = "2023"
}

@article{belle2anomaly1,
    author = "Kojima, Kazuki and on behalf of the Belle II Collaboration",
    collaboration = "Belle II",
    title = "{Recent Belle II results on semileptonic ! decays and tests of lepton-flavor universality
}",
    journal = "31st International Symposium on Lepton Photon Interactions at High Energies",
    year = "2023"
}

@article{hpqcd,
    author = "Harrison, Judd and Davies, Christine T. H.",
    title = "{$B \rightarrow D^*$ vector, axial-vector and tensor form factors for the full $q^2$ range from lattice QCD}",
    eprint = "2304.03137",
    archivePrefix = "arXiv",
    primaryClass = "hep-lat",
    month = "4",
    year = "2023"
}

@article{helampref2,
    author = "Be\v{c}irevi\'c, Damir and Fedele, Marco and Ni\v{s}and\v{z}i\'c, Ivan and Tayduganov, Andrey",
    title = "{Lepton Flavor Universality tests through angular observables of $\overline{B} \to D^{(\ast)} \ell \overline{\nu}$ decay modes}",
    eprint = "1907.02257",
    archivePrefix = "arXiv",
    primaryClass = "hep-ph",
    reportNumber = "DO-TH 19/10, LPT-Orsay-19-26, QFET-2019-06, TTP19-022",
    month = "7",
    year = "2019"
}

@article{hflav,
    author = "Amhis, Yasmine Sara and others",
    collaboration = "HFLAV",
    title = "{Averages of b-hadron, c-hadron, and \ensuremath{\tau}-lepton properties as of 2021}",
    eprint = "2206.07501",
    archivePrefix = "arXiv",
    primaryClass = "hep-ex",
    doi = "10.1103/PhysRevD.107.052008",
    journal = "Phys. Rev. D",
    volume = "107",
    number = "5",
    pages = "052008",
    year = "2023"
}

@book{kallenfunction,
    author = {K\"all\'en, Gunnar},
    title = "{Elementary particle physics}",
    publisher = "Addison-Wesley",
    address = "Reading, MA",
    year = "1964"
}

@article{bdltaunudatta,
    author = "Bhattacharya, Bhubanjyoti and Datta, Alakabha and Kamali, Saeed and London, David",
    title = "{A measurable angular distribution for $ \overline{B}\to {D}^{\ast }{\tau}^{-}{\overline{v}}_{\tau } $ decays}",
    eprint = "2005.03032",
    archivePrefix = "arXiv",
    primaryClass = "hep-ph",
    reportNumber = "UdeM-GPP-TH-20-277; UMISS-HEP-2020-01",
    doi = "10.1007/JHEP07(2020)194",
    journal = "JHEP",
    volume = "07",
    number = "07",
    pages = "194",
    year = "2020"
}

@article{bdlnumypaper,
author = "Kapoor, Tejhas and Huang, Zhuo-Ran and Kou, Emi",
    title = "{New physics search via angular distribution of $B \rightarrow D^* \ell {\nu}_{\ell}$ decay in the light of the new lattice data}",
    eprint = "2401.11636",
    archivePrefix = "arXiv",
    primaryClass = "hep-ph",
    doi = "10.1007/JHEP02(2025)053",
    journal = "JHEP",
    volume = "02",
    pages = "053",
    year = "2025"
}

@article{Blanke:2018yud,
    author = "Blanke, Monika and Crivellin, Andreas and de Boer, Stefan and Kitahara, Teppei and Moscati, Marta and Nierste, Ulrich and Ni\v{s}and\v{z}i\'c, Ivan",
    title = "{Impact of polarization observables and $ B_c\to \tau \nu$ on new physics explanations of the $b\to c \tau \nu$ anomaly}",
    eprint = "1811.09603",
    archivePrefix = "arXiv",
    primaryClass = "hep-ph",
    reportNumber = "PSI-PR-18-16; TTP-18-42, PSI-PR--18--16, TTP--18--42",
    doi = "10.1103/PhysRevD.99.075006",
    journal = "Phys. Rev. D",
    volume = "99",
    number = "7",
    pages = "075006",
    year = "2019"
}

@article{Fedele:2023ewe,
    author = "Fedele, Marco and Blanke, Monika and Crivellin, Andreas and Iguro, Syuhei and Nierste, Ulrich and Simula, Silvano and Vittorio, Ludovico",
    title = "{Discriminating B\textrightarrow{}D*\ensuremath{\ell}\ensuremath{\nu} form factors via polarization observables and asymmetries}",
    eprint = "2305.15457",
    archivePrefix = "arXiv",
    primaryClass = "hep-ph",
    reportNumber = "PSI-PR-23-15, PSI-PR-23-12, ZU-TH 22/23, TTP23-019, P3H-23-033, LAPTH-020/23",
    doi = "10.1103/PhysRevD.108.055037",
    journal = "Phys. Rev. D",
    volume = "108",
    number = "5",
    pages = "055037",
    year = "2023"
}

@article{Blanke:2019qrx,
    author = "Blanke, Monika and Crivellin, Andreas and Kitahara, Teppei and Moscati, Marta and Nierste, Ulrich and Ni\v{s}and\v{z}i\'c, Ivan",
    title = "{Addendum to \textquotedblleft{}Impact of polarization observables and $B_c\to \tau \nu$ on new physics explanations of the $b\to c \tau \nu$ anomaly''}",
    eprint = "1905.08253",
    archivePrefix = "arXiv",
    primaryClass = "hep-ph",
    reportNumber = "PSI-PR-19-09; ZU-TH 26/19; TTP-19-012; P3H-19-011",
    doi = "10.1103/PhysRevD.100.035035",
    month = "5",
    year = "2019",
    note = "[Addendum: Phys.Rev.D 100, 035035 (2019)]"
}

@article{Bhattacharya:2022bdk,
    author = "Bhattacharya, Bhubanjyoti and Browder, Thomas E. and Campagna, Quinn and Datta, Alakabha and Dubey, Shawn and Mukherjee, Lopamudra and Sibidanov, Alexei",
    title = "{Implications for the \ensuremath{\Delta}AFB anomaly in B\textasciimacron{}0\textrightarrow{}D*+\ensuremath{\ell}-\ensuremath{\nu}\textasciimacron{} using a new Monte~Carlo event generator}",
    eprint = "2206.11283",
    archivePrefix = "arXiv",
    primaryClass = "hep-ph",
    doi = "10.1103/PhysRevD.107.015011",
    journal = "Phys. Rev. D",
    volume = "107",
    number = "1",
    pages = "015011",
    year = "2023"
}

@article{Nierste:2008qe,
    author = "Nierste, Ulrich and Trine, Stephanie and Westhoff, Susanne",
    title = "{Charged-Higgs effects in a new B ---\ensuremath{>} D tau nu differential decay distribution}",
    eprint = "0801.4938",
    archivePrefix = "arXiv",
    primaryClass = "hep-ph",
    reportNumber = "TTP08-06, SFB-CPP-08-11",
    doi = "10.1103/PhysRevD.78.015006",
    journal = "Phys. Rev. D",
    volume = "78",
    pages = "015006",
    year = "2008"
}

@article{Alonso:2016gym,
    author = "Alonso, Rodrigo and Kobach, Andrew and Martin Camalich, Jorge",
    title = "{New physics in the kinematic distributions of $\bar B\to D^{(*)}\tau^-(\to\ell^-\bar\nu_\ell\nu_\tau)\bar\nu_\tau$}",
    eprint = "1602.07671",
    archivePrefix = "arXiv",
    primaryClass = "hep-ph",
    doi = "10.1103/PhysRevD.94.094021",
    journal = "Phys. Rev. D",
    volume = "94",
    number = "9",
    pages = "094021",
    year = "2016"
}

@article{Han:2022uho,
    author = "Han, Tao and Liao, Jiajun and Liu, Hongkai and Marfatia, Danny",
    title = "{Right-handed Dirac and Majorana neutrinos at Belle II}",
    eprint = "2207.07029",
    archivePrefix = "arXiv",
    primaryClass = "hep-ph",
    reportNumber = "PITT-PACC-2208",
    doi = "10.1007/JHEP04(2023)013",
    journal = "JHEP",
    volume = "04",
    pages = "013",
    year = "2023",
    note = "[Erratum: JHEP 09, 016 (2023)]"
}

@article{delAguila:2008ir,
    author = "del Aguila, Francisco and Bar-Shalom, Shaouly and Soni, Amarjit and Wudka, Jose",
    title = "{Heavy Majorana Neutrinos in the Effective Lagrangian Description: Application to Hadron Colliders}",
    eprint = "0806.0876",
    archivePrefix = "arXiv",
    primaryClass = "hep-ph",
    reportNumber = "UG-FT-230-08, CAFPE-100-08, UCI-TR-2008-21, BNL-HET-08-14",
    doi = "10.1016/j.physletb.2008.11.031",
    journal = "Phys. Lett. B",
    volume = "670",
    pages = "399--402",
    year = "2009"
}

@article{Bhattacharya:2015vja,
    author = "Bhattacharya, Subhaditya and Wudka, Jos{\'e}",
    title = "{Dimension-seven operators in the standard model with right handed neutrinos}",
    eprint = "1505.05264",
    archivePrefix = "arXiv",
    primaryClass = "hep-ph",
    doi = "10.1103/PhysRevD.94.055022",
    journal = "Phys. Rev. D",
    volume = "94",
    number = "5",
    pages = "055022",
    year = "2016",
    note = "[Erratum: Phys.Rev.D 95, 039904 (2017)]"
}

@article{Buchmuller:1985jz,
    author = "Buchmuller, W. and Wyler, D.",
    title = "{Effective Lagrangian Analysis of New Interactions and Flavor Conservation}",
    reportNumber = "CERN-TH-4254/85",
    doi = "10.1016/0550-3213(86)90262-2",
    journal = "Nucl. Phys. B",
    volume = "268",
    pages = "621--653",
    year = "1986"
}

@article{Grzadkowski:2010es,
    author = "Grzadkowski, B. and Iskrzynski, M. and Misiak, M. and Rosiek, J.",
    title = "{Dimension-Six Terms in the Standard Model Lagrangian}",
    eprint = "1008.4884",
    archivePrefix = "arXiv",
    primaryClass = "hep-ph",
    reportNumber = "IFT-9-2010, TTP10-35",
    doi = "10.1007/JHEP10(2010)085",
    journal = "JHEP",
    volume = "10",
    pages = "085",
    year = "2010"
}

@article{Belle:2018ezy,
    author = "Waheed, E. and others",
    collaboration = "Belle",
    title = "{Measurement of the CKM matrix element $|V_{cb}|$ from $B^0\to D^{*-}\ell^ {+} \nu_\ell$ at Belle}",
    eprint = "1809.03290",
    archivePrefix = "arXiv",
    primaryClass = "hep-ex",
    doi = "10.1103/PhysRevD.100.052007",
    journal = "Phys. Rev. D",
    volume = "100",
    number = "5",
    pages = "052007",
    year = "2019",
    note = "[Erratum: Phys.Rev.D 103, 079901 (2021)]"
}

@article{BaBar:2014omp,
    author = "Bevan, A. J. and others",
    collaboration = "BaBar, Belle",
    title = "{The Physics of the B Factories}",
    eprint = "1406.6311",
    archivePrefix = "arXiv",
    primaryClass = "hep-ex",
    reportNumber = "SLAC-PUB-15968, KEK-PREPRINT-2014-3, FERMILAB-PUB-14-262-T",
    doi = "10.1140/epjc/s10052-014-3026-9",
    journal = "Eur. Phys. J. C",
    volume = "74",
    pages = "3026",
    year = "2014"
}

@article{Colquhoun:2015oha,
    author = "Colquhoun, B. and Davies, C. T. H. and Dowdall, R. J. and Kettle, J. and Koponen, J. and Lepage, G. P. and Lytle, A. T.",
    collaboration = "HPQCD",
    title = "{B-meson decay constants: a more complete picture from full lattice QCD}",
    eprint = "1503.05762",
    archivePrefix = "arXiv",
    primaryClass = "hep-lat",
    doi = "10.1103/PhysRevD.91.114509",
    journal = "Phys. Rev. D",
    volume = "91",
    number = "11",
    pages = "114509",
    year = "2015"
}

@article{Kapoor:2024ufg,
    author = "Kapoor, Tejhas and Huang, Zhuo-Ran and Kou, Emi",
    title = "{New physics search via angular distribution of B {\textrightarrow} D$^{*}${\ensuremath{\ell}}{\ensuremath{\nu}}$_{\ell}$ decay in the light of the new lattice data}",
    eprint = "2401.11636",
    archivePrefix = "arXiv",
    primaryClass = "hep-ph",
    doi = "10.1007/JHEP02(2025)053",
    journal = "JHEP",
    volume = "02",
    pages = "053",
    year = "2025"
}

@article{Robinson:2018gza,
    author = "Robinson, Dean J. and Shakya, Bibhushan and Zupan, Jure",
    title = "{Right-handed neutrinos and R(D$^{(*)}$)}",
    eprint = "1807.04753",
    archivePrefix = "arXiv",
    primaryClass = "hep-ph",
    reportNumber = "LCTP-18-19",
    doi = "10.1007/JHEP02(2019)119",
    journal = "JHEP",
    volume = "02",
    pages = "119",
    year = "2019"
}

\end{document}